\documentclass[twocolumn,english,aps,prb]{revtex4-1}
\usepackage{color}
\usepackage{amsmath}
\usepackage{graphicx}
\usepackage{amssymb}
\usepackage{pdfpages}

\makeatletter
\@ifundefined{textcolor}{}
{%
 \definecolor{BLACK}{gray}{0}
 \definecolor{WHITE}{gray}{1}
 \definecolor{RED}{rgb}{1,0,0}
 \definecolor{GREEN}{rgb}{0,1,0}
 \definecolor{BLUE}{rgb}{0,0,1}
 \definecolor{CYAN}{cmyk}{1,0,0,0}
 \definecolor{MAGENTA}{cmyk}{0,1,0,0}
 \definecolor{YELLOW}{cmyk}{0,0,1,0}
 }

\@ifundefined{definecolor}
 {\usepackage{color}}{}
\@ifundefined{definecolor}{\@ifundefined{definecolor}
 {\usepackage{color}}{}
}{}

\newcommand{\sgn}{\operatorname{sgn}}

\newcommand{\ba}{\begin{eqnarray*}}
\newcommand{\ea}{\end{eqnarray*}}
\newcommand{\baa}{\begin{eqnarray}}
\newcommand{\eaa}{\end{eqnarray}}
\newcommand{\bea}{\begin{eqnarray}}
\newcommand{\eea}{\end{eqnarray}}
\newcommand{\be}{\begin{equation}}
\newcommand{\ee}{\end{equation}}

\newcommand{\sm}{SmB$_6$}
\newcommand{\pu}{PuB$_6$}

\DeclareMathOperator{\ii}{i\hspace{-1pt}}
\newcommand{\kdp}{$\bk\cdot\bp$ }
\newcommand{\C}{\mathcal{C}}

\newcommand{\bk}{\mathbf{k}}
\newcommand{\bp}{\mathbf{p}}

\makeatother

\usepackage{babel}

\begin{document}

\title{
Spin textures on general surfaces of the correlated topological insulator {\sm}
}

\author{Pier Paolo Baruselli}
\author{Matthias Vojta}
\affiliation{Institut f\"ur Theoretische Physik, Technische Universit\"at Dresden, 01062 Dresden, Germany}


\begin{abstract}
Employing the \kdp expansion for a family of tight-binding models for {\sm}, we analytically compute topological surface states on a generic $(lmn)$ surface. We show how the Dirac-cone spin structure depends on model ingredients and on the angle $\theta$ between the surface normal and the main crystal axes.
We apply the general theory to $(001)$, $(110)$, $(111)$, and $(210)$ surfaces, for which we provide concrete predictions for the spin pattern of surface states which we also compare with tight-binding results.
As shown in previous work, the spin pattern on a $(001)$ surface can be related to the value of mirror Chern numbers, and we explore the possibility of topological phase transitions between states with different mirror Chern numbers and the associated change of the spin structure of surface states.
Such transitions may be accessed by varying either the hybridization term in the Hamiltonian or the crystal-field splitting of the low-energy $f$ multiplets, and we compute corresponding phase diagrams.
\end{abstract}

\date{\today}


\maketitle

\section{Introduction}

The material {\sm} has triggered a large body of research activities recently, given the proposal\cite{takimoto,lu_smb6_gutz, tki_cubic} that it realizes a three-dimensional (3D) topological Kondo insulator (TKI). In general, TKIs are strongly correlated systems with $f$-electron local moments in which a topologically non-trivial bandstructure emerges at low temperature via Kondo screening.\cite{tki1} In addition, strong interactions may lead to novel phenomena not present in weakly correlated topological insulators (TIs) such as Bi$_2$Se$_3$ etc.

On the experimental front there is a growing body of results -- in particular from transport\cite{wolgast_smb6,fisk_smb6_topss, smb6_junction_prx} and photoemission studies\cite{neupane_smb6,mesot_smb6, smb6_arpes_feng,smb6_arpes_reinert,smb6_past_allen,smb6_arpes_mesot_spin} -- which appear consistent with the hypothesis the {\sm} indeed realizes a TKI. However, doubts have been raised about the proper interpretation of experimental data,\cite{sawatzky_smb6,smb6_prx_arpes,smb6_trivial} and recent quantum oscillation measurements have raised a puzzle.\cite{smb6_sebastian}

Theoretically, bandstructure calculations \cite{takimoto,lu_smb6_gutz} confirm {\sm} to be a strong TI, with $\mathbb{Z}_2$ indices $(\nu_0,\nu_1\nu_2\nu_3)=(1,111)$. In addition, it has been argued \cite{smb6_tci} that {\sm} is also a topological crystalline insulator,\cite{fu_tci} having three non-zero mirror Chern numbers (MCNs), denoted as $\C^+_{k_z=0}$, $\C^+_{k_z=\pi}$, $\C^+_{k_x=k_y}$. While Ref.~\onlinecite{smb6_tci} showed that $\C^+_{k_z=0}\!=\!2\!\mod\!4$, $\C^+_{k_z=\pi}\!=\!1\!\mod\!4$, $\C^+_{k_x=k_y}\!=1\!\mod\!2$ independent of bandstructure details,
recent work \cite{smb6_prl_io,sigrist_tki} has demonstrated that the exact values of these mirror Chern numbers depend on the details of the bandstructure,
with $\C^+_{k_z=\pi}\!=\!+1$, $\C^+_{k_z=0}\!=\!\pm 2$ and $\C^+_{k_x=k_y}\!=\!\pm 1$, giving four possible topological crystalline phases.

Given that the presence of surface states with spin-momentum locking is one of the most characteristic observable properties of TIs, a thorough characterization of these states, also for different surface orientations, is of crucial importance.
For {\sm} previous work has mostly focussed on the simplest $(001)$ surface, but the combined effects of parity invariants and MCNs promise rich physics on other surfaces, which is mostly unexplored -- with notable exceptions in Refs.~\onlinecite{smb6_tci,smb6_magnres, smb6_sebastian, smb6_stm_110,smb6_arpes_110}.

The aim of this paper is to close this gap on the theory side: We shall characterize the dispersion and spin structure of the surface states of {\sm} (and similar materials) for flat surfaces of {\em general} orientation. To this end we employ the \kdp approach, in which an effective Hamiltonian is obtained around the point $X=(0,0,\pi)$ of band inversion. This approach is particularly suitable because it allows to obtain fully analytical results for surface states induced by parity invariants in the limit of small momenta.\cite{liu_ti_model} An obstacle is that the inverted subspace of orbitals couples to the non-inverted one due to the low symmetry of a generic $(lmn)$ surface,
leading to large matrices which cannot be easily diagonalized.
To deal with this, we develop a method which allows one to approximatively compute the effective Hamiltonian on a general surface: by a careful choice of the quantization axis one can find a coordinate system in which the non-inverted subspace can be neglected. We compare the results of the \kdp approach with those from numerical tight-binding calculations and find excellent agreement.
In addition, we also discuss the possibility of topological transitions between states with different MCNs.\cite{smb6_prl_io,sigrist_tki}


\subsection{Summary of results}

\begin{figure}[tb]
\includegraphics[width=0.48\textwidth]{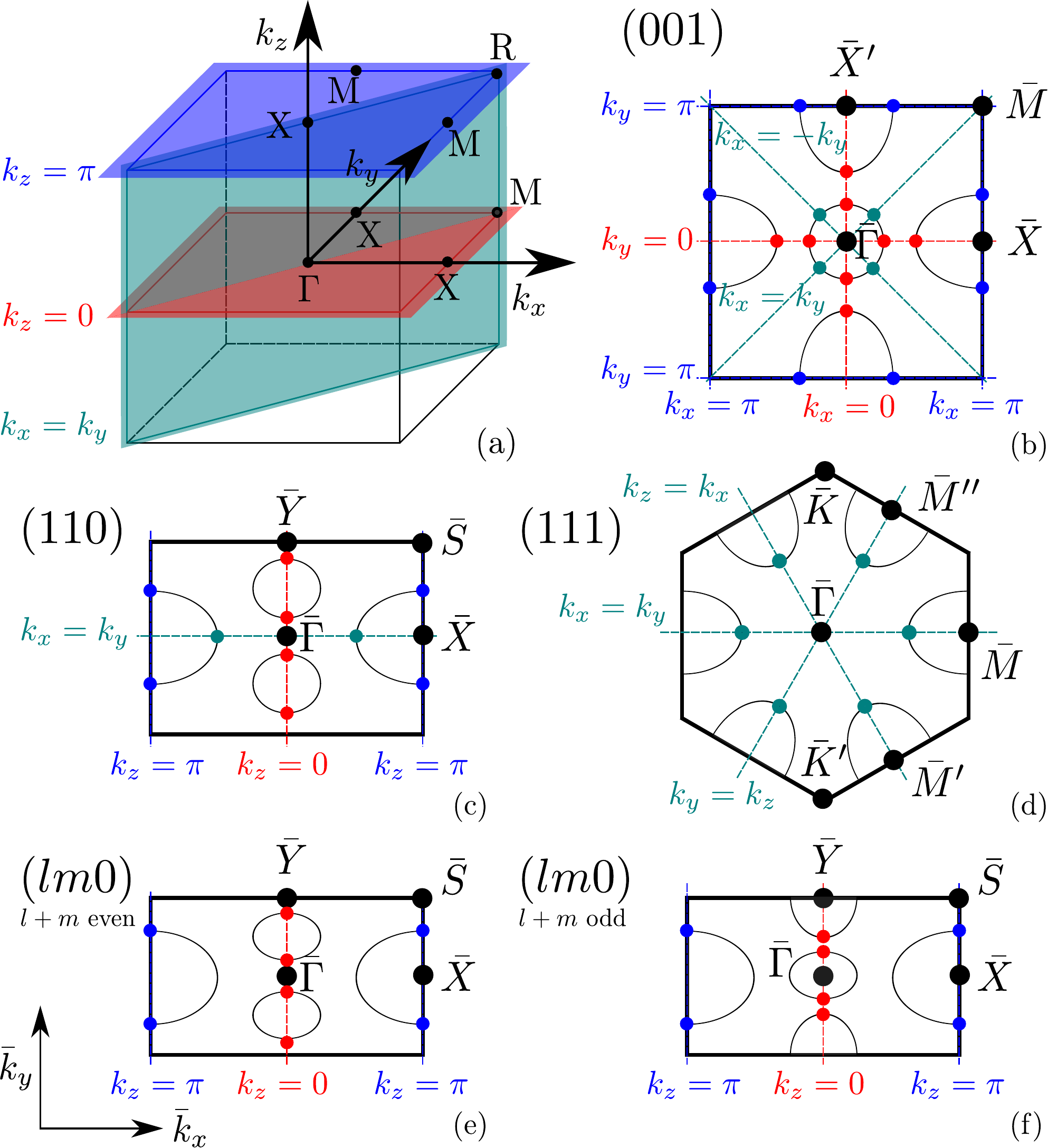}
\caption{
(a) 3D BZ and mirror planes $k_z=0$, $k_z=\pi$, $k_x=k_y$;
2D BZ for (b) $(001)$ surface (c) $(011)$ surface (d)$(111)$ surface (e) $(lm0)$ surface, $(l+m)$ even, (f) $(lm0)$ surface, $(l+m)$ odd, and their mirror planes.
On all surfaces $\bar\Gamma=(0,0)$;
on the $(001)$ $\bar X=(\pi,0)$, $\bar X'=(0,\pi)$, $\bar M=(\pi,\pi)$;
on the $(lm0)$ $\bar X=(\pi,0)$, $\bar Y=(0,\pi/\sqrt{m^2+n^2})$, $\bar S=(\pi,\pi/\sqrt{m^2+n^2})$;
on the $(111)$ $\bar K,\bar K'=(0,\pm 2\pi\sqrt{2}/3)$, $\bar M=(2\pi/\sqrt{6},0)$, $\bar M',\bar M''=(\pi/\sqrt{6},\pm \pi/\sqrt{2})$.
}\label{fig_bz}
\end{figure}

The first part of the paper is devoted to deriving the low-energy theory for {\sm} surface states for generic surface orientation.
Surface Dirac cones arise from the projection of time-reversal-invariant bulk momenta with inverted bands onto the 2D surface Brillouin zone (BZ). In {\sm} bands are inverted at the three bulk $X$ points, yielding in general three Dirac cones of surface states on a $(lmn)$ surface, Fig.~\ref{fig_bz}. To each cone we can assign an angle $\theta\equiv \arctan{\sqrt{l^2+m^2}}/{n}$ (and cyclic permutations of $l$, $m$, $n$) with $0\le\theta\le \pi/2$.
For given $\theta$ the effective surface-state Hamiltonian takes the generalized Dirac form
\bea
H^{eff}_{\theta}&=& -v_y\bar k_y \hat\sigma_x + v_x\bar k_x \hat \sigma_y + v_\perp\bar k_y \hat\sigma_z \equiv \epsilon_{\bar\bk}\vec n_{\bar\bk} \cdot \vec{\hat\sigma},
\label{heff_intro}
\eea
where $\hat\sigma_{x,y,z}$ denote pseudospin operators, $\bar k_x$ and $\bar k_y$ are momenta parallel to the surface, $\epsilon_{\bar\bk}$ is the surface-state dispersion,
and $v_x$, $v_y$, $v_\perp$ are velocities which depend on microscopic parameters and on $\theta$.
The unit vector $\vec n_{\bar\bk}$ encodes the direction of the pseudospin for states with positive energy. Its winding number\cite{smb6_prl_io,sigrist_tki} $\bar w_d\equiv \sgn(v_x v_y)$, which takes a value $\pm 1$ according to the sense of rotation of the pseudospin with respect to momentum, is related to microscopic parameters as
\begin{equation}
\bar w_d(\theta)\equiv \sgn(v_x v_y)=\sgn(w|h_1^v|\sin^2\theta+|f_1^v|\cos^2\theta),\label{wbar_d_intro}
\end{equation}
where $f_1^v$ and $h_1^v$ are hybridization parameters of the \kdp Hamiltonian at $X=(0,0,\pi)$ between $\Gamma_8^{(1)}$ and $d_{x^2-y^2}$ orbitals,
respectively, along the $z$ and $x$--$y$ directions,
while $w=\sgn(f_1^v h_1^v)=\sgn(\C^+_{k_z=0}\C^+_{k_z=\pi})$ characterizes the topological crystalline phase.
We find that the pseudospin acquires an out-of plane component when $\theta\ne 0,\pi/2$, since $v_\perp\propto \sin\theta\cos\theta$.
Formulas involving the physical spin become slightly more complicated but are not qualitatively different.






Since experimental results\cite{smb6_arpes_mesot_spin} suggest\cite{smb6_prl_io} that $w=+1$ in {\sm},
from Eq. \eqref{wbar_d_intro} we find that all Dirac cones have a positive winding number $\bar w_d=+1$; the same holds for the physical spin.
Were the phase $w=-1$ realized instead, a critical angle $\theta_c$ would exist,
marked by the vanishing of the argument of the $\sgn$ function in Eq. \eqref{wbar_d_intro},
and separating regimes with positive and negative winding number.

In the second part of the paper we demonstrate that how to access topological phase transition between phases with different $w=\pm 1$ by tuning the hybridization term or the crystal-field splitting. We illustrate this scenario by use of numerical diagonalization of tight-binding models, derive relevant phase diagrams, and discuss the observable signatures of the transitions in terms of changes of the surface states. Interestingly, the relevant models also admit phases with higher MCNs, albeit in small windows of parameters.

\subsection{Outline}

The remainder of the paper is organized as follows.
General aspects of the employed bandstructure model are described in Sec. \ref{sec:basis}.
In Section \ref{sec_kp} we derive the \kdp Hamiltonian which we use to compute surface states on the $(001)$ surface in Sections \ref{sec:gamma} and \ref{sec_x}.
In Section \ref{sec:other_surf} we apply the \kdp treatment to general surfaces.
Section \ref{sec_tpt} is devoted to topological phase transitions between phases with opposite $w$.
The paper closes with concluding remarks in Section \ref{sec:concl}.
Longer derivations are relegated to the supplemental material.


\section{Modelling}\label{sec:basis}

In this Section we provide information about the orbital basis which our model is built upon, as well as definitions of the pseudospin and of related quantitites.

\subsection{Orbital basis}\label{subsec:basis}

{\sm} crystallizes in the simple cubic (SC) structure, with a SC BZ, see Fig. \ref{fig_bz}(a).
Ab-initio calculations\cite{lu_smb6_gutz, smb6_korea,pub6} show that only bands arising from Sm orbitals are close to the Fermi energy, and, in particular, a total of 10 rare-earth orbitals per site are needed for a correct tight-binding description,\cite{prbr_io_smb6} namely the spin-degenerate $E_g$ ($d_{x^2-y^2}$ and $d_{z^2}$) quadruplet and the lowest-lying $f$-shell $j=5/2$ multiplet.
Other orbitals, including the Sm $j=7/2$ multiplet and all B$_6$ states, are excluded, since their energies are far away from the Fermi level.
The cubic crystal field splits the $j=5/2$ multiplet into a $\Gamma_8$ quadruplet and a $\Gamma_7$ doublet, which can be expressed in terms of $|j_z\rangle$ states as
$|\Gamma_8^{(1)}\pm\rangle = \sqrt\frac{5}{6}|\pm\frac{5}{2}\rangle + \sqrt\frac{1}{6}|\mp\frac{3}{2}\rangle$,
$|\Gamma_8^{(2)}\pm\rangle = |\pm\frac{1}{2}\rangle$,
$|\Gamma_7\pm\rangle = \sqrt\frac{1}{6}|\pm\frac{5}{2}\rangle -
\sqrt\frac{5}{6}|\mp\frac{3}{2}\rangle$ where $\pm$ denotes a pseudo-spin index.
For the $d$ states the effect of spin-orbit coupling will be neglected.

In what follows, we abbreviate $d^1\equiv d_{x^2-y^2}$, $d^2\equiv d_{z^2}$, $f^1\equiv \Gamma_8^{(1)}$, $f^2\equiv \Gamma_8^{(2)}$, $f^7\equiv \Gamma_7$,
so the basis of our Hamiltonian is $|d^1\uparrow\rangle$, $|d^1\downarrow\rangle$, $|d^2\uparrow\rangle$, $|d^2\downarrow\rangle$, $|f^1+\rangle$, $|f^1-\rangle$, $|f^2+\rangle$, $|f^2-\rangle$, $|f^7+\rangle$, $|f^7-\rangle$.

\subsection{Symmetries}

The time-reversal operator $T\equiv-2 \ii \hat S_y K$, where $\hat S_y$ acts on the spin variable and $K$ is the complex conjugation, acts on this basis as:
\bea
T|d^m\sigma\rangle&=&-\ii\sigma_y|d^m \sigma'\rangle ,\hspace{2pt}m=1,2;\sigma,\sigma'=\uparrow,\downarrow\\
T|f^m\sigma\rangle&=&+\ii\sigma_y|f^m \sigma'\rangle ,\hspace{2pt}m=1,2,7; \sigma,\sigma'=\pm,
\eea
where $\sigma_y\equiv(\sigma_y)_{\sigma\sigma'}$ acts on the subspace spanned by $\sigma,\sigma'=\uparrow,\downarrow$ for $d$ states, or by $\sigma,\sigma'=+,-$ for $f$ states.


Mirror-symmetry operators $M_l\equiv P C_2(l)$,
where $P$ is the inversion and $C_2(l)$ is a two-fold rotation around axis $l$, act as \cite{smb6_prl_io}:
\bea
M_l |d^m \sigma \rangle&=&-\ii \sigma_l |d^m \sigma' \rangle, \hspace{5pt} m=1,2,\label{mld}\\
M_l |f^m \sigma \rangle&=&+ \ii\sigma_l |f^m \sigma' \rangle, \hspace{5pt} m=1,2,7,\label{mlf}
\eea
where $l=x,y,z$; moreover
\bea
M_{x \pm y}|d^1 \sigma \rangle&=&+\ii\sigma_{x\pm y}|d^1 \sigma' \rangle,\hspace{10pt}\\
M_{x \pm y}|d^2 \sigma \rangle&=&-\ii\sigma_{x\pm y}|d^2 \sigma' \rangle,\\
M_{x \pm y}|f^m \sigma \rangle&=&-\ii\sigma_{x\pm y}|f^m \sigma' \rangle,\hspace{5pt}m=1,7,\\
M_{x \pm y}|f^m \sigma \rangle&=&+\ii\sigma_{x\pm y}|f^m \sigma' \rangle,\hspace{5pt}m=2,
\eea
where $\sigma_{x\pm y}=(\sigma_x\pm \sigma_y)/{\sqrt{2}}$.

Mirror Chern numbers are defined as:\cite{smb6_tci,smb6_prl_io,sigrist_tki}
\be\label{chern_number}
\C^\pm_{\overline{BZ}}=\frac{\ii}{2\pi}\sum_{a,b=1}^2\epsilon_{ab}\sum_{n=1}^N\int_{\overline{BZ}}d^2\bk \langle \partial_a u_n^\pm(\bk)| \partial_b u_n^\pm(\bk)\rangle,
\ee
with $M|u_n^\pm(\bk)\rangle=\pm \ii|u_n^\pm(\bk)\rangle$, and where $\bk$ lies in the plane $\overline{BZ}$  which is invariant with respect to the symmetry operator $M$
($M$=$M_z$ when $\overline{BZ}$ is $k_z=0$ or $k_z=\pi$, $M=M_{x-y}$ when $\overline{BZ}$ is $k_x=k_y$),
and we sum over all $N$ occupied bands.
We note that $\C^+_{\overline{BZ}}+\C^-_{\overline{BZ}}=0$ and, by cubic symmetry, $\C^+_{k_z=0}=\C^+_{k_x=0}=\C^+_{k_y=0}$ etc.

As remarked, four phases are possible, each characterized by the triplet of numbers $(\C^+_{k_z=0}, \C^+_{k_z=\pi}, \C^+_{k_x=k_y})$: $(\pm 2 ,+1,\pm 1)$.
Since the sign of MCNs fix the mirror-symmetry eigenvalues of surface states,\cite{kane_bisb} each phase has different surface-state properties as shown in Refs.~\onlinecite{smb6_prl_io,sigrist_tki}, see Section~\ref{sec:other_surf} below.
For {\sm}, the results of Ref. \onlinecite{smb6_prl_io} can be written in a concise form using $v$ and $w$:
\bea
v&\equiv&\sgn(\C^+_{k_z=0}\C^+_{k_x=k_y}),\label{vdef}\\
w&\equiv&\sgn(\C^+_{k_z=0}\C^+_{k_z=\pi}).\label{wdef}
\eea
For $(001)$ surface states $|\phi^+(\bk)\rangle$ of positive energies, i.e. above the Dirac energy, on the $\bar\Gamma$ cone the following relations are then satisfied:
\bea
M_y|\phi^+_{\bar\Gamma}(|k_x|,0)\rangle&=&-\ii w|\phi^+_{\bar\Gamma}(|k_x|,0)\rangle,\label{mg1}\\
M_{x-y}|\phi^+_{\bar\Gamma}(k_x\!=\!k_y\!>\!0)\rangle&=&+\ii vw|\phi^+_{\bar\Gamma}(k_x\!=\!k_y\!>\!0)\rangle,\label{mg2}\\
M_x|\phi^+_{\bar\Gamma}(0,|k_y|)\rangle&=&+\ii w|\phi^+_{\bar\Gamma}(0,|k_y|)\rangle,\label{mg3}
\eea
while on the $\bar X$ cone:
\bea
M_x|\phi^{+}_{\bar X}(0,|k_y|)\rangle&=&+\ii|\phi^{+}_{\bar X}(0,|k_y|)\rangle,\label{mx1}\\
M_y|\phi^{+}_{\bar X}(-|k_x|,0)\rangle&=&+\ii w|\phi^{+}_{\bar X}(-|k_x|,0)\rangle,\label{mx2}
\eea
and on the $\bar X'$ one:
\bea
M_y|\phi^{+}_{\bar X'}(|k_x|,0)\rangle&=&-\ii|\phi^{+}_{\bar X'}(|k_x|, 0)\rangle,\label{mxp1}\\
M_x|\phi^{+}_{\bar X'}(0, -|k_y|)\rangle&=&-\ii w|\phi^{+}_{\bar X'}(0, -|k_y|)\rangle.\label{mxp2}
\eea


\subsection{Relation between spin and pseudospin}

We start with remarks on notation.
With $\sigma_i$ ($i=x,y,z,0$) we denote the standard Pauli matrices for (pseudo)spin indices, i.e., for $d$ states they act in the space of $\uparrow$, $\downarrow$, while for $f$ states they act in the space of $+$, $-$.
%
%
With $\hat s_i$ ($i=x,y,z,0$) we denote Pauli matrices for operators acting into an arbitrary two-dimensional space, which is typically the space spanned by the doublet of surface states at $k_\parallel=0$.
With $\hat S_i$ ($i=x,y,z$)
we denote physical spin operators, with separate contributions from the $d$ ($ \hat S_i^d$) and $f$ shells ($ \hat S_i^f$).
With $\hat \sigma_i$ ($i=x,y,z$) we denote pseudospin operators, still with separate contributions from the $d$ ($\hat \sigma_i^d$) and $f$ shells ($\hat \sigma_i^f$), and defined as follows.

For $d$ electrons, with spin-orbit coupling neglected, we take the pseudospin to coincide with the physical spin, apart from a factor $2$: $2 \hat S_i^d=\hat\sigma_i^d$.
For $d_{x^2-y^2}$ states we have in the $|d^1\uparrow\rangle$, $|d^1\downarrow\rangle$ basis:
\be
\hat\sigma_x^d=\left(\!
\begin{array}{cc}
0&1\\
1&0
  \end{array}\!\right),
\hspace{10pt}
\hat\sigma_y^d=\left(\!
\begin{array}{cc}
0&-\ii\\
\ii&0
  \end{array}\!\right),
\hspace{10pt}
\hat\sigma_z^d=\left(\!
\begin{array}{cc}
1&0\\
0&-1
  \end{array}\!\right).
\ee

For $f$ electrons, the physical spin in the $\Gamma_8^{(1)}$ - $\Gamma_8^{(2)}$ - $\Gamma_7$ basis is given in the supplement.\cite{suppl_tci_long} If we restrict to $\Gamma_8^{(1)}$ states, we have in the $|f^1+\rangle$, $|f^1-\rangle$ basis:
\be
2 \hat S_x^f=\frac{5}{21}\left(\!
\begin{array}{cc}
0&-1\\
-1&0
  \end{array}\!\right),
\hspace{10pt}
2 \hat S_y^f=\frac{5}{21}\left(\!
\begin{array}{cc}
0&\ii\\
-\ii&0
  \end{array}\!\right),
\nonumber
\ee
\be
2 \hat S_z^f=\frac{11}{21}\left(\!
\begin{array}{cc}
-1&0\\
0&1
  \end{array}\!\right).
\label{sprefac}
\ee
In the same basis we take as pseudospin:
\be
\hat\sigma_x^f=\left(\!
\begin{array}{cc}
0&-1\\
-1&0
  \end{array}\!\right),
\hspace{10pt}
\hat\sigma_y^f=\left(\!
\begin{array}{cc}
0&\ii\\
-\ii&0
  \end{array}\!\right),
\hspace{10pt}
\hat\sigma_z^f=\left(\!
\begin{array}{cc}
-1&0\\
0&1
  \end{array}\!\right),
\label{sigprefac}
\ee
that is, we get rid of the prefactors with respect to the real spin. As a consequence, for $\Gamma_8^{(1)}$ states, the spin is parallel to the pseudospin with direction-dependent coefficients $5/21$ or $11/21$.
For $\Gamma_7$ states Eq.~\eqref{sigprefac} continues to apply, whereas all the prefactors in Eq.~\eqref{sprefac} become $-5/21$, such that spin and pseudospin are antiparallel.
We note that the minus sign appearing in our definition of the pseudospin with respect to the standard Pauli matrices is linked to the different behaviour of $d$ and $f$ states under mirror operators,
see Eqs. \eqref{mld}, \eqref{mlf}. Indeed, we can write $M_l=-\ii\hat\sigma_l$, and, as shown in Ref. \onlinecite{sigrist_tki}, once the pseudospin has this defined mirror-symmetry property,
MCNs fix its texture on surface states.

We also stress that the expectation value of the pseudospin $|\langle \vec{\hat\sigma} \rangle |$ is always normalized to unity on surface states, while that of the physical spin $\vec  {\hat S}$ has no definite normalization, but $|\langle \vec  {\hat S} \rangle |\le 1/2 $ holds, with the equal sign for pure $d$ states. Hence, it is often useful to use pseudospin rather than spin operators. In the course of the paper we will always refer to both quantities, bearing in mind that experiments must be compared to results for the physical spin.

It is also possible to take into account the expectation value of the orbital angular momentum; as shown in the supplement,\cite{suppl_tci_long} this is zero in the $d$ shell, and equal to $(-8)$ times the spin expectation value in the $f$ shell. We will not refer to this quantity in the rest of the paper.


\section{\kdp Hamiltonian}\label{sec_kp}

In this section we derive the \kdp Hamiltonian by expanding the tight-binding Hamiltonian of Ref. \onlinecite{prbr_io_smb6} around $X=(0,0,\pi)$. We measure bulk momenta $\bk$ relative to $X$ and keep all first-order terms in $k_x$, $k_y$, $k_z$ and all mixed terms up to second order.
Even though the lattice is cubic, the \kdp Hamiltonian has tetragonal symmetry, as dictated by the momentum-space location of $X$.

We will exclusively work in a renormalized single-particle picture, based on the assumption that many-body effects can be captured by proper renormalizations of single-particle terms, in particular the $f$ kinetic energy and hybridization.\cite{hewson} For band structures, this assumption has been confirmed by many-body numerical techniques.\cite{assaad_review,smb6_korea2}

\subsection{Full orbital basis}

In the 10-dimensional basis $|d^1\uparrow\rangle$, $|d^1\downarrow\rangle$, $|d^2\uparrow\rangle$, $|d^2\downarrow\rangle$, $|f^1+\rangle$, $|f^1-\rangle$, $|f^2+\rangle$, $|f^2-\rangle$, $|f^7+\rangle$, $|f^7-\rangle$ of bulk Bloch states the result is as follows:
\onecolumngrid
\be
H=
\left(\!
\begin{array}{cccc|cccccc}
{\epsilon_1^d(\bk)} & 0 & 0 & 0 & -\ii  V {f_1^v} {k_z} & -\ii V{h_1^v} k_- & 0 & -\ii V{h_{12}^v} k_+ & -\ii V{f_7^v} {k_z} & -\ii V{h_7^v} k_- \\
 0 & {\epsilon_1^d(\bk)} & 0 & 0 & -\ii V{h_1^v} k_+ & \ii V{f_1^v} {k_z} & -\ii V{h_{12}^v} k_- & 0 & -\ii V{h_7^v} k_+ & \ii V{f_7^v} {k_z} \\
 0 & 0 & {\epsilon_2^d(\bk)} & 0 & 0 & -\ii V{h_{21}^v} k_+ & -\ii V{f_2^v} {k_z} & -\ii V{h_2^v} k_- & 0 & -\ii Vh_{72}^v k_+ \\
 0 & 0 & 0 & {\epsilon_2^d(\bk)} & -\ii V{h_{21}^v} k_- & 0 & -\ii V{h_2^v} k_+ & \ii V{f_2^v} {k_z} & -\ii Vh_{72}^vk_-  & 0 \\\hline
 \ii V{f_1^v} {k_z} & \ii V{h_1^v} k_- & 0 & \ii V{h_{21}^v} k_+ & {\epsilon_1^f(\bk)} & 0 & 0 & 0 & {m_{78}} & 0 \\
 \ii V{h_1^v} k_+ & -\ii V{f_1^v} {k_z} & \ii V{h_{21}^v} k_- & 0 & 0 & {\epsilon_1^f(\bk)} & 0 & 0 & 0 & {m_{78}} \\
 0 & \ii V{h_{12}^v} k_+ & \ii V{f_2^v} {k_z} & \ii V{h_2^v} k_- & 0 & 0 & {\epsilon_2^f(\bk)} & 0 & 0 & 0 \\
 \ii V{h_{12}^v} k_- & 0 & \ii V{h_2^v} k_+ & -\ii V{f_2^v} {k_z} & 0 & 0 & 0 & {\epsilon_2^f(\bk)} & 0 & 0 \\
 \ii V{f_7} {k_z} & \ii V{h_7^v} k_- & 0 & \ii Vh_{72} k_+  & {m_{78}} & 0 & 0 & 0 & {\epsilon_7^f(\bk)} & 0 \\
 \ii V{h_7^v} k_+ & -\ii V{f_7} {k_z} & \ii Vh_{72} k_-  & 0 & 0 & {m_{78}} & 0 & 0 & 0 & {\epsilon_7^f(\bk)}
\end{array}\label{hkdotp}
\!\right)
\ee
\twocolumngrid
\noindent
with $k_\pm\equiv k_x\pm \ii k_y$.
Kinetic-energy terms are diagonal, and given by:
\bea
\epsilon_i^d(\bk)&=&\epsilon^d_i-t_d[k_z^2 g_i^d+k_\parallel^2 l_i^d], \hspace{5pt} i=1,2,\label{epsd}\\
\epsilon_i^f(\bk)&=&\epsilon^f_i-t_f[k_z^2 g_i^f+k_\parallel^2 l_i^f], \hspace{5pt} i=1,2,7,\label{epsf}
\eea
with $k^2_\parallel\equiv k_x^2+k_y^2$,
$t_d>0$ (electron-like), $t_f<0$ (hole-like).
The $g_i$ and $l_i$ represent combinations of tight-binding parameters and are defined in the Appendix.
At zeroth order we have a coupling $m_{78}$ between $\Gamma_8^{(1)}$ and $\Gamma_7$:
\be
m_{78}=-4\eta_{78}^{f1}+8\eta_{x7}^{f2}>0.\label{m_78}
\ee
Hybridization terms are non-diagonal; we will need the following ones:
\bea
f_1^v&=& 2\eta_x^{v1}+2\eta_x^{v2}+6\eta_z^{v2}>0,\label{f1etav}\\
f_2^v&=& 2\eta_z^{v1}+6\eta_x^{v2}+2\eta_z^{v2}<0,\label{f2etav}\\
f_7^v&=&2\eta_7^{v1}-4\sqrt{3}\eta_7^{v2}-4\eta_{x7}^{v2}<0\label{f7etav},\\
h_1^v&=& -\frac{1}{2}\eta_x^{v1}-\frac{3}{2}\eta_z^{v1}-3\eta_x^{v2}+3\eta_z^{v2}>0,\label{h1etav}\\
h_2^v&=& -\frac{3}{2}\eta_x^{v1}-\frac{1}{2}\eta_z^{v1}+3\eta_x^{v2}-3\eta_z^{v2}<0,\label{h2etav}\\
h_7^v&=&\eta_7^{v1}+2\sqrt{3}\eta_7^{v2}-6\eta_{x7}^{v2}>0.\label{h7etav}
\eea
The explicit form of other terms appearing in the Hamiltonian \eqref{hkdotp},
which are not needed in what follows, is given in the Appendix,
together with the meaning of different tight-binding parameters $\eta$.
We also provide numerical values for some of these parameters, extracted from density-functional-theory (DFT) calculations of Refs. \onlinecite{pub6,prbr_io_smb6} for \pu, which has a bandstructure very similar to \sm.
In the rest of the paper we will not rely on the exact parameter values, but will often make use of (relative) signs, as indicated in Eqs. \eqref{m_78}-\eqref{h7etav}.

\subsection{Reduced orbital basis}
\label{sec:reduced}

To enable analytical calculations, we will need to work with matrices of dimension (at most) $4\times 4$. This requires a further basis reduction (and associated approximations), and we discuss different possible choices in turn. The best choice will depend on microscopic parameters, in particular the crystal-field splitting between $\Gamma_7$ and $\Gamma_8$ multiplets.

\begin{figure}[tb]
\includegraphics[width=0.48\textwidth]{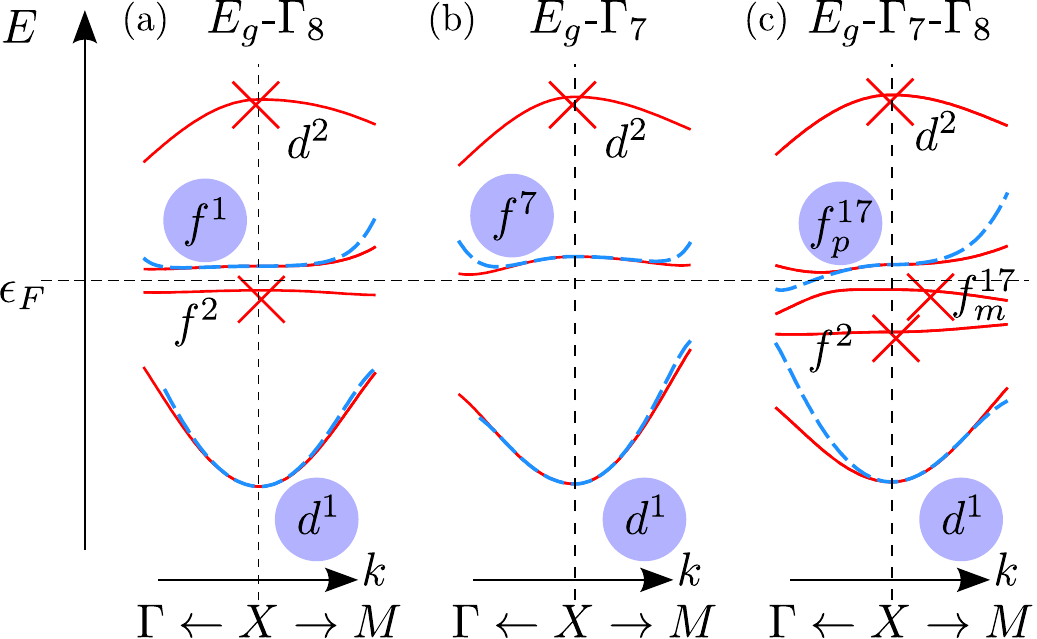}
\caption{
Schematic bulk bandstructure around $X$ illustrating the approximation schemes; shown are tight-binding dispersions (solid red) and the \kdp approximation with 4 states (dotted blue), for details see text.
(a) Out of $E_g$ and $\Gamma_8$ (8 total states), we keep $d^1\equiv d_{x^2-y^2}$ and $f^1\equiv \Gamma_8^{(1)}$, and neglect $d^2\equiv d_{z^2}$ and $f^2\equiv \Gamma_8^{(2)}$, which are not inverted.
(b) Out of $E_g$ and $\Gamma_7$ (6 total states), we keep $d^1$ and $f^7 \equiv \Gamma_7$, and neglect $d^2$.
(c) Out of $E_g$, $\Gamma_7$ and $\Gamma_8$ (10 total states), we keep $d^1$ and the linear combination $f^{17}_p$ of $f^1$ and $f^7$ of higher energy,
and neglect states $d^2$, $f^2$, and the linear combination $f^{17}_m$
of lower energy, which is inverted, but remains below the Fermi energy $\epsilon_F$.
All states are twice (pseudo)spin degenerate.
}\label{fig_kdp}
\end{figure}

We may either retain $\Gamma_8$ states and work in subspace~1, spanned by $|d^1\uparrow\rangle$, $|d^1\downarrow\rangle$, $|f^1+\rangle$, $|f^1-\rangle$ states, see Fig. \ref{fig_kdp}(a). The Hamiltonian becomes:
\be
H^{(1)}=
\left(
\begin{array}{cc|cc}
 {\epsilon_1^d(\bk)} & 0  & -\ii V {f_1^v} {k_z} & -\ii V{h_1^v} k_-  \\
 0 & {\epsilon_1^d(\bk)} &  -\ii V{h_1^v} k_+ & \ii V{f_1^v} {k_z}  \\\hline
 \ii V{f_1^v} {k_z} & \ii V{h_1^v} k_- &  {\epsilon_1^f(\bk)} & 0  \\
 \ii V{h_1^v} k_+ & -\ii V{f_1^v} {k_z} &  0 & {\epsilon_1^f(\bk)}
 \end{array}\label{hkdotp_1}
\right).
\ee
Alternatively, we may retain $\Gamma_7$ states yielding subspace~1', spanned by $|d^1\uparrow\rangle$, $|d^1\downarrow\rangle$, $|f^7+\rangle$, $|f^7-\rangle$ , see Fig. \ref{fig_kdp}(b). The Hamiltonian is:
\be
H^{(1)'}=
\left(
\begin{array}{cc|cc}
 {\epsilon_1^d(\bk)} & 0  & -\ii V {f_7^v} {k_z} & -\ii V{h_7^v} k_-  \\
 0 & {\epsilon_1^d(\bk)} &  -\ii V{h_7^v} k_+ & \ii V{f_7^v} {k_z}  \\\hline
 \ii V{f_7^v} {k_z} & \ii V{h_7^v} k_- &  {\epsilon_7^f(\bk)} & 0  \\
 \ii V{h_7^v} k_+ & -\ii V{f_7^v} {k_z} &  0 & {\epsilon_7^f(\bk)}
 \end{array}\label{hkdotp_7}
\right).
\ee
When using both $\Gamma_7$ and $\Gamma_8$ states, we will just retain the linear combination $|f^{17}_p\pm\rangle$ of $|f^{1}\pm\rangle$ and $|f^{7}\pm\rangle$ of higher energy, giving:
\be
H^{(1)''}=
\left(
\begin{array}{cc|cc}
 {\epsilon_1^d(\bk)} & 0  & -\ii V {f_{17}^v} {k_z} & -\ii V{h_{17}^v} k_-  \\
 0 & {\epsilon_1^d(\bk)} &  -\ii V{h_{17}^v} k_+ & \ii V{f_{17}^v} {k_z}  \\\hline
 \ii V{f_{17}^v} {k_z} & \ii V{h_{17}^v} k_- &  {\epsilon_{17}^f(\bk)} & 0  \\
 \ii V{h_{17}^v} k_+ & -\ii V{f_{17}^v} {k_z} &  0 & {\epsilon_{17}^f(\bk)}
 \end{array}\label{hkdotp_17}
\right)
\ee
with explicit expressions of $f_{17}^v$, $h_{17}^v$, ${\epsilon_{17}^f(\bk)}$ given in Section \ref{ssec_g78}; see Fig. \ref{fig_kdp}(c).

Matrices \eqref{hkdotp_1}, \eqref{hkdotp_7}, \eqref{hkdotp_17} have the same cylindrical symmetry and the same form, but different parameters. In the kinetic energy sector they are all similar, in that parameters $g_1^d$, $l_{1,7}^d$ have always a negative sign, indicating the curvature of the band at $X$: upward for $d$ states, since $(-t_d)$ is negative, and downward for $f$ states, since $(-t_f)$ is positive, see Eqs. \eqref{epsd}, \eqref{epsf}.
However, they differ in the hybridization sector, since parameters $f_i^v$ and $h_i^v$ can have all different signs, that we will later link to the topological properties of {\sm} and to the spin pattern of surface states.

It is also sometimes useful to consider the Hamiltonian in subspace~2, spanned by $|d^2\uparrow\rangle$, $|d^2\downarrow\rangle$, $|f^2+\rangle$, $|f^2-\rangle$:
\be
H^{(2)}=
\left(
\begin{array}{cc|cc}
 {\epsilon_2^d(\bk)} & 0  & -\ii V {f_2^v} {k_z} & -\ii V{h_2^v} k_-  \\
 0 & {\epsilon_2^d(\bk)} &  -\ii V{h_2^v} k_+ & \ii V{f_2^v} {k_z}  \\\hline
 \ii V{f_2^v} {k_z} & \ii V{h_2^v} k_- &  {\epsilon_2^f(\bk)} & 0  \\
 \ii V{h_2^v} k_+ & -\ii V{f_2^v} {k_z} &  0 & {\epsilon_2^f(\bk)}
 \end{array}\label{hkdotp_2}
\right).
\ee
Subspace~2 is not inverted, hence not relevant for topological properties.
However, Hamiltonian parameters can be tuned\cite{tki_cubic} to achieve band inversion in this subspace instead of subspace~1, thus it is instructive to see how this (experimentally irrelevant) situation compares to the others.

We note that subspaces~1 and 1' together, including both $\Gamma_7$ and $\Gamma_8$ states, form a  six-dimensional space corresponding to $j_z=\pm 3/2$, while subspace~2 is four-dimensional and corresponds to $j_z=\pm 1/2$.

\subsection{Relation to earlier work}

A Hamiltonian similar to Eq. \eqref{hkdotp} was introduced in Ref.~\onlinecite{yu_smb6_qpi}, with a few differences.
First, the $d_{z^2}$ orbital was neglected, reducing the Hilbert space to 8 orbitals -- this is a meaningful approximation since this orbital is far from the Fermi energy and not involved in the band inversion.
Second, instead of working with $\Gamma_7$ and $\Gamma_8$ states, the authors used eigenstates of the $j_z$ operator, which is just a basis rotation.
Finally, the spirit is different: here we derive the Hamiltonian from a tight-binding model constructed from DFT results, 
and we stress how different parameters affect the effective Hamiltonian, rather than taking numerical values directly from DFT.
This leads to a better understanding of how different tight-binding terms affect the topological properties of {\sm}.
Taking into account these differences, our approach is compatible with that of Ref.~\onlinecite{yu_smb6_qpi}, even though we reach a different conclusion about the spin structure on the $(001)$ surface,
most likely due to quantitative difference in the numerical value of parameters; see Section \ref{sec_x}.

An approach similar to ours is followed in Ref. \onlinecite{dzero_pert}, where, however, the starting tight-binding model is different;
we will return to this point in Section \ref{sec:gamma}.
In addition, the authors do not focus on the spin structure of surface states, which instead is our primary goal.

\section{Expansion around $\bar\Gamma$ for $(001)$ surface}\label{sec:gamma}

In this Section we demonstrate the usage of the \kdp Hamiltonian of Section~\ref{sec_kp} to compute the effective surface-state Hamiltonian for the Dirac cone at $\bar\Gamma$ of a $(001)$ surface. The method consists of finding surface states in the form of a Kramers doublet exactly at $\bar\Gamma$, i.e. setting $k_{x,y}=0$, and then expanding in $k_{x,y}$ to build an effective Hamiltonian for finite $k_{x,y}$ onto the $k_{x,y}=0$ basis. This is a standard approach in the theory of weakly correlated TIs, see e.g. Ref.~\onlinecite{liu_ti_model}.
We will always assume ideal surfaces, i.e., a confining potential which is zero inside the crystal and infinite outside.\cite{suppl_tci_long}

\subsection{$E_g$ - $\Gamma_8$ basis}
First we concentrate on $E_g$ and $\Gamma_8$ states, i.e., we neglect the $\Gamma_7$ states in columns and rows 9 and 10 of Eq. \eqref{hkdotp}.
Setting $k_x=k_y=0$ defines an unperturbed Hamiltonian $H_0$. It splits into four $2\times 2$ blocks, that we call
$H_0^{+(1)}$ with basis $|d^1\uparrow\rangle$, $|f^1+\rangle$,
$H_0^{-(1)}$ with basis $|d^1\downarrow\rangle$, $|f^1-\rangle$,
$H_0^{+(2)}$ with basis $|d^2\uparrow\rangle$, $|f^2+\rangle$,
$H_0^{-(2)}$ with basis $|d^2\downarrow\rangle$, $|f^2-\rangle$.
The fact that there are no terms connecting $H_0^{+/-(1)}$ to $H_0^{+/-(2)}$ 
is a direct consequence of tetragonal symmetry along the $\Gamma$--$X$ direction, and will always be true, even keeping more terms in the Hamiltonian.
After finding surface states for $H_0^{(1)}$, we take the Hamiltonian $H_P\equiv H-H_0$, containing all terms in $k_{x,y}$, as a perturbation to get the effective Hamiltonian as a function of $k_{x,y}$.

For the unperturbed Hamiltonian $H_0^{(1,2)}$ blocks, within our model, we get:
\begin{align}\label{h1pp}
H_0^{+(1,2)}=
\left( \begin{array}{ll}
\epsilon_{1,2}^d-t_dk_z^2 g_{1,2}^d & -\ii Vk_z f_{1,2}^v\\
\ii Vk_z f_{1,2}^v &\epsilon_{1,2}^f-t_fk_z^2 g_{1,2}^f
  \end{array}\right)&,
\end{align}
while the other two blocks $H_{0}^{-(1,2)}$ are obtained by substituting $V$ with $(-V)$.
As shown in the supplement,\cite{suppl_tci_long} it is possible to analytically compute surface states at $k_{x,y}=0$. Those exist only in the subspace where band inversion is achieved,
that is in subspace~1 as defined in Section~\ref{sec:reduced}, since $\epsilon_1^d<\epsilon_1^f$ and $\epsilon_2^d>\epsilon_2^f$. After tracing out the $z$ coordinate of the wavefunction they read:
\bea
|\psi_+\rangle&=&\alpha|d^1\uparrow\rangle+\beta|f^1+\rangle,\nonumber\\
|\psi_-\rangle&=&\alpha|d^1\downarrow\rangle-\beta|f^1-\rangle,\label{psipm}
\eea
where the ket vectors are now Bloch states carrying a two-dimensional surface momentum, and
\bea
\alpha&=&\sqrt{\frac{t_fg_1^f}{t_fg_1^f-t_dg_1^d}},\label{alpha}\\
\beta&=&-\sgn(Vf_1^v)\sqrt{\frac{t_dg_1^d}{t_dg_1^d-t_fg_1^f}}.\label{beta}
\eea
We stress that $\alpha$ and $\beta$ can both be chosen real, with $\alpha^2+\beta^2=1$, and $\beta^2/\alpha^2\sim |t_d/t_f|\gg 1$, hence surface states have mainly $f$ character.

After finding surface states at $k_{x,y}=0$, we can consider the perturbing Hamiltonian\cite{suppl_tci_long} $H_P$
and build an effective $2\times 2$ Hamiltonian via
\begin{align}
H^{eff}_{\bar\Gamma}=
\left( \begin{array}{ll}
\langle \psi_+|H_P|\psi_+\rangle & \langle \psi_+|H_P|\psi_-\rangle\\
\langle \psi_-|H_P|\psi_+\rangle & \langle \psi_-|H_P|\psi_-\rangle
  \end{array}\right),\label{hpheff}&
\end{align}
to find:
\bea
H^{eff}_{\bar\Gamma}&=&|v_0|w'(k_x\hat s_y-k_y\hat s_x),\label{heffvw}\\
|v_0|&\equiv&2|Vh_1^v|\frac{\sqrt{-t_dt_fg_1^dg_1^f}}{t_fg_1^f-t_dg_1^d},\label{v_0}\\
w'&\equiv&-\sgn(\alpha\beta V h_1^v)=\sgn(f_1^vh_1^v)=\pm 1,\label{w}
\eea
where $\hat s_x$, $\hat s_y$, $\hat s_z$ are Pauli matrices in the $|\psi_\pm\rangle$ basis.

Evaluating the spin expectation values (SEV) in the basis \eqref{psipm} yields:
\bea
2\langle \vec  {\hat S} \rangle = 2\langle ( \hat S_x,  \hat S_y,  \hat S_z) \rangle=\left( \gamma_5^+\hat s_x, \gamma_5^+\hat s_y, \gamma_{11}^-\hat s_z\right),
\eea
where we introduce
\be
\gamma_5^\pm=\alpha^2\pm \frac{5}{21}\beta^2, \hspace{5pt} \gamma_{11}^\pm=\alpha^2\pm \frac{11}{21}\beta^2,\label{gamma511}
\ee
with the properties $\gamma_5^+,\gamma_{11}^+>0 $ and $\gamma_5^-,\gamma_{11}^-<0$ since $\alpha^2\ll \beta^2$.
For the pseudospin we find:
\bea
\langle \vec {\hat\sigma} \rangle = \langle (\hat\sigma_x, \hat\sigma_y, \hat\sigma_z) \rangle=\left( \hat s_x, \hat s_y,\gamma^- \hat s_z\right)\label{psbasis_gamma},
\eea
with $\gamma^-=\alpha^2-\beta^2<0$.
This shows that
we can substitute $\hat s_x$ and $\hat s_y$ in Eq. \eqref{heffvw} exactly with pseudospin operators $\hat\sigma_x$ and $\hat\sigma_y$,
or approximatively, with spin operators $ \hat S_x$ and $ \hat S_y$.
This justifies what we did in Ref. \onlinecite{smb6_prl_io}, where we wrote the effective Hamiltonian directly in term of spin operators 
once knowing the SEV from mirror-symmetry eigenvalues.

We can now easily diagonalize Eq. \eqref{heffvw};
the state $|\phi^+(\bk)\rangle$ at a given momentum $\bk=(k_x,k_y)$ with positive energy $\epsilon_k=+|v_0|k$ ($k=|\bk|$)
has the SEV:
\bea
2\langle \phi^+(\bk)| \vec  {\hat S} |\phi^+(\bk)\rangle=w' \gamma_5^+ (-\sin\theta_\bk,\cos\theta_\bk,0), \label{sev_gamma}
\eea
with $\cos\theta_\bk=k_x/k$, $\sin\theta_\bk=k_y/k$, and the pseudospin:
\bea
\langle \phi^+(\bk)| \vec{\hat\sigma} |\phi^+(\bk)\rangle=w'(-\sin\theta_\bk,\cos\theta_\bk,0). \label{psev_gamma}
\eea
These equations show that  $w'$ dictates the sense of rotation of the SEV (i.e. the chirality) and of the pseudospin.
Moreover $|\phi^+(\bk)\rangle$ is such that:
\bea
M_y|\phi^+(|k_x|,0)\rangle&=&-\ii w'|\phi^+(|k_x|,0)\rangle,\label{mg1b}\\
M_{x-y}|\phi^+(k_x\!=\!k_y\!>\!0)\rangle&=&-\ii w'|\phi^+(k_x\!=\!k_y\!>\!0)\rangle,\label{mg2b}\\
M_x|\phi^+(0,|k_y|)\rangle&=&+\ii w'|\phi^+(0,|k_y|)\rangle,\label{mg3b}
\eea
which, when comparing to equations \eqref{mg1}-\eqref{mg3}, shows that $w'$ from Eq. \eqref{w} is actually equal to $w\equiv\sgn(\C^+_{k_z=0}\C^+_{k_z=\pi})$ from Eq. \eqref{wdef}:
as a consequence, in what follows, we will simply put $w'=w$;
we also obtain $v\equiv\sgn (\C^+_{k_z=0}\C^+_{k_x=k_y})=-1$ from Eqs. \eqref{vdef}, \eqref{mg2}, \eqref{mg2b}.

As noticed in Ref.~\onlinecite{yu_smb6_qpi}, due to time-reversal and $C_{4v}$ symmetry -- the surface symmetry group must contain a rotation by $\pi$ along an axis perpendicular to the surface --
the SEV along $z$ on this surface is always zero, even beyond the small-momentum expansion.

Equations \eqref{psipm}, \eqref{heffvw}, \eqref{w} represent the most important results of this section. They show that the $\bar\Gamma$ cone only exists in subspace~1, and its chirality $w$ depends on the relative sign of the hybridization term in $H_0$ through $f_1^v$ \eqref{f1etav} and in $H_P$ through $h_1^v$ \eqref{h1etav}. This is in agreement with the results of Ref.~\onlinecite{sigrist_tki}.

We remark that for the model 
used in Refs. \onlinecite{tki_cubic,dzero_pert} the kinetic energy is such that $\epsilon_1^d>\epsilon_1^f$, $\epsilon_2^d<\epsilon_2^f$. As a result, the minimum in the conduction band at $X$ has $d_{z^2}$ character ($X_6^+$ symmetry representation instead of $X_7^+$) and surface states near $\bar\Gamma$ only exist in subspace~2, with basis $d_{z^2}$ and $\Gamma_8^{(2)}$.
As noted in Ref. \onlinecite{smb6_prl_io}, this leads to $v=+1$  (instead of $v=-1$).
Moreover, with the same procedure as above, we can show that in this case the chirality $w=\sgn(f_2^vh_2^v)$, with $f_2^v$ from Eq. \eqref{f2etav} and $h_2^v$ from Eq. \eqref{h2etav};
this shows that hybridization parameters $\eta_z^{v1}$ and $\eta_z^{v2}$, if dominant, lead to $w=-1$, while dominant $\eta_x^{v2}$ to $w=+1$ \cite{smb6_prl_io}.
Most of our equations are formally equivalent to the ones in Ref. \onlinecite{dzero_pert}, once a proper replacements of quantities from subspace~1 to subspace~2 is performed, while the final results for spin structures are different due to the different expressions for $w$.

\subsection{$E_g$ - $\Gamma_7$ basis}

We repeat the calculation, now retaining the $\Gamma_7$ doublet together with the $E_g$ quartet, i.e., neglecting rows and columns from 5 to 8 in Eq. \eqref{hkdotp}.

Along $\Gamma$--$X$ the $\Gamma_7$ doublet can only hybridize with $d_{x^2-y^2}$, that we therefore assume to be inverted, and we only consider subspace 1' from Section~\ref{sec:reduced}, see Eq. \eqref{hkdotp_7}. (Note that in the opposite case, with the inversion in $d_{z^2}$, no insulator would be obtained.)
We can repeat the same steps of the previous subsection with the substitutions
\bea
\epsilon_1^f&\rightarrow&  \epsilon_7^f>\epsilon_1^d,\hspace{10pt}
g_1^f\rightarrow  g_7^f<0,\hspace{10pt}
l_1^f \rightarrow l_7^f<0,\nonumber\\
 f_1^v&\rightarrow&  f_7^v,\hspace{10pt} h_1^v\rightarrow h_7^v,
\eea
so we obtain the effective surface Hamiltonian \eqref{heffvw} with $w=\sgn(f_7^vh_7^v)$
and the basis:
\bea
|\psi_+\rangle&=&\alpha|d^1\uparrow\rangle+\beta|f^7+\rangle,\nonumber\\
|\psi_-\rangle&=&\alpha|d^1\downarrow\rangle-\beta|f^7-\rangle.\label{psipm7}
\eea
For the pseudospin expectation value we obtain the same results as in the previous subsection, Eqs. \eqref{psbasis_gamma}, \eqref{psev_gamma};
the SEV in the basis \eqref{psipm7} is
\bea
2\langle \vec  {\hat S} \rangle = \left( \gamma_5^-\hat s_x, \gamma_5^-\hat s_y, \gamma_5^+\hat s_z\right),
\eea
and the wavefunction $|\phi^+(\bk)\rangle$ has the SEV
\bea
2\langle \phi^+(\bk)| \vec  {\hat S} |\phi^+(\bk)\rangle=w\gamma_5^-(-\sin\theta_\bk,\cos\theta_\bk,0),
\eea
which is reversed with respect to the case with the $\Gamma_8$ quadruplet, since $\gamma_5^-<0$.
This is a consequence of the fact that, for $\Gamma_8$ states, SEV and pseudospin are parallel, while
for $\Gamma_7$ states they are antiparallel.\cite{sigrist_tki}
We also notice that $|\phi^+(\bk)\rangle$ has the same mirror-symmetry eigenvalues as before, Eqs. \eqref{mg1b}-\eqref{mg3b},
so, given the same MCNs, $\Gamma_7$ states give an opposite SEV pattern with respect to $\Gamma_8$ states \cite{smb6_prl_io}.
This is due to the fact that, strictly speaking, MCNs denote the sense of rotation of the pseudospin on surface states,\cite{sigrist_tki}
while relations for the SEV can be in general more complicated.\cite{smb6_prl_io,sigrist_tki}

\subsection{$E_g$ - $\Gamma_7$ - $\Gamma_8$ basis}
\label{ssec_g78}

The two different choices of the previous subsections correspond to large values of the crystal-field splitting, such as we can ignore
either $\Gamma_7$ or $\Gamma_8$ states.
If we have to take into account both $\Gamma_7$ and $\Gamma_8$ multiplets, the $\bar\Gamma$ cone is composed by $d_{x^2-y^2}$, $\Gamma_8^{(1)}$ and $\Gamma_7$ states,
Hamiltonian $H_0^{+(1)}$ \eqref{h1pp} becomes a $3\times 3$ matrix,
and no analytic solution can be found anymore.
However, when $k_z=0$, i.e. exactly at $X$, we can diagonalize the $f$ block:
\begin{align}\label{h_17}
H_0^{1,7}=
\left(\! \begin{array}{lll}
\epsilon_1^f&m_{78}\\
m_{78} & \epsilon_7^f
  \end{array}\!\right)=\frac{ \epsilon_1^f+\epsilon_7^f}{2}+
\left(\! \begin{array}{lll}
-\frac{\Delta}{2}&m_{78}\\
m_{78} & \frac{\Delta}{2}
  \end{array}\,\right),
\end{align}
with $\Delta=\epsilon_7^f-\epsilon_1^f$, out of which we pick the state which is mostly responsible for the band inversion, i.e. the one of higher energy, 
which is (we do the same for $H_0^{-(1)}$):
\bea
|f^{17}_p\pm\rangle&=&\beta_1 |f^1\pm\rangle+\beta_7| f^7\pm\rangle\label{f17},\\
\sgn(\beta_1\beta_7)&=&\sgn(m_{78}),\hspace{5pt} \beta_1^2+\beta_7^2=1.\label{beta_17_sign}
\eea
Keeping only $|d^1\uparrow\rangle$ and $|f^{17}_p+\rangle$ states, we now get
\begin{align}\label{h1pp_df+}
H_0^{+(1)}=
\left( \begin{array}{ll}
\epsilon_1^d-t_dk_z^2 g_1^d & -\ii Vk_z f_{17}^v\\
-\ii Vk_z f_{17}^v &\epsilon_+^f-t_fk_z^2 g_{17}^f
  \end{array}\right)&,
\end{align}
which is Eq. \eqref{h1pp} with the substitutions
$f_1^v\rightarrow \beta_1 f_1^v+ \beta_7 f_7^v\equiv f_{17}^v$,
$g_1^f\rightarrow \beta_1^2 g_1^f+ \beta_7^2 g_7^f\equiv g_{17}^f<0$.
With this approach we neglect the other linear combination of $|f^1\rangle$ and $|f^7\rangle$ states, called $|f^{17}_m\rangle$
in Fig. \ref{fig_kdp}(c),
which at $X$ remains below the Fermi energy. 

With this approximation, whose validity we will assess in Section \ref{sec_tpt}, the problem is now solvable by hand.
We get the doublet in the form:
\bea
|\psi_+\rangle&=&\alpha|d^1\uparrow\rangle+\beta\beta_1|f^1+\rangle+\beta\beta_7|f^7+\rangle,\nonumber\\
|\psi_-\rangle&=&\alpha|d^1\downarrow\rangle-\beta\beta_1|f^1-\rangle-\beta\beta_7|f^7-\rangle,\label{psipm17}
\eea
with $\sgn(\alpha\beta)=-\sgn(Vf_{17}^v)$.
When we project the perturbing Hamiltonian, which is now a $6\times 6$ matrix, we get Eq. \eqref{heffvw}
with the substitution $h_1^v\rightarrow \beta_1 h_1^v+ \beta_7 h_7^v\equiv h_{17}^v$.
The chirality $w$, as a consequence, is now given by:
\bea
w&=&\sgn[f_{17}^v h_{17}^v]\nonumber\\
&=&\sgn[(\beta_1 f_1^v+ \beta_7 f_7^v)(\beta_1 h_1^v+ \beta_7 h_7^v)].\label{w_17}
\eea
This approach is equivalent to starting from Eq. \eqref{hkdotp_17} with the given expressions of $f_{17}^v$ and $g_{17}^v$,
and $\epsilon_{17}^f(\bk)=\epsilon_+^f-t_f(k_z^2 g_{17}^f+k_\parallel^2 l_{17}^f)$, $l_{17}^f=\beta_1^2 l_1^f+ \beta_7^2 l_7^f$.

Relations for the pseudospin, Eqs. \eqref{psbasis_gamma}, \eqref{psev_gamma}, remain invariant,
with a redefinition of $w$ according to Eq. \eqref{w_17}.
For spin operators we find:
\bea
2\langle \vec  {\hat S} \rangle &=& \left( \gamma_5^{+'}\hat s_x, \gamma_5^{+'}\hat s_y, \gamma_{11}^{-'}\hat s_z\right),\\
\gamma_{5}^{\pm'}&=&\alpha^2\pm\frac{5}{21}\beta^2\beta_1^2\mp\frac{5}{21}\beta^2\beta_7^2\mp\frac{4\sqrt{5}}{21}\beta^2\beta_1\beta_7,\label{gamma5p}\\
\gamma_{11}^{\pm'}&=&\alpha^2\pm\frac{11}{21}\beta^2\beta_1^2\mp\frac{5}{21}\beta^2\beta_7^2\pm\frac{8\sqrt{5}}{21}\beta^2\beta_1\beta_7,\label{gamma11p}
\eea
and for the SEV on the state of positive energy:
\bea
2\langle \phi^+(\bk)| \vec  {\hat S} |\phi^+(\bk)\rangle=w\gamma_{5}^{+'}(-\sin\theta_\bk,\cos\theta_\bk,0).
\eea
While the pseudospin is the same as before,
the SEV, on the other hand, can be parallel or antiparallel to the pseudospin according to the sign of $\gamma_{5}^{+'}$, which depends on the relative weights $\beta_1$ and $\beta_7$.
We also note that the sign of the  interference term in Eq. \eqref{gamma5p} depends on the relative sign of $\beta_1$ and $\beta_7$, which is the sign of $m_{78}$, see Eq. \eqref{beta_17_sign}.
Ab-initio calculations indicate $m_{78}>0$
for {\sm}, see Eq. \eqref{m_78}, favoring antiparallel spin and pseudospin.

\section{Expansion around $\bar X$ for $(001)$ surface}\label{sec_x}

Here we employ the same technique as in the previous Section, but for the effective Hamiltonian at the surface momenta $\bar X$, $\bar X'$ on the $(001)$ surface.

\subsection{$E_g$ - $\Gamma_8$ basis}

One route to obtain surface states around $\bar X$ is to project $X'=(\pi,0,0)$ onto the $(001)$ surface.\cite{dzero_pert} In this case $H_0^{+(1)}$ is a $4\times 4$ matrix which does not admit a simple analytical solution.
An alternative route is to project $X=(0,0,\pi)$ onto the (100) surface: In this case $k_y$ and $k_z$ remain good quantum numbers, and we obtain surface states near $k_y=0$, $k_z=\pi$.
To get the effective Hamiltonian for the $(001)$ surface, we then must perform a rotation of the coordinate system.

We will follow this second route, even if apparently more involved, as it allows for a good approximation which makes the problem solvable by hand.
Upon retaining the $\Gamma_8$ quadruplet, the Hamiltonian $H_{0}^+$, obtained by setting $k_y=k_z=0$, has now as a basis $|d^1\uparrow\rangle$, $|d^2\uparrow\rangle$, $|f^1-\rangle$, $|f^2-\rangle$,
but does not decouple any more into two $2\times 2$ blocks, due to the terms in $h_{12}^v$ and $h_{21}^v$.
However, we may adopt the approximation to neglect these couplings between subspace~1 and 2,
such that $|\psi_+\rangle$ and $|\psi_-\rangle$ live entirely into subspace~1,
which is reasonable since it is where band inversion is achieved:  
\bea
|\psi_+\rangle&\simeq&\alpha|d^1\uparrow\rangle+\beta|f^1-\rangle,\nonumber\\
|\psi_-\rangle&\simeq&\alpha|d^1\downarrow\rangle+\beta|f^1+\rangle.\label{psipm100}
\eea
This is found to be an excellent approximation when compared to the tight-binding results in the limit of small momenta: the weight of states in subspace~2 rarely exceeds a few percent.
We remark that such a simplification can be achieved only with our choice of projecting $X=(0,0,\pi)$ onto the (100) surface. As shown below, the weight of states in subspace~2 on the $(001)$ surface is 75\%, i.e., projecting $X'=(\pi,0,0)$ onto the $(001)$ surface does not admit any obvious approximation.

Neglecting the couplings between subspaces 1 and 2, $H_0^+$ in the basis $|d^1\uparrow\rangle$, $|f^1-\rangle$ reads
\begin{align}\label{h_0_x}
H_0^+=
\left( \begin{array}{ll}
\epsilon_1^d-t_dk_x^2 l_1^d & -\ii Vk_x h_1^v\\
\ii Vk_x h_1^v &\epsilon_1^f-t_fk_x^2 l_1^f
  \end{array}\right)&,
\end{align}
which corresponds to Eq. \eqref{h1pp} with the substitution $g_1^d \rightarrow l_1^d$, $g_1^f \rightarrow l_1^f$, $k_z\rightarrow k_x$, $f_1^v\rightarrow h_1^v$.
We can take coefficients $\alpha$ and $\beta$ as real, with $\sgn(\alpha\beta)=-\sgn(Vh_1^v)$ from Eq. \eqref{beta},
as a calculation similar to the one of the previous section shows, and which also leads
to the effective Hamiltonian for small momenta.

We have, however, to go back to the $(001)$ surface through a coordinate rotation $k_x\rightarrow k_z \rightarrow k_y \rightarrow k_x$,
which changes $d$ and $f$ states according to Wigner U-matrices,\cite{tki_cubic,prbr_io_smb6,suppl_tci_long}
and a basis rotation $\hat s_x\rightarrow \hat s_z \rightarrow \hat s_y \rightarrow \hat s_x$,
to get the effective Hamiltonian around $\bar X'$: 
\bea
H^{eff}_{\bar X'}= -|v_1| w k_y\hat s_x+|v_2|k_x\hat s_y, \label{heffxp}
\eea
with basis:
\bea
|\psi_+'\rangle&=&\frac{\alpha}{2}\left(|d^1\!\uparrow\rangle-\sqrt{3}|d^2\!\uparrow\rangle\right)+\frac{\beta}{2}\left(|f_1+\rangle-\sqrt{3}|f_2+\rangle\right),\nonumber\\
|\psi_-'\rangle&=&\frac{\alpha}{2}\left(|d^1\!\downarrow\rangle-\sqrt{3}|d^2\!\downarrow\rangle\right)-\frac{\beta}{2}\left(|f_1-\rangle-\sqrt{3}|f_2-\rangle\right),\label{psipmp}
\eea
and velocities:
\bea
v_1&=&-2\alpha\beta V f_1^v = 2w|Vf_1^v|\frac{\sqrt{-t_dt_fl_1^dl_1^f}}{t_fl_1^f-t_dl_1^d},\label{v_1}\\
v_2&=&-2\alpha\beta V h_1^v=2 |V h_1^v|\frac{\sqrt{-t_dt_fl_1^dl_1^f}}{t_fl_1^f-t_dl_1^d}>0.\label{v_2}
\eea

It can be observed that the relative sign of the two Dirac velocities, that is the winding number of the $\bar X$ cone, is given by $\sgn(v_1v_2)=w=\sgn(f_1^vh_1^v)$,
that is, the same expression which gives the chirality of the $\bar\Gamma$ cone.
Thus, we recover, through a different derivation, the results of Ref. \onlinecite{sigrist_tki}.

\begin{figure}[tb]
\includegraphics[width=0.48\textwidth]{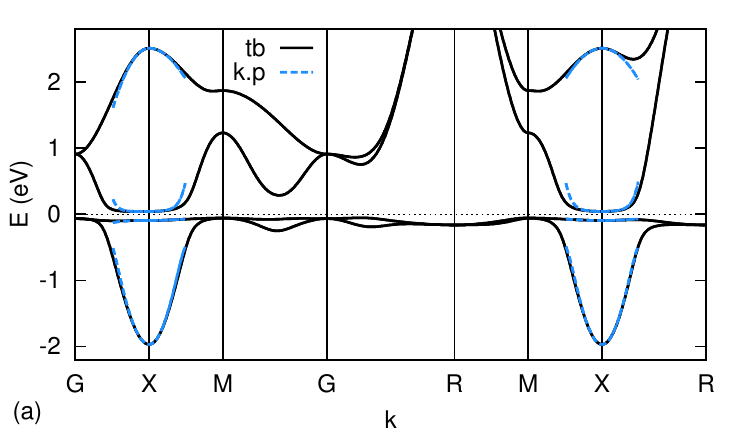}
\includegraphics[width=0.48\textwidth]{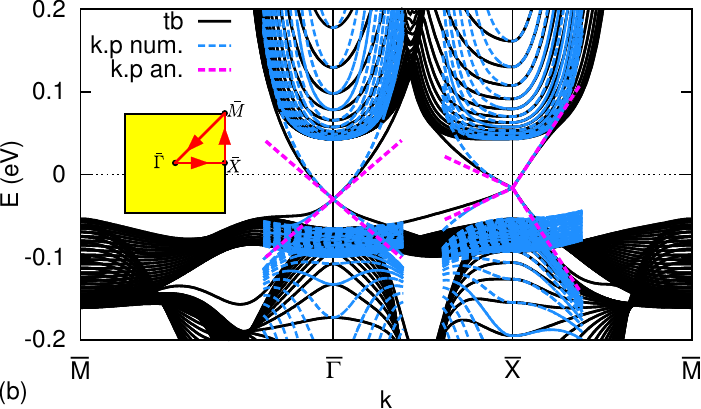}
\caption{
(a) Bulk tight-binding bandstructure for parameters $t_c=0.8$eV, $t_f=-0.002$eV, $V=0.063$eV, $\eta_z^{d1}=\eta_z^{f1}=0.8$, $\eta_z^{d2}=-0.3$, $\eta_z^{f2}=-0.45$, $\epsilon_d-\epsilon_8=1.45$eV,
$\eta_z^{v1}=-2.1$, $\eta_z^{v2}=0.6$.
Also shown is the result of the \kdp approximation around $X$, see Fig. \ref{fig_kdp}(a).
(b) Tight-binding bandstructure for a $(001)$ slab of 30 layers, together with the analytical \kdp approximation of surface states described in the text, and
the numerical solution of the  \kdp model including bulk states.\cite{suppl_tci_long}
We see that both \kdp solutions give Dirac energies and velocities in excellent agreement with the tight-binding solution;
obtaining accurate velocities requires to take into account third-order terms in $\bk$ for the hybridization.\cite{suppl_tci_long}
}\label{fig_bands_001}
\end{figure}

In Fig. \ref{fig_bands_001} we present a comparison between the analytical \kdp result and the numerical tight-binding solution. We note that obtaining the exact value of the velocities $v_0$, $v_1$, $v_2$ requires to take into account higher-order terms in the small-momentum expansion;\cite{suppl_tci_long} for our choice of the parameters, it is enough to take the hybridization up to third order in $\bk$.

Spin  and pseudospin operators become:
\bea
2\langle \vec  {\hat S} \rangle &= &\left( \gamma_5^+\hat s_x, \gamma_{11}^+\hat s_y, \gamma_5^-\hat s_z\right),\\
\langle \vec {\hat\sigma} \rangle &=&\left( \hat s_x, \hat s_y, \gamma^-\hat s_z\right), \label{psbasis_xp}
\eea
which shows that operators $\hat s_x$ and $\hat s_y$ in Eq. \eqref{heffxp} can be substituted exactly by pseudopin operators $\hat\sigma_x$ and $\hat\sigma_y$, or,
up to a constant factor, by spin operators $ \hat S_x$ and $ \hat S_y$,
once again justifying writing the effective Hamiltonian in terms of spin operators.\cite{smb6_prl_io}
For the state $|\phi^{+}_{\bar X'}(\bk)\rangle$ with positive energy $\epsilon_\bk=\sqrt{v_1^2k_y^2+v_2k_x^2}$ at a given momentum $\bk$ for Eq. \eqref{heffxp} 
the SEV and pseudospin are:
\bea
2\langle \phi^{+}_{\bar X'}(\bk)| \vec  {\hat S} |\phi^{+}_{\bar X'}(\bk)\rangle&=&\left(-\gamma_5^+w\sin\theta_\bk,\gamma_{11}^+ \cos\theta_\bk,0\right),\label{sev_xp}\\
\langle \phi^{+}_{\bar X'}(\bk)| \vec{\hat\sigma} |\phi^{+}_{\bar X'}(\bk)\rangle&=&\left(-w\sin\theta_\bk, \cos\theta_\bk,0\right),\label{psev_xp}
\eea
with $\cos\theta_\bk=|v_2| k_x/\epsilon_\bk$, $\sin\theta_\bk=|v_1|k_y/\epsilon_\bk$.
Moreover we find:
\bea
M_y|\phi^{+}_{\bar X'}(|k_x|,0)\rangle&=&-\ii|\phi^{+}_{\bar X'}(|k_x|, 0)\rangle,\\
M_x|\phi^{+}_{\bar X'}(0, -|k_y|)\rangle&=&-\ii w|\phi^{+}_{\bar X'}(0, -|k_y|)\rangle,
\eea
which agrees with Eqs. \eqref{mxp1}, \eqref{mxp2}, confirming that $w=\sgn(f_1^vh_1^v)=\sgn(\C^+_{k_z=0}\C^+_{k_z=\pi})$.

After a $\pi/2$ rotation we also obtain the effective Hamiltonian around $\bar X$:
\bea
H^{eff}_{\bar X}= |v_1|w k_x\hat s_y-|v_2|k_y\hat s_x, \label{heffx}
\eea
and similar expressions hold for the SEV and the pseudospin:
\bea
2\langle \phi^{+}_{\bar X}(\bk)| \vec  {\hat S} |\phi^{+}_{\bar X}(\bk)\rangle&=&\left(-\gamma_{11}^+\sin\theta_\bk,\gamma_5^+ w\cos\theta_\bk,0\right),\label{sev_x}\\
\langle \phi^{+}_{\bar X}(\bk)| \vec{\hat\sigma} |\phi^{+}_{\bar X}(\bk)\rangle&=&\left(-\sin\theta_\bk, w\cos\theta_\bk,0\right),\label{psev_x}
\eea
where now
$\cos\theta_\bk={|v_1|k_x}/{\epsilon_\bk}$ and $\sin\theta_\bk={|v_2| k_y}/{\epsilon_\bk}$.
Moreover, Eqs. \eqref{mx1} and \eqref{mx2} hold.

%

Equations \eqref{heffxp}, \eqref{psipmp}, \eqref{heffx} are the most important results of this section: they show that Dirac cones at $\bar X$ and $\bar X'$ have two different velocities, whose relative sign is given by Eq. \eqref{w}, that is, the same expression which gives the chirality of the $\bar\Gamma$ cone.
Eq.~\eqref{psipmp} implies that in our approximation the Dirac cones at $\bar X$, $\bar X'$ are composed from $75\%$ of $\Gamma_8^{(2)}$ and $d_{z^2}$ states and from $25\%$ of $\Gamma_8^{(1)}$  and $d_{x^2-y^2}$ states. This is in good agreement with tight-binding results, which for typical values of the parameters shows these percentages to be $\gtrsim 70\%$ and $\lesssim 30\%$.
We recall that, for the present reduced basis, the $\bar\Gamma$ cone is entirely composed by $\Gamma_8^{(1)}$  and $d_{x^2-y^2}$ states.


\subsection{$E_g$ - $\Gamma_7$ basis}
When using the $\Gamma_7$ doublet instead of $\Gamma_8$ quadruplet the basis for $H_0^+$ is spanned by $|d^1\uparrow\rangle$, $|d^2\uparrow\rangle$, $|f^7-\rangle$;
with the same approximation of the previous subsection we can neglect term $h_{72}^v$ in Eq. \eqref{hkdotp}, hence $|d^2\uparrow\rangle$.
The basis for surface states at $k_y=0$, $k_z=\pi$ on the (100) surface becomes:
\bea
|\psi_+\rangle&=&\alpha|d^1\uparrow\rangle+\beta|f^7-\rangle,\nonumber\\
|\psi_-\rangle&=&\alpha|d^1\downarrow\rangle+\beta|f^7+\rangle.
\eea
After rotation, the effective Hamiltonian is Eq. \eqref{heffxp}
with basis:
\bea
|\psi_+'\rangle&=&\frac{\alpha}{2}\left(|d_1\uparrow\rangle-\sqrt{3}|d_2\uparrow\rangle\right)-{\beta}|f_7+\rangle,\nonumber\\
|\psi_-'\rangle&=&\frac{\alpha}{2}\left(|d_1\downarrow\rangle-\sqrt{3}|d_2\downarrow\rangle\right)+{\beta}|f_7-\rangle,\label{psipmp7}
\eea
and the usual substitutions $f_1^v,h_1^v, l_1^{d,f}\rightarrow f_7^v,h_7^v,l_7^{d,f}$;
the winding number has the same expression of the chirality at $\bar\Gamma$: $w=\sgn(f_7^vh_7^v)$.

The pseudospin behaves as in the previous subsection, Eqs. \eqref{psbasis_xp}, \eqref{psev_xp}.
Spin operators are:
\bea
2\langle \vec S \rangle &=& \left( \gamma_5^-\hat s_x, \gamma_5^- \hat s_y, \gamma_5^+ \hat s_z\right),
\eea
and the SEV on the state with positive energy is:
\bea
2\langle \phi^{+}_{\bar X'}(\bk)| \vec  {\hat S} |\phi^{+}_{\bar X'}(\bk)\rangle=\gamma_5^-\left(-w\sin\theta_\bk,\cos\theta_\bk,0\right),
\eea
which is antiparallel with respect to the $\Gamma_8$ case since $\gamma_5^-<0$.

We see that when performing the rotation to go back to the $(001)$ surface, $\Gamma_7$ goes into itself, Eq.~\eqref{psipmp7},
so now surface states at $\bar X$ are mostly $\Gamma_7$ in character, just like surface states at $\bar\Gamma$.
In this case -- when projecting along the $x$ direction -- the approximation of neglecting subspace 2  is always reliable, since here we are only discarding a $d$ state ($d_{z^2}$), which contributes a small weight to surface states; this is confirmed by tight-binding results.


\subsection{$E_g$ - $\Gamma_7$- $\Gamma_8$ basis}
With the choice of the basis
\bea
|\psi_+\rangle&=&\alpha|d^1\uparrow\rangle+\beta\beta_1|f^1-\rangle+\beta\beta_7|f^7-\rangle,\nonumber\\
|\psi_-\rangle&=&\alpha|d^1\downarrow\rangle+\beta\beta_1|f^1+\rangle+\beta\beta_7|f^7+\rangle,\label{psipm100_17}
\eea
we obtain the same Hamiltonian at $\bar X'$, Eq. \eqref{heffxp},
with the new definition of the winding number, Eq. \eqref{w_17}.
The relations for the pseudospin, Eqs. \eqref{psbasis_xp}, \eqref{psev_xp}, remain invariant.
For the spin we find:
\bea
2\langle \vec  {\hat S} \rangle &=& \left( \gamma_{5}^{+'}\hat s_x, \gamma_{11}^{+'}\hat s_y, \gamma_{5}^{-'}\hat s_z\right),
\eea
and for the SEV on the state of positive energy:
\bea
2\langle \phi^+(\bk)| \vec  {\hat S} |\phi^+(\bk)\rangle=(-w\gamma_5^{+'}\sin\theta_\bk,\gamma_{11}^{+'}\cos\theta_\bk,0).
\eea
Similar relations hold at $\bar X$.

We note that terms $\gamma_5^{+'}$ and $\gamma_{11}^{+'}$, Eqs. \eqref{gamma5p}, \eqref{gamma11p}, can be positive or negative according to the relative weight of $\beta_1$ and $\beta_7$ in $|f^{17}_p\pm\rangle$.
Even if unlikely, it may also happen that, when $|\beta_1|\sim|\beta_7|$, they carry different signs; in this case the winding number of the SEV would be different from the winding number of the pseudospin, with only the latter directly related to the topological phase.
In Ref. \onlinecite{smb6_prl_io} we assumed this scenario not to occur, which should be a safe assumption for most of parameter space.

\subsection{Comparison with DFT and experiments}

We now relate our model-dependent analysis to concrete results for {\sm} found in the literature.
As we have shown, the winding number $w\equiv\sgn(\C^+_{k_z=0}\C^+_{k_z=\pi})$ depends on the retained $f$ multiplet ($\Gamma_7$ or $\Gamma_8$), the symmetry of the inverted subspace ($d_{x^2-y^2}$ - subspace~1 - symmetry representation $X_7^+$, or $d_{z^2}$ - subspace~2 - symmetry representation $X_6^+$), and the hybridization.
Based on DFT results \cite{smb6_korea_dft,pub6, prbr_io_smb6}, we can safely state that the band inversion happens in subspace~1, which is spanned by $d_{x^2-y^2}$, $\Gamma_8^{(1)}$ and $\Gamma_7$. This leads to $v\equiv\sgn (\C^+_{k_z=0}\C^+_{k_x=k_y})=-1$,
and, when we take into account the $\Gamma_8$ quadruplet only, to $w=\sgn(f_1^v h_1^v)$; as a consequence, $\eta_z^{v2}$ leads to $w=+1$, while $\eta_x^{v1}$ and $\eta_x^{v2}$ to $w=-1$,
see Eqs. \eqref{f1etav}, \eqref{h1etav}.
Ab-initio calculations\cite{pub6, prbr_io_smb6} show the largest hybridization term to be $\eta_z^{v1}$, which, however,
does not lead alone to a gap by symmetry mismatch (actually, it does not appear in $f_1^v$);
the second most important term is $\eta_z^{v2}$, hence $w=+1$; numerical solutions by keeping many hybridization terms show $w=+1$ to remain the correct solution.
When we use the $\Gamma_7$ doublet, instead, $w=\sgn(f_7^v h_7^v)$,  which gives $w=+1$ for $\eta_7^{v1}$ and $\eta_{x7}^{v2}$, and $w=-1$ for $\eta_7^{v2}$, see Eqs. \eqref{f7etav}, \eqref{h7etav},
which is the largest term in ab-initio calculations, leading to $w=-1$ even when keeping more hybridization terms.
Hence, retaining different multiplets leads to different values of $w$,
and to different topological phases.

Spin-resolved photoemission data\cite{smb6_arpes_mesot_spin} indicate a winding number $w=+1$ on the $\bar X$ cone, leading to a preference toward our $\Gamma_8$ model;
this is in agreement with Ref. \onlinecite{smb6_dmft_slab}, which finds surface states to be mostly $\Gamma_8$.
While early theory papers\cite{takimoto,lu_smb6_gutz, tki_cubic} have not discussed the Dirac-cone spin structure, it was shown in Refs. \onlinecite{smb6_prl_io,sigrist_tki} that a full characterization of the {\sm} electronic structure requires the knowledge of the exact value of mirror Chern numbers, which directly influence the spin structure of surface states.
In a few ab-initio calculations the spin structure is addressed: in Ref. \onlinecite{yu_smb6_qpi} it seems to contradict experimental results, rather suggesting $w=-1$;
while the one of Ref. \onlinecite{smb6_korea2} agrees with experiments,
as well as the one of Ref. \onlinecite{prbr_io_smb6}, which is, however, based on {\pu} ab-initio calculations.\cite{pub6}
We thus believe that the question deserves further consideration;
from our point of view it reduces to understanding if $\Gamma_7$ ($w=-1$) or $\Gamma_8$ ($w=+1$) states are mostly responsible for the bulk gap and the composition of surface states;
in Section \ref{sec_tpt} we show that varying their relative energy leads to a topological phase transition $w=-1\leftrightarrow +1$
between these two possibilities, where the latter one should be realized in ``clean'' {\sm}.


\section{Generic flat surface}
\label{sec:other_surf}

So far we have studied surface states on the $(001)$ surface, well studied already in previous papers. The power of the \kdp approach is, however, that we can obtain analytical results also for a generic flat $(lmn)$ surface without much additional effort. Here we will in particular consider $(110)$, $(111)$, and $(210)$ surfaces. We note that, by construction, the \kdp approach only yields results for surface Dirac cones protected by parity invariants, because those arise from bands near time-reversal-invariant momenta. In contrast, Dirac cones only protected by mirror symmetries are not accessible -- this will be relevant for the $(110)$ surface of {\sm}.

In the following we denote surface momenta as $\bar k_x$ and $\bar k_y$ to distinguish them from bulk momenta, a distinction which on the $(001)$ surface is not needed, since there $\bar k_x=k_x$, $\bar k_y=k_y$.


\subsection{Surface states from parity invariants and mirror Chern numbers}

Since there are three bulk $X$ points with band inversion, parity invariants predict in general three Dirac ones in the 2D BZ. Mirror symmetries might complicate the situation, and we discuss a number of surfaces explicitly.

\begin{figure}[tb]
\includegraphics[width=0.46\textwidth]{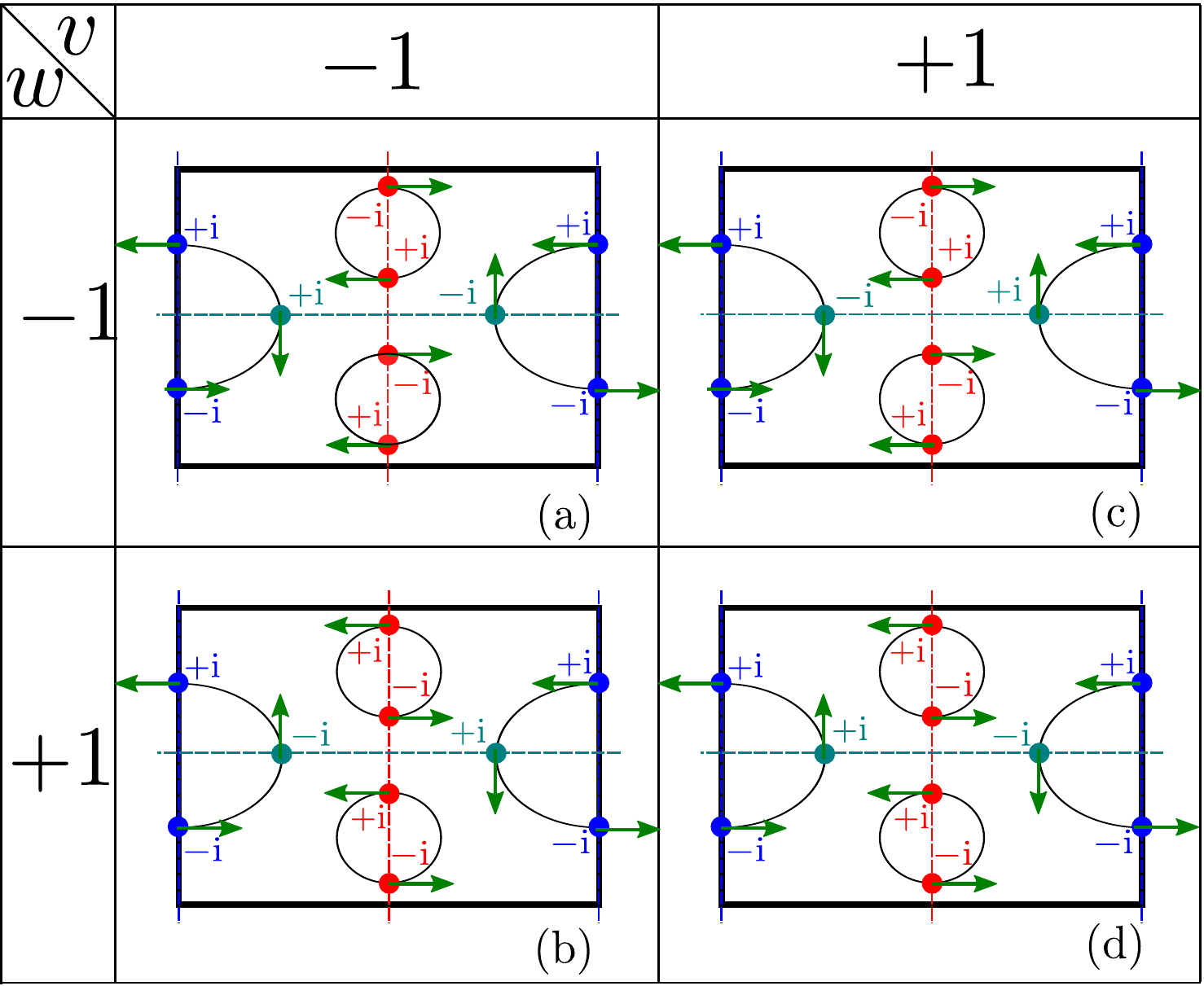}
\caption{Mirror-symmetry eigenvalues and SEV on the BZ for a $(110)$ surface for the same cases as in Fig. \ref{fig_chern110}.
}\label{fig_chern110}
\end{figure}

On the $(110)$ surface, one $X$ point is projected onto the $\bar X=(\pi,0)$ point of the rectangular surface BZ,
while the other two onto $\bar Y=(0,\pi/\sqrt{2})$, which will then hybridize and gap out:
as a consequence, only a single Dirac cone is predicted by parity invariants at $\bar X$.
As shown in Ref. \onlinecite{smb6_tci}, mirror Chern numbers predict the presence of two additional Dirac cones along the $\bar\Gamma$--$\bar Y$ direction.
Indeed, the $k_z=0$ plane is projected to $\bar k_x=0$,
while $k_z=\pi$  to $\bar k_x=\pi$,
and $k_x=k_y$ to $\bar k_y=0$; see Fig. \ref{fig_bz}(c).
We must therefore have a Dirac cone along $\bar X$--$\bar S$ and $\bar X$--$\bar \Gamma$,
which is simply the cone at $\bar X$ predicted by parity invariants,
and two new cones along $\bar \Gamma$--$\bar Y$ as a consequence of $\C_{k_x=k_y}^+=\pm 2$;
these cones are protected by mirror symmetry only.
In addition to this, we can characterize the $\bar X$ cone with a winding number;
its SEV is fixed along the $\bar X$--$\bar S$ direction by $\C_{k_z=\pi}^+=+1$,
but it changes along the $\bar X$--$\bar \Gamma$ direction according to $\C_{k_x=k_y}^+$ and $\C_{k_z=\pi}^+$: while the first one fixes mirror eigenvalues, the second one tells what the SEV for a given mirror eigenvalue is. As a result, the winding number of the $\bar X$ cone is $w$, the same as on the $(001)$ surface. This is shown in Fig. \ref{fig_chern110}.
We note, however, that the winding number on the $\bar X$ cone of the $(110)$ surface is only fixed at low energies\cite{smb6_prl_io}, when we can neglect subspace~2; at higher energies the spin direction along $\bar k_y=0$ can thus in principle be reversed, and so the winding number. This is different from the $(001)$ surface where the winding number is constrained by the symmetry operation $M_z$.

On the $(210)$ surface, the situation is different: one $X$ point is projected onto the $\bar X=(\pi,0)$ point of the rectangular surface BZ, as for the $(110)$ surface;
however, the other two $X$ points are projected to $\bar\Gamma=(0,0)$ and $\bar Y=(0,\pi/\sqrt{5})$.
Hence, parity invariants predict three Dirac cones, with no additional cones protected by mirror symmetry only.
Since no Dirac cone is crossed by two mirror planes, see Fig. \ref{fig_bz}(e), we cannot make any general predictions on the
winding number.

\begin{figure}[tb]
\includegraphics[width=0.46\textwidth]{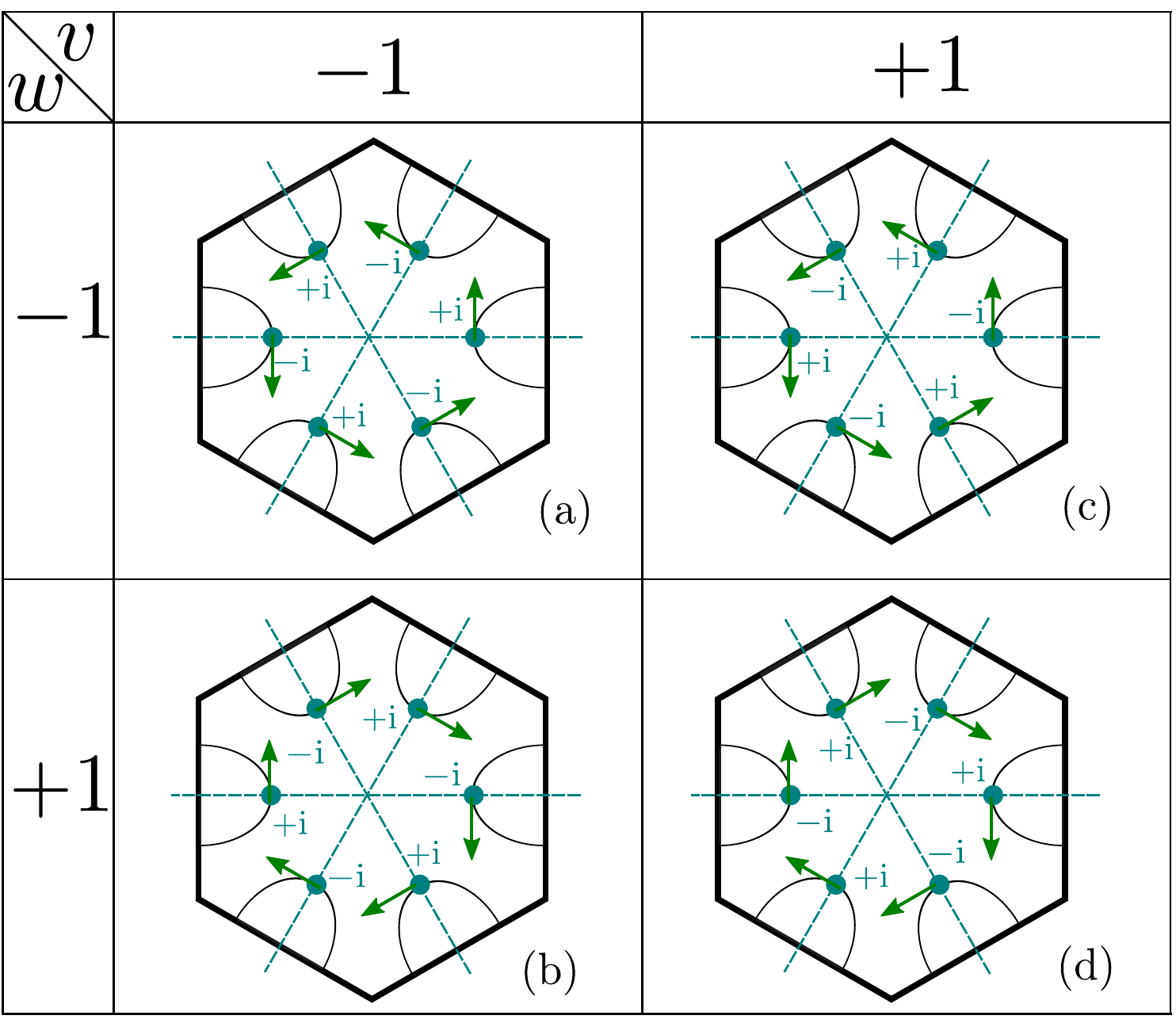}
\caption{Mirror-symmetry eigenvalues and SEV on the BZ for a $(111)$ surface for the same cases as in Fig. \ref{fig_chern111}.
}\label{fig_chern111}
\end{figure}

On the $(111)$ surface, the three bulk $X$ points are projected to the three inequivalent $\bar M$ points of the hexagonal surface BZ.
Mirror planes $k_x=k_y$, $k_y=k_z$, $k_z=k_x$ are projected along the three $\bar \Gamma$ - $\bar M$ directions, fixing mirror-symmetry eigenvalues of Dirac cones along those lines;
the only information we get is that the SEV is antiparallel at the two extrema of each cone,
and nothing can be said about winding numbers using mirror eigenvalues only.
These results are shown in Fig.~\ref{fig_chern111}.

\begin{figure}[tb]
\includegraphics[width=0.46\textwidth]{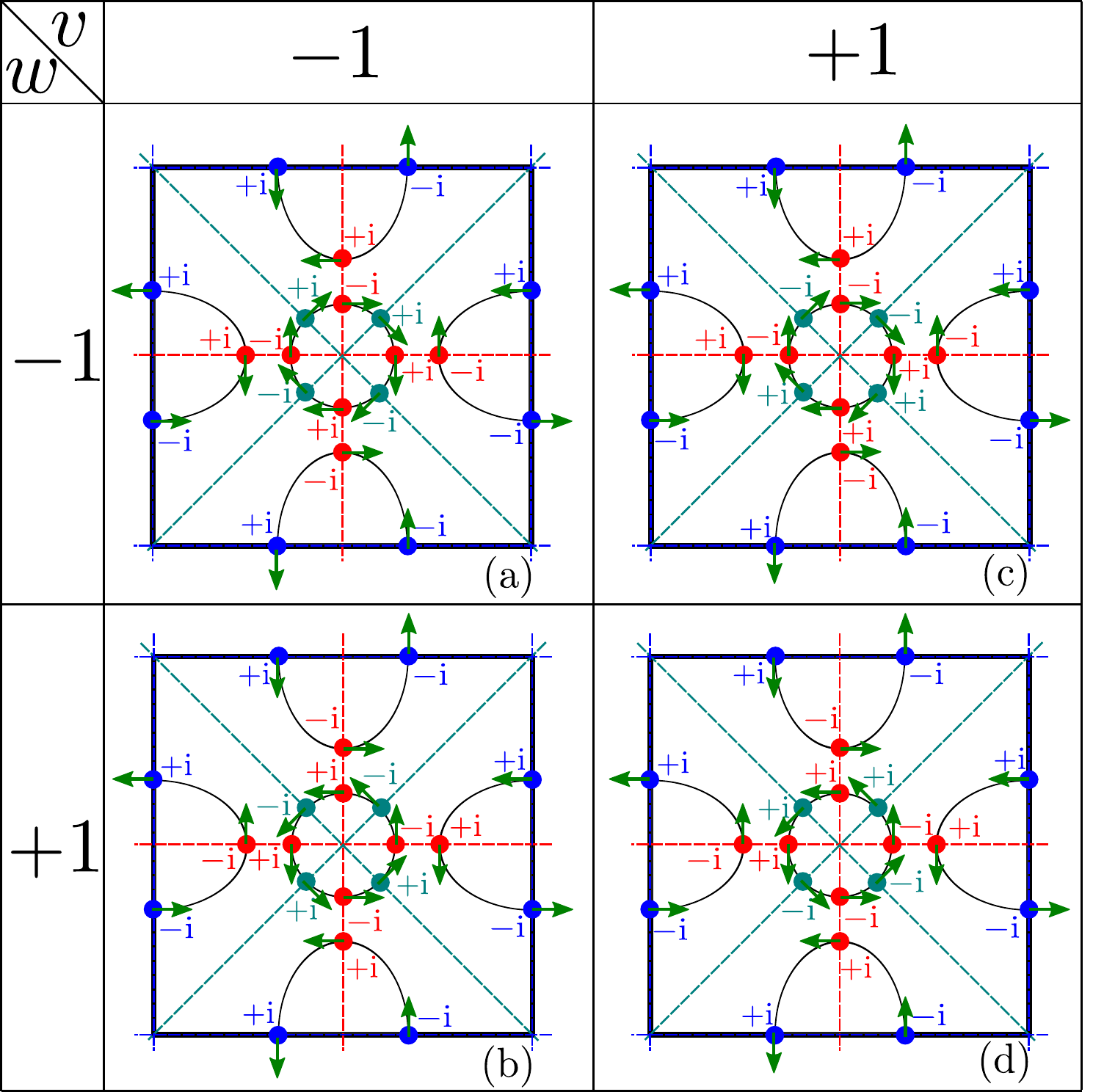}
\caption{Mirror-symmetry eigenvalues and SEV (green arrows) on the surface BZ for a $(001)$ surface, as a function of
$v\equiv\sgn (\C^+_{k_z=0}\C^+_{k_x=k_y})$ and $w\equiv\sgn(\C^+_{k_z=0}\C^+_{k_z=\pi})$. The panels correspond to MCNs $(\C^+_{k_z=0}, \C^+_{k_z=\pi},\C^+_{k_x=k_y})$ as follows: (a) $(-2,+1,+1)$, (b) $(+2,+1,-1)$, (c) $(-2,+1,-1)$, (d) $(+2,+1,+1)$.
To draw the SEV we assume $\Gamma_8$ states; for $\Gamma_7$ states the SEV is reversed.
From Ref. \onlinecite{smb6_prl_io}.
}\label{fig_chern100}
\end{figure}

We finally stress that for all the surfaces, the qualitative spin structure along high-symmetry directions only depends on $w$
as one can realize by comparing in each of the Figures \ref{fig_chern100},
\ref{fig_chern110}, \ref{fig_chern111}, the pairs of panels (a)-(c) and (b)-(d),
which share the same $w$, differ by $v$, and still have the same SEV.
The number $v$, on the other hand, dictates the orbital composition of the different cones,
and does not give any information on the spin.

\subsection{Geometrical considerations}

We want to develop a general theory for the spin structure of a Dirac cone on a given surface,
following what we did in Sections \ref{sec:gamma} and \ref{sec_x}.

Given a generic $(lmn)$ surface (without loss of generality, we will only consider nonnegative $l$, $m$, $n$ integers), the three indices are equivalent due to cubic symmetry. However, when we choose to expand around $X=(0,0,\pi)$, the resulting \kdp Hamiltonian has tetragonal symmetry, with the $z$ direction inequivalent to $x$ and $y$. Consequently, the third index, which fixes the new $z$ direction, is inequivalent to the first two:
so we introduce the $(lm/n)$ notation, to stress that index $n$ is inequivalent from $l$ and $m$.
So, given $(lmn)$, we have in general three inequivalent triplets $(lm/n)=(ml/n)$, $(mn/l)=(nm/l)$, and $(ln/m)=(nl/m)$ which correspond to the three possible choices for the $z$ axis, or, alternatively, to the \kdp expansion at each of the three different $X$ points.

The $(lm/n)$ triplet describes the direction along which $\bar k_z$ points, with polar angles:
\bea
\theta&\equiv&\arctan\frac{\sqrt{l^2+m^2}}{n},\label{theta_lmn}\\
\phi&\equiv&\arctan\frac{m}{l},\label{phi_lmn}
\eea
and $\bar k_z\rightarrow -id/dz$, while $\bar k_x$ and $\bar k_y$ will remain good quantum number.

We can thus perform a rotation in momentum space with Euler angles $\omega$, $\omega'=\theta$, $\omega''=\phi$ (we adopt the $zyz$ convention),
where $\omega$, which is for the moment arbitrary, corresponds to a rotation in the $\bar k_x$, $\bar k_y$ plane;
details are given in the supplement.\cite{suppl_tci_long}
When $\omega=0$ we find that the $X=(0,0,\pi)$ point is projected at
\be
\bar\bk_X=(-\pi\sin\theta,0),\label{barbk}
\ee
so $\bar k_x$ is the direction which joins $\bar \Gamma$ to the position of the cone, unless $\theta=0$, which corresponds to the $\bar\Gamma$ cone on the $(001)$ surface,
for which $\bar k_x$ and $\bar k_y$ directions are equivalent.
Also, to the $X$ point we can assign the $(lm/n)$ triplet, and an angle $\theta$ as defined in Eq. \eqref{theta_lmn}.

We can also find
that $X'=(\pi,0,0)$ and $X''=(0,\pi,0)$
are projected respectively to:
\bea
\bar\bk_{X'}&=&\pi(\cos\theta\cos\phi,-\sin\phi),\label{barbkp}\\
\bar\bk_{X''}&=&\pi(\cos\theta\sin\phi,\cos\phi),\label{barbkpp}
\eea
which are the positions of the two other Dirac cones in the 2D BZ when $\omega=0$.

\begin{figure}[tb]
\includegraphics[width=0.48\textwidth]{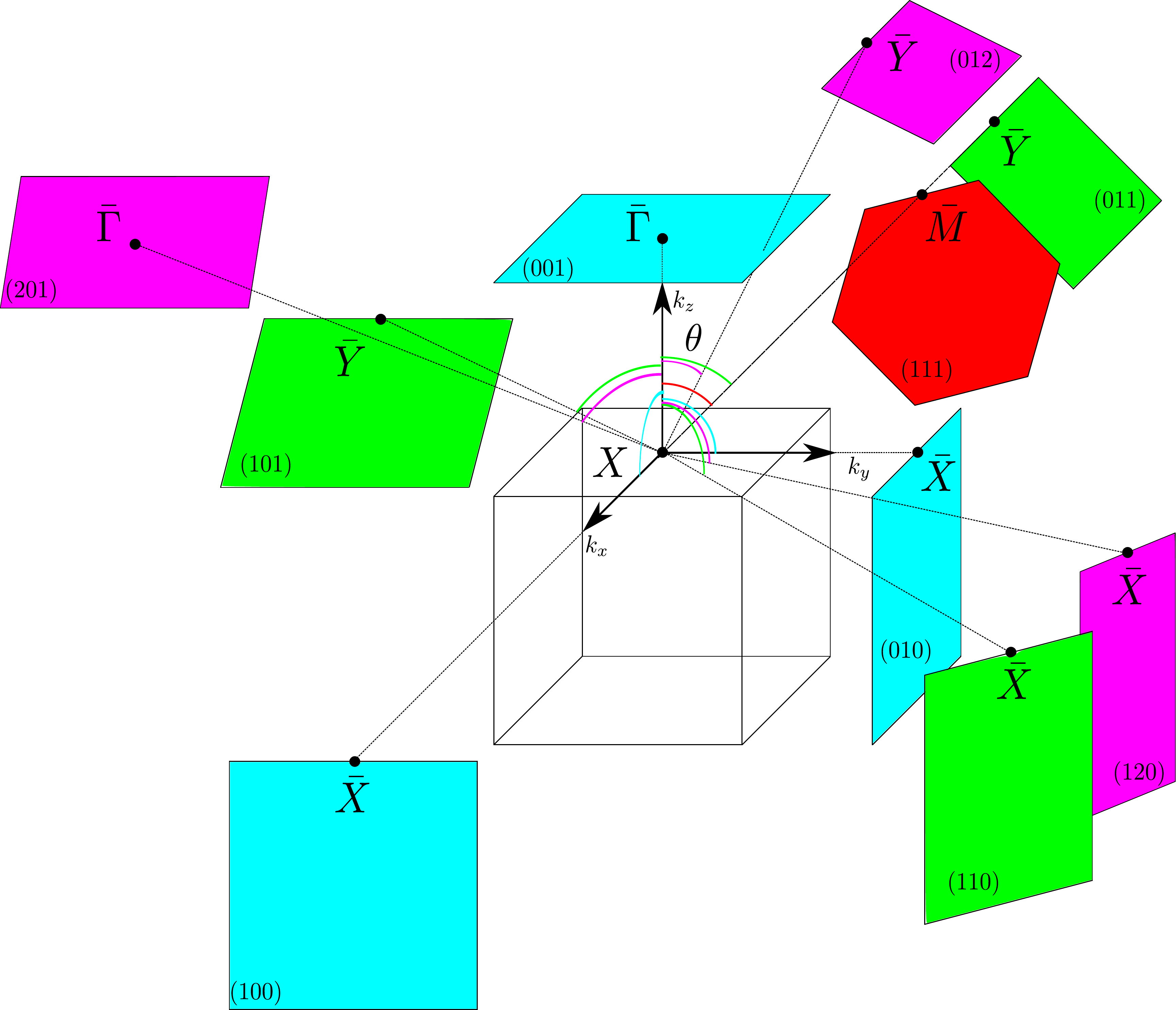}
\caption{The projection along different directions gives rise to different surface BZs, and the $X=(0,0,\pi)$ point is projected on different surface HSP.
For a given $(lmn)$ surface, we project the $X$ point along the $(lmn)$, $(mnl)$ and $(nlm)$ directions,
(we use the same color for each triplet $(lmn)$),
giving rise to three surface Dirac cones.
The only relevant parameter for each surface cone is the angle $\theta$ between $k_z$ and the projection direction, which enters the expression for the winding number Eq. \eqref{wbar}.
}\label{fig_proj}
\end{figure}

However, as argued in Section \ref{sec_x}, it is advantageous to always project $X$, since this allows to safely neglect subspace~2. Hence, instead of considering three distinct $X$ points and projecting them onto the same $(lmn)$ surface, we follow the equivalent procedure of only considering the single $X=(0,0,\pi)$ point which we project onto the three $(lmn)$, $(mnl)$, $(nlm)$ surfaces, as sketched in Fig. \ref{fig_proj}. Eqs. \eqref{theta_lmn} and \eqref{phi_lmn} continue to apply, but with a permutation of the indices $l$, $m$, $n$ in such a way that, given a $(lmn)$ surface, the triplets $(lm/n)$, $(mn/l)$, $(nl/m)$ correspond each to one of the cones.\cite{suppl_tci_long}
Specifically, on the $(001)$ surface the $\bar\Gamma$ cone corresponds to the triplet $(00/1)$ while the $\bar X$ and $\bar X'$ cones to $(10/0)$ and $(01/0)$, respectively. Similarly, the $\bar X$ cone on the $(110)$ surface corresponds to $(11/0)$, while $(10/1)$ and $(01/0)$ correspond to the two $\bar Y$ cones. Finally, on the $(111)$ surface all cones are equivalent.
We stress that, in our approximation, all Dirac cones will live in subspace~1, but with a cone-dependent orbital quantization axis.


\subsection{Results for a generic surface}

In this Subsection we find the effective surface Hamiltonian for a given $(lm/n)$ triplet;
this Hamiltonian is valid for small momenta around the surface point $\bar\bk_X$ on which the bulk $X=(0,0,\pi)$ point is projected; details are given in the supplement.\cite{suppl_tci_long}
As before, we focus on subspace~1, so ignoring any coupling to subspace~2, and for the moment use only $\Gamma_8$ states.

We find that the effective Hamiltonian, up to the linear term in $\bar \bk_\parallel$, is:
\bea
H^{eff}_{\theta} 
&=&v_1 \bar k_x \hat s_y - v_2 \bar k_y \hat s_x\nonumber\\
&=&|v_1|w  \bar k_x \hat s_y - |v_2| \bar k_y \hat s_x,\label{hefflmn}
\eea
where $v_1$ and $v_2$ depend on $\theta$:
\bea
v_1(\theta)&=&2|Vh_1^v|\frac{f_1^v}{\bar f_1^v(\theta)}\frac{\sqrt{-t_dt_f\bar g_1^d(\theta)\bar g_1^f(\theta)}}{t_f\bar g_1^f(\theta)-t_d \bar g_1^d(\theta)},\label{v_1_theta}\\
v_2(\theta)&=&2|Vh_1^v|\frac{\sqrt{-t_dt_f\bar g_1^d(\theta)\bar g_1^f(\theta)}}{t_f\bar g_1^f(\theta)-t_d \bar g_1^d(\theta)}>0,\label{v_2_theta}
\eea
with
\bea
\bar g_1^a(\theta)&=&g_1^a\cos^2\theta+l_1^a \sin^2\theta<0, \hspace{5pt} a=d/f,\label{bar_g1a}\\
\bar f_{1}^v(\theta)&=&\sqrt{(f_1^v)^2\cos^2\theta+(h_1^v)^2\sin^2\theta}\sgn(h_1^v).\label{barf1v}
\eea

Limiting cases are $|v_1(0)|=|v_2(0)|=|v_0|$ from Eq. \eqref{v_0}, $v_1(\pi/2)=v_1$ from Eq. \eqref{v_1} and $v_2(\pi/2)=v_2$ from Eq. \eqref{v_2}.
Moreover, the centre of the cone is at the energy:
\bea
E(\theta)=\frac{\epsilon_1^f t_d \bar g_1^d(\theta)-\epsilon_1^d t_f \bar g_1^f(\theta)}{t_d \bar g_1^d(\theta)-t_f \bar g_1^f(\theta)}.\label{e_theta}
\eea

Eq. \eqref{hefflmn} has the same form as Eq. \eqref{heffx} for the $\bar X$ cone on the $(001)$ surface,
but the basis is in general different, as well as the values of the velocities $v_1$, $v_2$.
It also has formally the same spectrum $E_{\bar\bk}=\pm \epsilon_{\bar\bk}=\pm \sqrt{v_1^2 \bar k_x^2 + v_2^2 \bar k_y^2}$,
which gives rise to elliptic isoenergy contours.

When we look at pseudospin operators, we discover that $\hat s_x$ is in general not simply proportional to $\hat \sigma_x$, but contains a $\hat\sigma_z$ component as well:
\bea
\langle \vec {\hat\sigma} \rangle &=&\left(A_\theta \hat s_x+ B_\theta \gamma^- \hat s_z, \hat s_y,  -B_\theta \hat s_x+A_\theta \gamma^- \hat s_z\right),
\eea
with
\bea
A_\theta&=&\frac{|h_1^v| \sin^2\theta +w|f_1^v| \cos^2\theta }{|\bar f_1^v|},\label{a_theta}\\
B_\theta&=&\frac{|h_1^v|  -w|f_1^v|  }{|\bar f_1^v|}\sin\theta\cos\theta.\label{b_theta}
\eea

As a consequence, we find:
\be
\hat s_x= A_\theta \hat \sigma_x - B_\theta \hat \sigma_z,\hspace{10pt}\hat s_y=\hat \sigma_y,
\ee
since $A_\theta^2+B_\theta^2=1$.
Inserting these expressions into Eq. \eqref{hefflmn} we obtain the Hamiltonian in terms of pseudospin operators:
\bea
&&H^{eff}_{\theta}=v_1 \bar k_x \hat\sigma_y - v_2 \bar k_y A_\theta \hat \sigma_x + v_2 \bar k_y B_\theta \hat\sigma_z\\
&&=|v_1|w \bar k_x \hat \sigma_y-|v_2 A_\theta| w \bar w_d(\theta) \bar k_y \hat \sigma_x+ v_2 B_\theta \bar k_y \hat \sigma_z\nonumber\\
&&\equiv|v_x|w \bar k_x \hat \sigma_y - |v_y| w \bar w_d(\theta) \bar k_y \hat \sigma_x + v_\perp \bar k_y \hat \sigma_z,\label{hefflmn_sigma}
\eea
with $v_x=v_1$, $v_y=v_2 A_\theta$, $v_\perp=v_2 B_\theta$, and
\bea
\bar w_d(\theta)&=&\sgn\left(|f_1^v|\cos^2\theta  + w|h_1^v|\sin^2\theta \right) \nonumber\\
&=&\sgn\left[|f_1^v|n^2  + w|h_1^v|(l^2+m^2)\right]. \label{wbard}
\eea
Equivalently this can be written as:
\bea
H^{eff}_{\theta}&=&\epsilon_{\bar\bk} (n^x_{\bar\bk} \hat\sigma_x + n^y_{\bar\bk} \hat \sigma_y + n^z_{\bar\bk} \hat\sigma_z)=\epsilon_{\bar\bk}\vec n_{\bar\bk} \cdot \vec{\hat\sigma},\label{hefflmn_versor}\\
\vec n_{\bar\bk}&=&(-v_y\bar k_y ,v_x\bar k_x,v_\perp\bar k_y )/\epsilon_{\bar\bk},
\eea
with $\vec n_{\bar\bk}$ a unit vector; this is Eq.~\eqref{heff_intro} quoted in the introduction.
Hence, surface states are eigenstates of the pseudospin operator $\vec n_{\bar\bk} \cdot \vec{\hat\sigma}$; due to spin-orbit coupling, surface states are never eigenstates of the physical spin operator $\vec  {\hat S}$.

We can read off that the pseudospin of the state $|\phi^{+}(\bar\bk)\rangle$ with positive energy $\epsilon_{\bar\bk}$ is $\vec n_{\bar\bk}$, or:
\bea
&&\langle \phi^{+}(\bar\bk)| \vec{\hat\sigma} |\phi^{+}(\bar\bk)\rangle=\vec n_{\bar\bk}\nonumber\\
&&=\left(-\sin\theta_{\bar\bk} |A_\theta| w\bar w_d(\theta),w\cos\theta_{\bar\bk} ,\sin\theta_{\bar\bk} B_\theta \right),\label{psevlmn}
\eea
where we have defined
$\sin\theta_{\bar\bk}=|v_2|\bar k_y/\epsilon_{\bar\bk}$, $\cos\theta_{\bar\bk}=|v_1|\bar k_x/\epsilon_{\bar\bk}$.
For the SEV we find:
\bea
&&2\langle \phi^{+}(\bar\bk)| \vec  {\hat S} |\phi^{+}(\bar\bk)\rangle\nonumber\\
&&=\left(-\sin\theta_{\bar\bk} |A_\theta^+| w\bar w(\theta),\gamma_5^+w\cos\theta_{\bar\bk} ,\sin\theta_{\bar\bk} B_\theta^+ \right),\label{sevlmn}
\eea
with
\bea
A_\theta^+&=&\frac{|h_1^v| \gamma_{11}^+\sin^2\theta +w|f_1^v|\gamma_5^+ \cos^2\theta }{|\bar f_1^v|},\label{a_theta_p}\\
B_\theta^+&=&\frac{|h_1^v| \gamma_{11}^+-w|f_1^v| \gamma_5^+ }{|\bar f_1^v|}\sin\theta\cos\theta,\label{b_theta_p}
\eea
and
\bea
\bar w(\theta)&=&\sgn\left(|f_1^v| \gamma_5^+ \cos^2\theta  + w|h_1^v|\gamma_{11}^+ \sin^2\theta \right)\nonumber \\
&=&\sgn\left[|f_1^v| \gamma_5^+ n^2  + w|h_1^v|\gamma_{11}^+(l^2+m^2)\right]. \label{wbar}
\eea

Eqs. \eqref{psevlmn}, \eqref{sevlmn} constitute central results of this section, to be analyzed in the following.
We first notice that there is no $\phi$ dependence,
as a consequence of the cylindrical symmetry of the \kdp Hamiltonian Eq. \eqref{hkdotp_1}.
%
For the in-plane SEV, we can define a $\theta$-dependent winding number $\bar w(\theta)$.
The winding number can be simply found by looking at the relative sign of the in-plane SEV component along $\bar k_x$ and $\bar k_y$, to get $\bar w(\theta)=\sgn(w A_\theta^+)$, leading to Eq. \eqref{wbar}.
%

First, we will consider for simplicity $\bar w_d(\theta)$ from Eq. \eqref{wbard},
which represents the winding number of the pseudospin, see Eq. \eqref{psevlmn}.
This winding number $\bar w_d(\theta)$ depends both on the surface geometry via the angle $\theta$ as well as on microscopic details of the material via \kdp parameters $f_1^v$ and $h_1^v$.
As a consequence, $\bar w_d(\theta)$ is in general not uniquely determined by $w\equiv\sgn(\C^+_{k_z=0}\C^+_{k_z=\pi})$, the latter characterizing the topological phase.
The only exceptions are $\theta=0$ and $\theta=\pi/2$.
In particular,
for $\theta=0$,  $\bar w_d(0)=+1$, which says that the $\bar\Gamma$ cone on the $(001)$ surface always has a positive winding number due to its high symmetry (see Section \ref{sec:gamma}).
When $\theta=\pi/2$, instead, $\bar w_d(\pi/2)=w$, which says that for the $\bar X$ cone on the (100) surface the winding number depends directly on the topological phase (see Section \ref{sec_x}). This also applies to the $\bar X$ cone of all $(lm0)$ surfaces (but only at small momenta, whereas on the $(001)$ surface it holds at any momenta as a consequence of mirror planes\cite{smb6_prl_io,sigrist_tki}).
We further notice that $\bar w_d(\theta)$ is always positive if $w=+1$. In contrast, for $w=-1$ there exist a critical angle $\theta_c=\arctan (f_1^v/h_1^v)^2$ such that $\bar w_d(\theta)=+1$ ($-1$) for $\theta<\theta_c$ ($\theta>\theta_c$), respectively.

The qualitative behaviour of the winding number of the SEV, $\bar w(\theta)$, is very similar to the one of the pseudospin, $\bar w_d(\theta)$.
It displays a different critical value $\theta_c$ if $w=-1$, dictated by $A^+_{\theta_c}=0$ instead of $A_{\theta_c}=0$, but remains $+1$ always if $w=+1$.

We then note that the SEV perpendicular to the surface is in general nonzero, unless $\theta=0$, $\theta=\pi/2$, or $f_1^v=h_1^v$ (which corresponds to the limiting case of a hybridization with cubic symmetry in the \kdp Hamiltonian).
Being proportional to $\bar k_y$, it will point along the positive $z$ direction on half of the cone,
and along the negative direction on the other half.
We stress that when $w=+1$, the out-of-plane component of the SEV and of the pseudospin is likely to be small, since the two terms of Eqs. \eqref{b_theta_p}, \eqref{b_theta}
tend to cancel each other ($|f_1^v| \approx |h_1^v|$),
while is expected to be large were the $w=-1$ phase realized, since in that case the two terms would sum.
In this case, the effect would be mostly visible close to $\theta_c$; in particular, exactly at $\theta_c$, at $\bar k_x=0$ the SEV would point perpendicular to the surface.

For $\Gamma_7$ states, results for the pseudospin are identical;
for the real spin, we have to substitute $\gamma_5^\pm, \gamma_{11}^\pm \rightarrow \gamma_5^\mp$:
as usual, this implies that the SEV has opposite sign with respect to $\Gamma_8$ states, so it is antiparallel to the pseudospin.

For the $\Gamma_7$-$\Gamma_8$ case, results for the pseudospin are identical, with the usual redefinition of $w$ according to Eq. \eqref{w_17};
for the real spin, we have to substitute $\gamma_5^\pm, \gamma_{11}^\pm \rightarrow \gamma_5^{\pm'},\gamma_{11}^{\pm'}$.
As already remarked in Section \ref{sec_x}, the SEV is usually either parallel (when $|\beta_1|\gg|\beta_7|$) or antiparallel (when $|\beta_7|\gg|\beta_1|$) to the pseudospin,
while for $|\beta_1|\sim|\beta_7|$ pathological situations can arise, in which the SEV is somewhere parallel, somewhere else antiparallel to the pseudospin in a momentum-dependent way;
we, however, ignore this (unlikely) situation.

\begin{figure}[tb]
\includegraphics[width=0.48\textwidth]{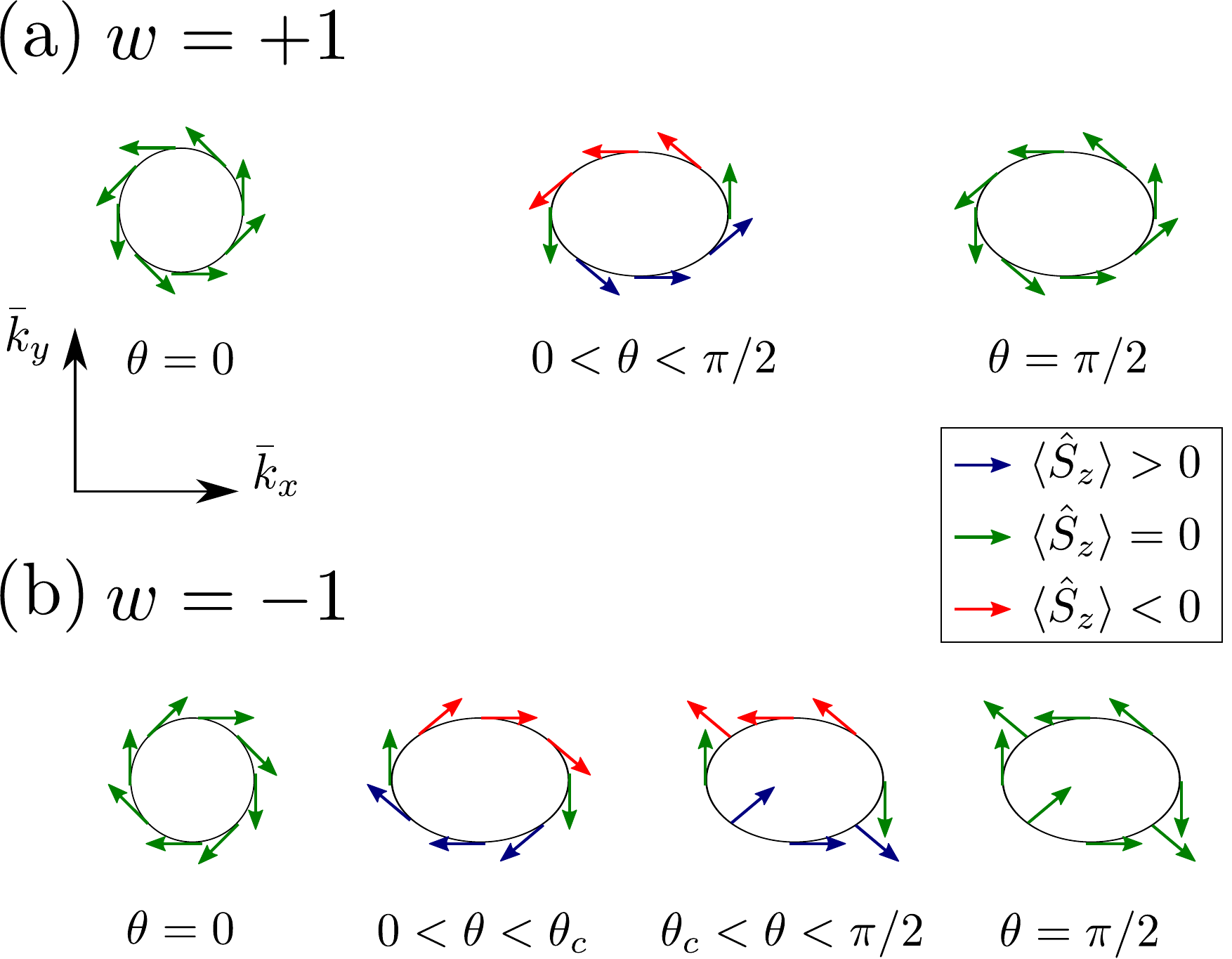}
\caption{Possible scenarios for the SEV of a Dirac cone at linear order in the \kdp expansion on a general $(lmn)$ surface,
as a function of $w\equiv\sgn(\C^+_{k_z=0}\C^+_{k_z=\pi})$ and of $\theta\equiv \arctan{\sqrt{l^2+m^2}}/{n}$.
(a) When $w=+1$ the winding number is positive for every $\theta$, $\bar w(\theta)=+1$.
When $\theta \ne 0, \pi/2$, the SEV acquires an out-of-plane component (in the figure encoded by the color of the arrows), which is proportional to $\bar k_y=0$, and depends on the parameter $B_\theta^+$ of Eq. \eqref{b_theta_p},
which is in general small.
(b) When $w=-1$, the winding number is negative when $\theta$ is larger than a critical value $\theta_c$, and in particular when $\theta=\pi/2$, and positive when $\theta <\theta_c$,
and in particular when $\theta=0$.
Like for $w=+1$, when $\theta \ne 0, \pi/2$, the SEV acquires an out-of-plane component, which in this case can be large.
}\label{fig_sev_kdp}
\end{figure}

The possible scenarios for the spin structure are summarized in Fig. \ref{fig_sev_kdp}.
We stress that these results refer to the $\omega=0$ case; for finite $\omega$ one has to rigidly rotate these patterns by an angle $\omega$.

We now apply this general theory to a few particular cases.
We note that, when compared to tight-binding results, the values of the velocities Eqs. \eqref{v_1_theta}, \eqref{v_2_theta} are not exact.
As explained in the supplement,\cite{suppl_tci_long} higher-order terms in $\bk$ need to be kept to reproduce these velocities exactly:
while those can be easily taken into account for the $(001)$ surface, on a general $(lmn)$ it is not straightforward, so in what follows we will stick to the simple theory of this Section.
We finally remark that keeping more terms will also in general break the cylindrical symmetry of the Hamiltonian \eqref{hkdotp_1} by introducing a $\phi$ dependence in the effective Hamiltonian.

\subsection{Results for $(001)$ surface}

First, we can obtain again the results of Sections \ref{sec:gamma} and \ref{sec_x} for a $(001)$ surface.
For the $\bar\Gamma$ cone, $\theta=0$, $\bar w(0)=\bar w_d(0)=+1$, $B_0=0$, $A^+_0=w\gamma_5^+$, $A_0=w$.
The effective Hamiltonian \eqref{hefflmn_sigma}
corresponds to Eq. \eqref{heffvw} with $v_x=v_y\equiv v_0$, (there we showed $\hat s_{x,y}=\hat \sigma_{x,y}$).
The SEV \eqref{sevlmn} becomes 
Eq. \eqref{sev_gamma};
the pseudospin \eqref{psevlmn} gives 
Eq. \eqref{psev_gamma}.

For the $\bar X$ cone, $\theta=\pi/2$, $\bar w(\pi/2)=\bar w_d(\pi/2)=w$, $B_{\pi/2}=0$, $A^+_{\pi/2}=\gamma_{11}^+$, $A_{\pi/2}=1$.
The effective Hamiltonian \eqref{hefflmn_sigma} becomes 
Eq. \eqref{heffx}.
The SEV \eqref{sevlmn} becomes 
Eq. \eqref{sev_x};
the pseudospin \eqref{psevlmn} gives 
Eq. \eqref{psev_x}.
Results for the $\bar X'$ cone can be achieved by rotating the results for $\bar X$ by the same angle $\omega=\pi/2$
in both momentum and spin space.

\begin{figure}[tb]
\includegraphics[width=0.48\textwidth]{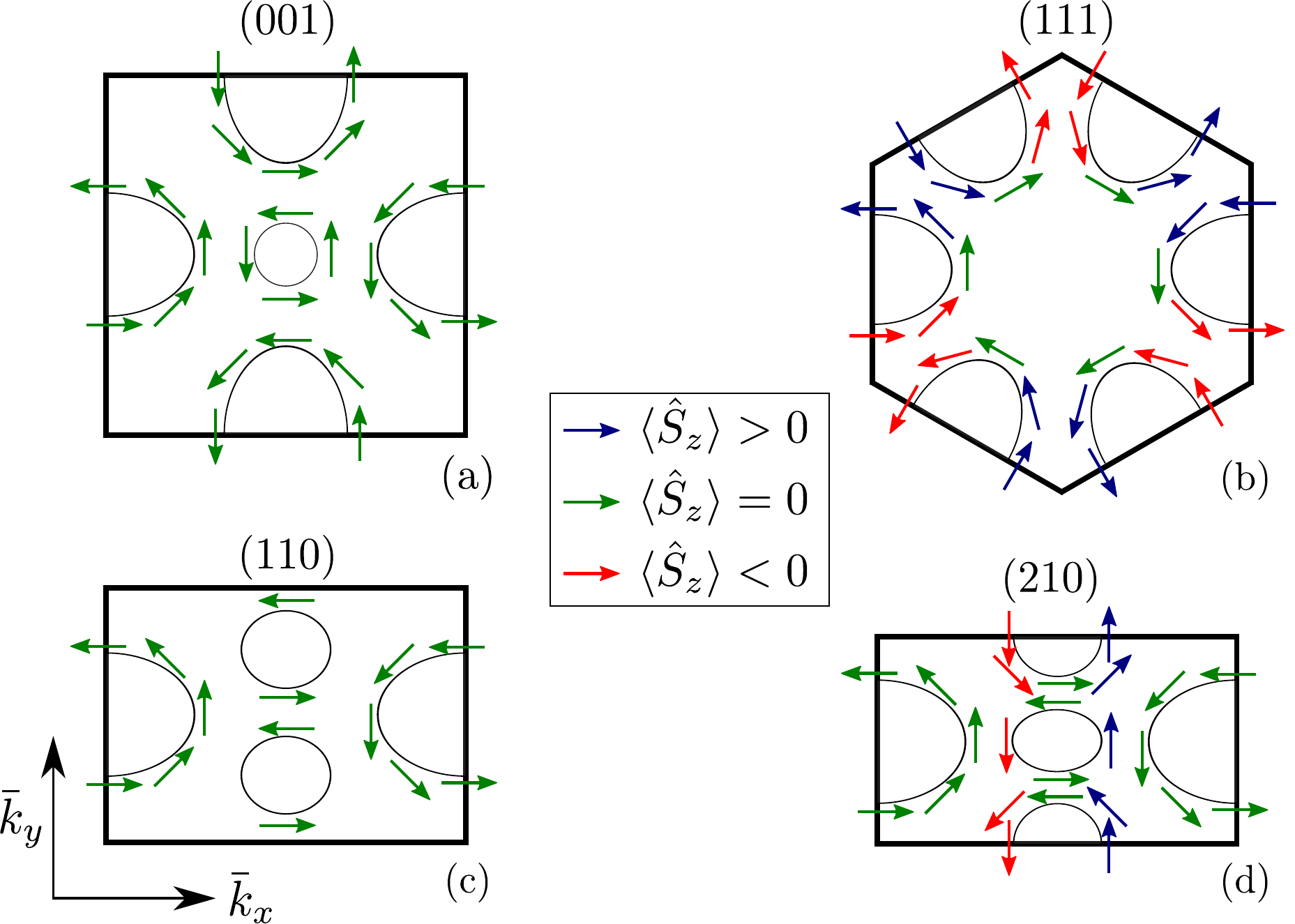}
\caption{Predicted SEV for surface states above the Dirac energies from the perturbative calculation in the small-momentum limit, when we assume $w=+1$, for all the surfaces considered in this paper: (a) $(001)$, (b) $(111)$, (c) $(110)$ and $(lm0)$ with $l+m$ even, (d) $(210)$ and $(lm0)$ with $l+m$ odd.
In all Dirac cones at high symmetry points the winding number is positive; for Dirac cones not at high symmetry points, i.e. the two central cones in case (c), our perturbative approach cannot be applied so we cannot make predictions,
except along the high symmetry direction $\bar k_x=0$.
The color of the arrows encode the expectation value of the spin perpendicular to the surface as in Fig. \ref{fig_sev_kdp};
this value is generally small.
}\label{fig_sev}
\end{figure}

Predictions for the SEV in the small-momentum limit for this surface, when $w=+1$, are shown in Fig. \ref{fig_sev}(a). As remarked, these predictions hold also for larger momenta along high-symmetry directions.

\subsection{Results for $(110)$ surface}

On this surface one bulk $X$ point is projected onto $\bar X$, while two $X$ points are projected onto $\bar Y$.
We start with $\bar X$ which corresponds to the triplet $(11/0)$, hence giving $\theta=\pi/2$. We observe that it is the same value of $\theta$ which describes the $\bar X$ cone on the $(001)$ surface; hence, we can apply most of the results of Section \ref{sec_x} and of the previous subsection. In particular, the effective Hamiltonian is:
\bea
H^{eff}_{\theta=\pi/2}=|v_1| w \bar k_x \hat \sigma_y - |v_2| \bar k_y \hat \sigma_x,\label{heff110}
\eea
and the SEV on eigenstates is given by
\bea
&&2\langle \phi^{+}(\bk)| \vec  {\hat S} |\phi^{+}(\bk)\rangle=\left(-\sin\theta_\bk\gamma_{11}^+,\cos\theta_\bk\gamma_5^+,0\right),
\eea
so the SEV lies in the surface plane, and $w$ still denotes the SEV winding number.
The only difference with respect to Section \ref{sec_x} is that here we cannot express our basis with a quantization axis perpendicular to the surface without enlarging the basis (as we did in Eq.~\eqref{psipmp}) because a $\phi=\pi/4$ rotation does not belong to cubic symmetry operations. We can only state that, in our approximation, surface states at $\bar X$ are composed of $\Gamma_8^{(1)}$ and $d_{x^2-y^2}$ states w.r.t. a quantization axis parallel to $\bar k_x$.

In fact, this same theory applies to all $(lm0)$ surfaces, which all have a Dirac cone at $\bar X$. For the $(110)$ surface, which has $C_{2v}$ symmetry, the spin remains within the surface plane
even beyond the present approximation, see Section \ref{sec_x}. However, other surfaces have only $C_s$ symmetry, not containing a rotation by $\pi$, and the SEV can point out of the surface beyond this approximation.

Turning to the cones at $\bar Y$, we note that in the low-order \kdp approximation they are projected exactly at the same energy, which is given by Eq. \eqref{e_theta} with $\theta=\pi/4$.
We also find $\omega'=-\pi/2$, $\omega''=\pi/2$, which implies that one cone is rotated by $\pi$ with respect to the other one. This means that their combined SEV perpendicular to the surface is zero in agreement with $C_{2v}$ symmetry.

\begin{figure}[tb]
\includegraphics[width=0.48\textwidth]{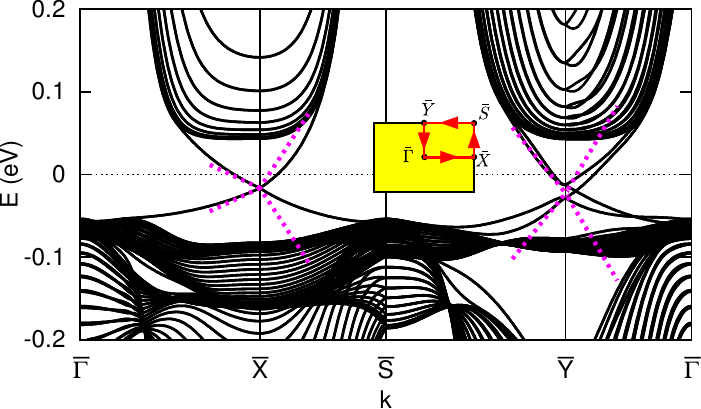}
\caption{
Tight-binding bandstructure for a $(110)$ slab of 30 layers, together with the analytical \kdp approximation of surface states for the same parameters as Fig. \ref{fig_bands_001}.
Note that the low-order \kdp approximation yields two identical cones at $\bar Y$, while in the full solution the two cones hybridize and gap out, except along the $\bar Y$--$\bar \Gamma$ direction, where two new cones protected by mirror symmetry appear (only one is shown).
}\label{fig_bands_110}
\end{figure}

However, when solving the tight-binding model\cite{smb6_tci} the two cones are projected at {\em different} energies. They then anticross, hence get gapped and become topological trivial,
except along the $\bar\Gamma$--$\bar Y$ direction where their crossing leads to two new cones protected by mirror symmetry; see Fig. \ref{fig_bands_110}. These cones are topological nontrivial but not originating from parity invariants, such that the \kdp method is not applicable as noted before.

Predictions for the SEV in the small-momentum limit for this surface, when $w=+1$, are shown in Fig. \ref{fig_sev}(c).

We note that, provided that the Fermi energy lies above the Dirac energy of all cones, our results are compatible with that of a very recent ARPES experiment\cite{smb6_arpes_110} on {\sm} where two surface states were observed, centered at $\bar X$ and $\bar Y$, respectively.
In that case, the signal at $\bar Y$ should arise from two nearly-degenerate cones.

\subsection{Results for $(111)$ surface}

In this case all indices are equal, and $\theta=\arctan\sqrt{2}$.
The $X$ point is projected at $\bar M=(2\pi/\sqrt{6},0)$.
%
This is the situation in which none of the terms in Eq. \eqref{hefflmn_sigma} vanishes. The SEV can point out of the surface, and the winding number depends on model details according to:
\be
\bar w_d(\theta=\arctan\sqrt{2})=\sgn(2w|h_1^v|+|f_1^v|).\label{wbar_d_111}
\ee
The two other cones, at $\bar M'$ and $\bar M''$, are equivalent to the one at $\bar M$, and their effective Hamiltonians and SEV
can be found after a $\omega',\omega''=\pm 2\pi/3$ rotation.

\begin{figure}[tb]
\includegraphics[width=0.48\textwidth]{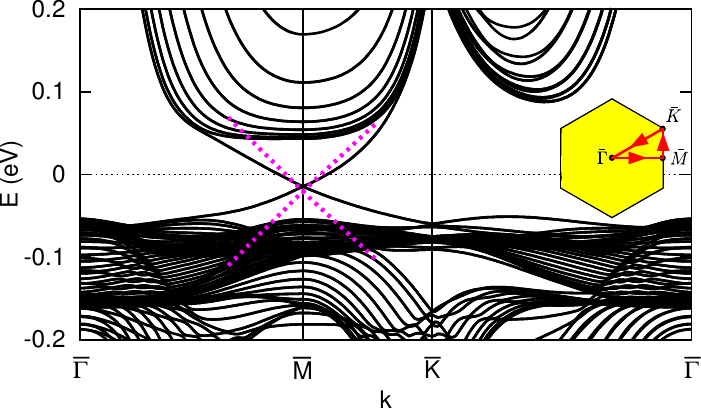}
\caption{
Tight-binding bandstructure for a $(111)$ slab of 30 layers, and analytical \kdp approximation of surface states
for the same parameters as Fig. \ref{fig_bands_001}.
We show just one of the three equivalent $\bar M$ points.
}\label{fig_bands_111}
\end{figure}

Predictions about the SEV in the small-momentum limit for this surface, when $w=+1$, are shown in Fig. \ref{fig_sev}(b);
an example of the bandstructure is given in Fig. \ref{fig_bands_111}.

\subsection{Results for $(210)$ surface}

Now we consider a $(lm0)$ surface for which $\theta=\pi/2$.
Eqs. \eqref{barbk}, \eqref{barbkp}, \eqref{barbkpp} yield: 
\bea
\bar \bk_X&=&(\pi,0),\\
\bar\bk_{X'}&=&(0,-\pi\sin\phi)=(0,-m\pi/\sqrt{l^2+m^2}),\\
\bar \bk_{X''}&=&(0,\pi\cos\phi)=(0,l\pi/\sqrt{l^2+m^2}).
\eea
The 2D BZ is defined by $\bk_1=(2\pi,0)$, $\bk_2=(0,2\pi/\sqrt{l^2+m^2})$, so $\bar \bk_{X''}-\bar \bk_{X'}=(0,(l+m)\pi/\sqrt{l^2+m^2})$ is zero up to a multiple of $\bk_2$ if $l+m$ is even,
while it is equal to $\bk_2/2=(0,\pi/\sqrt{l^2+m^2})$ when $l+m$ is odd.
We thus arrive at the conclusion that if $(l+m)$ is even, both $X'$ and $X''$ are projected onto the same 2D BZ point, and parity invariants predict a single Dirac point at $\bar X$,
while, if $(l+m)$ is odd, $X'$ and $X''$ are projected onto different 2D BZ points, and parity invariants predict three Dirac points at $\bar X$, $\bar Y$, $\bar \Gamma$.
In the first case, to which the $(110)$ surface belongs, mirror Chern numbers still predict two Dirac cones along the $\bar Y$--$\bar \Gamma$ direction;
while, in the second case, to which the $(210)$ surface belongs, there are already two Dirac cones along this direction predicted by parity invariants, so mirror Chern numbers do not predict any more cones.

In addition, when $l\ne m$, so, for all these surfaces except the $(110)$ one, the $k_x=k_y$ mirror plane is no more projected onto $\bar k_y=0$,
while $k_z=0,\pi$ is always projected onto $\bar k_x=0,\pi$, so two mirror planes survive.
However, no Dirac cone is cut by two mirror planes, so we cannot in general define winding numbers without resorting to a concrete model;
and, when it exists, the $\bar\Gamma$ cone is anisotropic, since $\bar k_x$ and $\bar k_y$ correspond to inequivalent bulk directions, except on the $(001)$ surface.


For the $(210)$ surface, the cone at $\bar\Gamma$ is described by $\theta=\arctan 2$,
and the cone at $\bar Y$ by $\theta=\arctan(1/2)$;
$\bar X$ corresponds to $\theta=\pi/2$, always leading to winding number $\bar w(\theta)=w$.
We can thus see that, varying $l$ and $m$, this class of surfaces allows to tune the winding number on the $\bar\Gamma$ and $\bar Y$ cones when $w=-1$.

\begin{figure}[tb]
\includegraphics[width=0.48\textwidth]{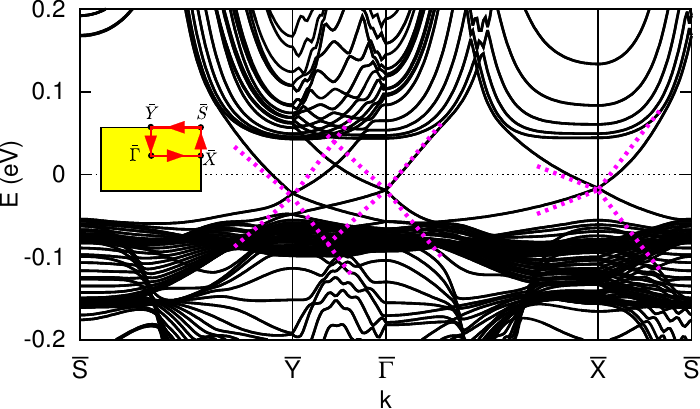}
\caption{
Tight-binding bandstructure for a $(210)$ slab of 30 layers, and analytical \kdp approximation of surface states
for the same parameters as Fig. \ref{fig_bands_001}.
}\label{fig_bands_210}
\end{figure}

We remark that, except for the $(110)$ one, these surfaces have $C_s$ symmetry, which does not contain rotations by $\pi$, so in general the SEV can point also out of plane:
this happens already in the low-order \kdp approximation for cones $\bar\Gamma$ and $\bar Y$, while for $\bar X$ we would need to keep more terms in the small-momentum expansion.
A numerical diagonalization of the tight-binding model shows that, with reference to Fig. \ref{fig_sev}(d) (where  we show
predictions about the SEV in the small-momentum limit),
when $w=+1$, $\langle  \hat S_z\rangle<0$ for $\bar k_x-\pi<0$, and $\langle  \hat S_z\rangle>0$ otherwise.
In Fig. \ref{fig_bands_210} we give an example for the bandstructure.

\begin{table*}
\centering
$$
\begin{array}{|c|c|c|c|c|c|c|}\hline
\mbox{Surface}& \mbox{Symm.} &\mbox{Triplet}& \bar \bk_X &\mbox{S. at }\bar \bk_X& \theta & \bar w_d(\theta)\\\hline
(001) &C_{4v}	& (00/1)& \bar \Gamma=(0,0)		&C_{4v}&  0 & 1\\
      &		& (01/0)& \bar X=(\pi,0)		&C_{2v}& \pi/2 & w\\
      &		& (10/0)& \bar X'=(0,\pi)		&C_{2v}& \pi/2 & w\\\hline
(110) &C_{2v}	& (11/0)& \bar X=(-\pi,0)		&C_{2v}& \pi/2 & w\\
      &		& (10/1)& \bar Y=(0,-\pi/\sqrt{2})		&C_{s}& \pi/4^* & \sgn(|f_1^v|+w|h_1^v|)^*\\
      &		& (01/1)& \bar Y=(0,\pi/\sqrt{2})		&C_{s}& \pi/4^* & \sgn(|f_1^v|+w|h_1^v|)^*\\\hline
(111) &C_{3v}	& (11/1)& \bar M=(-2\pi/\sqrt{6},0) 	&C_{s}& \arctan{\sqrt{2}} & \sgn(|f_1^v|+2w|h_1^v|)\\
      &		& (11/1)& \bar M'=(\pi/\sqrt{6},-\pi/\sqrt{2}) 	&C_{s}& \arctan{\sqrt{2}} & \sgn(|f_1^v|+2w|h_1^v|)\\
      &		& (11/1)& \bar M''=(\pi/\sqrt{6},\pi/\sqrt{2}) 	&C_{s}& \arctan{\sqrt{2}} & \sgn(|f_1^v|+2w|h_1^v|)\\\hline
(210) &C_{s}	& (21/0) & \bar X=(-\pi,0) 		&C_{s}& \pi/2	& w\\
      &		& (10/2) & \bar Y=(0,-\pi/\sqrt{5}) 	&C_{s}& \arctan(1/2)	& \sgn(4|f_1^v|+w|h_1^v|)\\
      &		& (02/1) & \bar \Gamma=(0,0) 		&C_{s}& \arctan(2)	& \sgn(|f_1^v|+4w|h_1^v|)\\\hline
(lm0) &C_{s}	& (lm/0) & \bar X=(-\pi,0) 		&C_{s}& \pi/2	& w\\
l \mbox{ even} && (m0/l) & \bar Y=(0,-\frac{\pi}{\sqrt{m^2+n^2}}) 		&C_{s}& \arctan(m/l)	& \sgn(l^2|f_1^v|+wm^2|h_1^v|)\\
m \mbox{ odd}  && (0l/m) & \bar \Gamma=(0,0) &C_{s}& \arctan(l/m)	& \sgn(m^2|f_1^v|+wl^2|h_1^v|)\\\hline
(lmn) &	/	& (lm/n) & \bar\bk_{X}=\pi(-\frac{\sqrt{l^2+m^2}}{\sqrt{l^2+m^2+n^2}},0)&/&	 \arctan(\sqrt{l^2+m^2})/n)	& \sgn(n^2|f_1^v|+w(l^2+m^2)|h_1^v|)\\
      &		& (mn/l) & \bar\bk_{X'}=\frac{\pi}{\sqrt{l^2+m^2}}(\frac{nl}{\sqrt{l^2+m^2+n^2}},-m)&/& \arctan(\sqrt{m^2+n^2})/l)	& \sgn(l^2|f_1^v|+w(m^2+n^2)|h_1^v|)\\
      &		& (nl/m) & \bar\bk_{X''}=\frac{\pi}{\sqrt{l^2+m^2}}(\frac{mn}{\sqrt{l^2+m^2+n^2}},l)&/& \arctan(\sqrt{n^2+l^2})/m)	& \sgn(m^2|f_1^v|+w(n^2+l^2)|h_1^v|)\\\hline
 \end{array}$$
\caption{For each $(lmn)$ surface we report its symmetry, the surface momenta $\bar \bk_X$ to which each of the $X$ points is projected,
the corresponding $(lm/n)$ triplet, which is obtained fixing the $z$ direction in the \kdp Hamiltonian, the symmetry at $\bar \bk_X$,
the angle $\theta$ at which the $X$ point is projected on the surface,
and the pseudospin winding number $\bar w_d(\theta)$ from Eq. \eqref{wbard};
for the physical spin a similar result is given by Eq. \eqref{wbar}.
For $(lm0)$ surfaces we assume $l$ odd, $m$ even.
In the definition of surface momenta $\bar\bk_X$ we keep minus signs to agree with the general $(lmn)$ formulas of the last three rows.
$^*$ On the $(110)$ surface we can formally apply our theory to the two cones at $\bar Y$, but those become topologically trivial since they come in pair.
 }\label{table_cones}
\end{table*}

\subsection{Summary of results on different surfaces}

At this point it is useful to summarize our main results for {\sm} surface states. All Dirac cones can be described by a generalized Dirac Hamiltonian, and the winding number of the SEV, defined ignoring the out-of-plane component of the spin, is $\bar w(\theta)$ in Eq. \eqref{wbar}.
For $\theta=\pi/2$, corresponding to the $\bar X$ cone on the $(001)$ surface, we get
\be
\bar w_d(\pi/2)=\bar w(\pi/2)=w=\sgn(f_1^vh_1^v)=\sgn(\C^+_{k_z=0}\C^+_{k_z=\pi}).
\ee
Making use of the experimental fact\cite{smb6_arpes_mesot_spin} that the winding number of this cone is $w=+1$ we deduce that
\be
\bar w_d(\theta)=\sgn(\sin^2 \theta|h_1^v|+\cos^2\theta|f_1^v|)=+1=\bar w(\theta).
\ee
Hence, the winding number $\bar w(\theta)$ is $+1$ for all Dirac cones on all surfaces; exceptions could by those Dirac cones protected by mirror symmetry only [being present on $(lm0)$ surface with $(l+m)$ even], as the \kdp approach is not applicable there.
A concise summary is in Table~\ref{table_cones} and Fig. \ref{fig_sev}.


To derive these results the following approximations were made, to be discussed in turn:
(i) we have used an effective single-particle approach,
(ii) we have worked in the small-momentum limit,
(iii) we have ignored the coupling to subspace~2,
(iv) we have treated the interplay between $\Gamma_7$ and $\Gamma_8$ states in an approximate way,
(v) we have ignored surface details.

Approximation (i) implies that, after renormalization effects due to the Hubbard repulsion taken into account, the SEV behaves as in the non-interacting picture. This is based on assumptions frequently made in the field of heavy-fermion metals,\cite{hewson} but would need to be verified in many-body calculations based e.g. on dynamical mean-field theory (DMFT).

Approximation (ii) is standard in the context of TI surface states; we remark that, on the $(001)$ surface, the presence of mirror planes allows to extend our results to larger momenta.\cite{smb6_prl_io}

Approximation (iii) can be verified within our model by comparing to tight-binding results, and we have found it justified in all the cases we analyzed.

Approximation (iv) is somewhat delicate, as both $\Gamma_7$ and $\Gamma_8$ states are known to be close to the Fermi energy,\cite{lu_smb6_gutz,smb6_korea2} and, as a consequence, both contribute to surface states. We have argued in Ref. \onlinecite{smb6_prl_io} that a minimal model can ignore $\Gamma_7$ states, but a definite answer requires more accurate ab-initio calculations which are not available at present.

Approximation (v) requires thorough consideration, especially because {\sm} surfaces are known not to cleave well.\cite{pnas_smb6_stm,hoffman_smb6} On the $(001)$ surface topological arguments can help making general claims,\cite{smb6_prl_io,sigrist_tki} but on other surfaces microscopic details may become important. We have recently studied effects of surface reconstruction and surface scattering potentials for the $(001)$ surface within tight-binding models in some detail,\cite{smb6_rec_io} showing that band backfolding and the possibly resulting crossings of Dirac cones are the main effects. Similar studies for other surfaces are left for future work.


\section{Topological phase transitions}
\label{sec_tpt}

The first part of the paper, together with previous work,\cite{smb6_prl_io,sigrist_tki} that the spin winding numbers $\bar w(\theta)$ for surface Dirac cones depend on the relative strength of different hybridization terms, with a central role played by the combination of MCNs $w\equiv\sgn(\C^+_{k_z=0}\C^+_{k_z=\pi})$.
This prompts us to study the possibility of bulk topological phase transitions between states with different $w=\pm 1$ which could be observed as a change in the spin structure of surface states -- this is the subject of the second part of the paper. We will consider the tuning of both hybridization terms and crystal-field splitting, noting that the latter is more likely to be accessible by pressure or doping.


\subsection{Varying the hybridization}
\label{sec:varyhyb}

Theoretically, the easiest way to induce a topological transition is to consider two different hybridization terms in a given model and to change their relative strength.\cite{sigrist_tki}

For example we can take the $E_g$-$\Gamma_8$ model with hybridization terms $\eta_x^{v2}$ and $\eta_z^{v2}$, both leading to a fully insulating phase, the first one with $w=-1$ and the second one with $w=+1$. We now study how the system evolves when we take $\eta_x^{v2}=\cos\xi$, $\eta_z^{v2}=\sin\xi$, so that $\xi=0$ yields $w=+1$ and $\xi=\pm\pi/2$ yields $w=-1$.
When we retain these two hybridization terms, we get:
$f_1^v(\xi)=2\cos\xi+6\sin\xi$, $h_1^v(\xi)=-3\cos\xi+3\sin\xi$,
so $w(\xi)=\sgn[f_1^v(\xi)h_1^v(\xi)]$, and we expect a topological phase transition for $w(\xi_c)=0$,
so when $2\cos\xi+6\sin\xi=0$, leading to $\xi_{c1}=-\arctan(1/3)$,
or when $-3\cos\xi+3\sin\xi=0$, leading to $\xi_{c2}=\pi/4$.

\begin{figure}[tb]
\includegraphics[width=0.49\textwidth]{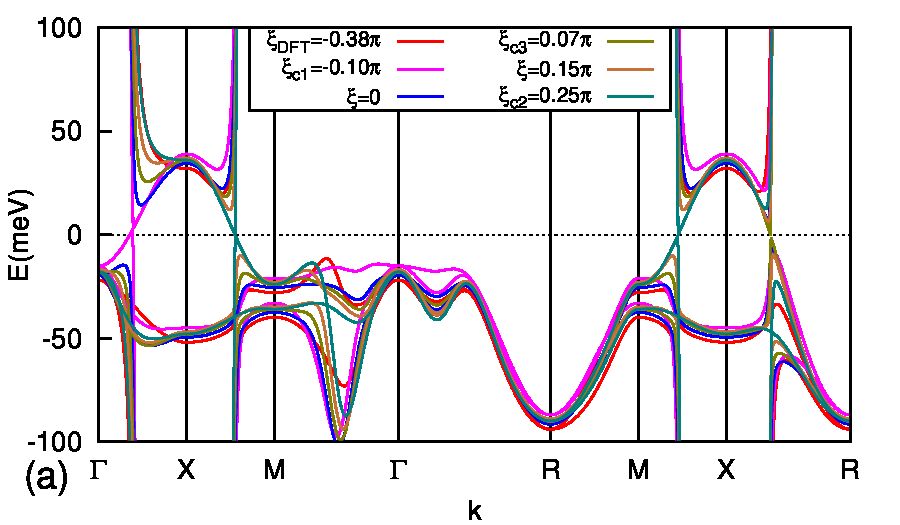}
\includegraphics[width=0.49\textwidth]{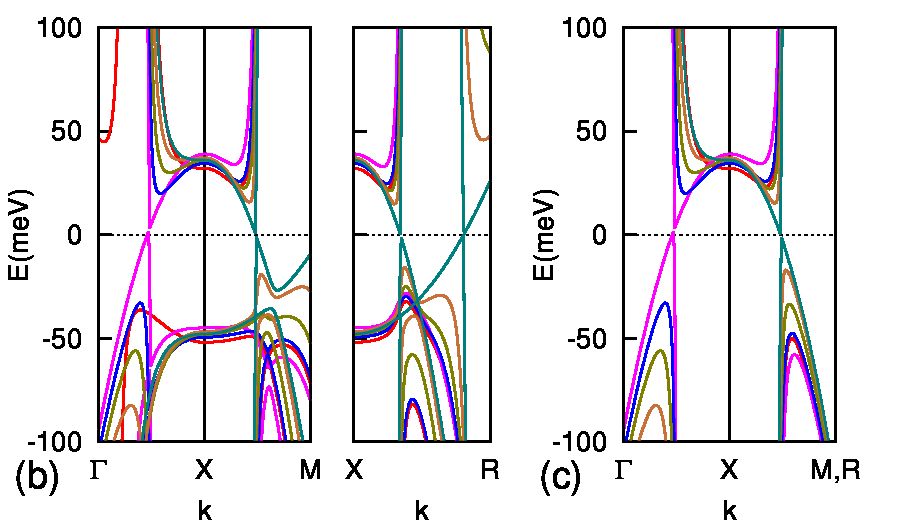}
\caption{Evolution of the bandstructure as a function of the hybridization for the $E_g$ - $\Gamma_8$ model when $\eta_x^{v2}=\cos\xi$, $\eta_z^{v2}=\sin\xi$,
(a) for the tight-binding model, (b) for the \kdp Hamiltonian, (c) for the \kdp Hamiltonian in subspace~1.
In (a) the gap closes along $X$-$\Gamma$ at $\xi_{c1}=-\arctan(1/3)\simeq -0.1\pi$, along $X$-$M$ at $\xi_{c2}=\pi/4$ and along $X$-$R$ at $\xi_{c3}\simeq 0.07\pi$ denoting the topological phase transitions among the three phases
$(+2,+1,-1)$ (red curve), $(-2,+1,+1)$ (blue curve), $(-2,-3,-1)$ (orange curve).
This is summarized in Fig. \ref{fig_tpt1b}(a).
In (b) and (c) $\xi_{c3}=\xi_{c2}$; in (b) the second pair of bands crossing along $X$-$R$ is in subspace~2 due to the vanishing of $h_2^v$. 
Other non-zero parameters are $t_c=0.8$eV, $t_f=-0.015$eV, $V=0.03$eV, $\eta_z^{d1}=\eta_z^{f1}=0.8$, $\eta_z^{d2}=\eta_z^{f2}=-0.3$, $\epsilon_d-\epsilon_8=1.45$eV,
which are chosen to reproduce qualitatively (but not quantitatively) the bandstructure of Refs. \onlinecite{pub6,prbr_io_smb6}.
}\label{fig_tpt1}
\end{figure}

By numerically diagonalizing the tight-binding model, with results shown in Fig.~\ref{fig_tpt1}(a),
we find that this prediction is partially true, but the situation is more involved.
When $f_1^v(\xi)=0$ the gap closes at $\xi=\xi_{c1}$ along the $X$--$\Gamma$ direction,
and when $h_1^v(\xi)=0$ the gap closes at $\xi=\xi_{c2}$ along the $X$--$M$ direction.
However, a third transition at $\xi=\xi_{c3}$, whose value is parameter-dependent, occurs along the $X$--$R$ direction.
To account for this third transition requires to take into account the full momentum dependence of the hybridization term\cite{sigrist_tki} or at least higher-order terms in the \kdp expansion.\cite{suppl_tci_long}
In general the closing of the gap along $X$--$M$ and $X$--$R$ does not happen
at the same energy; when $\xi_{c3}<\xi<\xi_{c2}$, we find a phase with MCNs $(-2,-3,-1)$.
For the \kdp Hamiltonian  with all 8 orbitals, Fig.~\ref{fig_tpt1}(b), and with the 4 orbitals in subspace~1, Fig.~\ref{fig_tpt1}(c), the gap closes for the predicted values of $\xi_{c1}$ and $\xi_{c2}$, but the $(-2,-3,-1)$ phase is not described.

\begin{figure}[tb]
\includegraphics[width=0.49\textwidth]{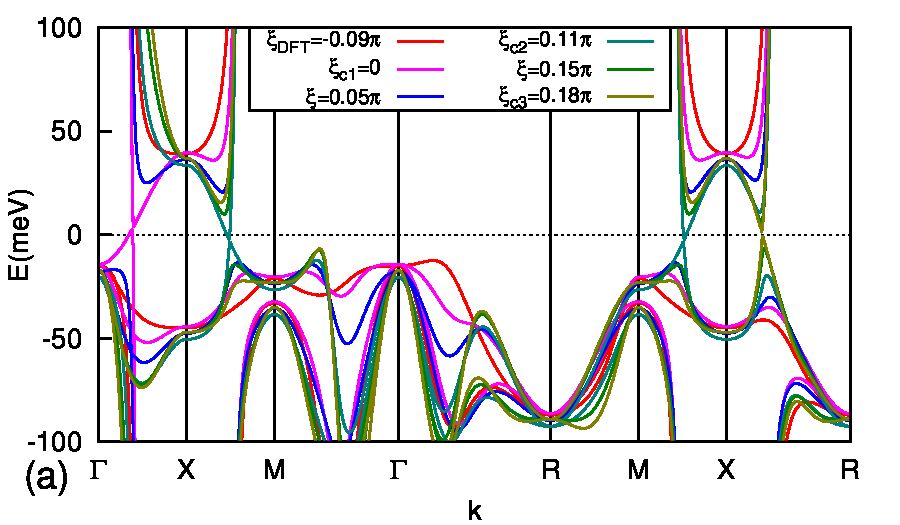}
\includegraphics[width=0.49\textwidth]{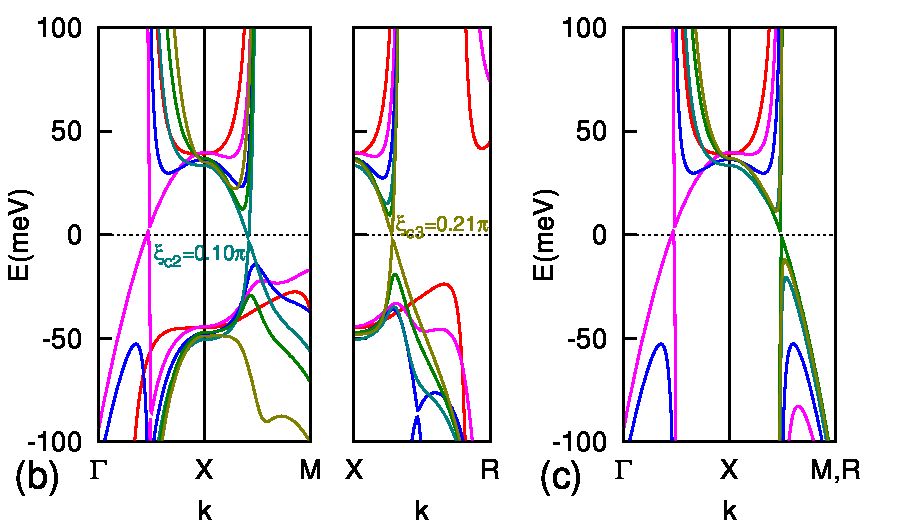}
\caption{Same as Fig. \ref{fig_tpt1} but with $\eta_z^{v1}=\cos\xi$, $\eta_z^{v2}=\sin\xi$.
In (a) the gap closes along $X$-$\Gamma$ at $\xi_{c1}=0$, along $X$--$M$ at $\xi_{c2}\simeq 0.11\pi$ and along $X$--$R$ at $\xi_{c3}\simeq 0.18\pi$ denoting the topological phase transitions among the three phases
$(+2,+1,-1)$ (red curve), $(-2,+1,+1)$ (blue curve), $(+2,-3,+1)$ (green curve).
This is summarized in Fig. \ref{fig_tpt1b}(b).
In (b) $\xi_{c2}=0.10\pi$, $\xi_{c3}=0.21\pi$; in (c) $\xi_{c3}=\xi_{c2}=\arctan(1/2)\simeq 0.15\pi$.
}\label{fig_tpt1-2}
\end{figure}

\begin{figure}[tb]
\includegraphics[width=0.48\textwidth]{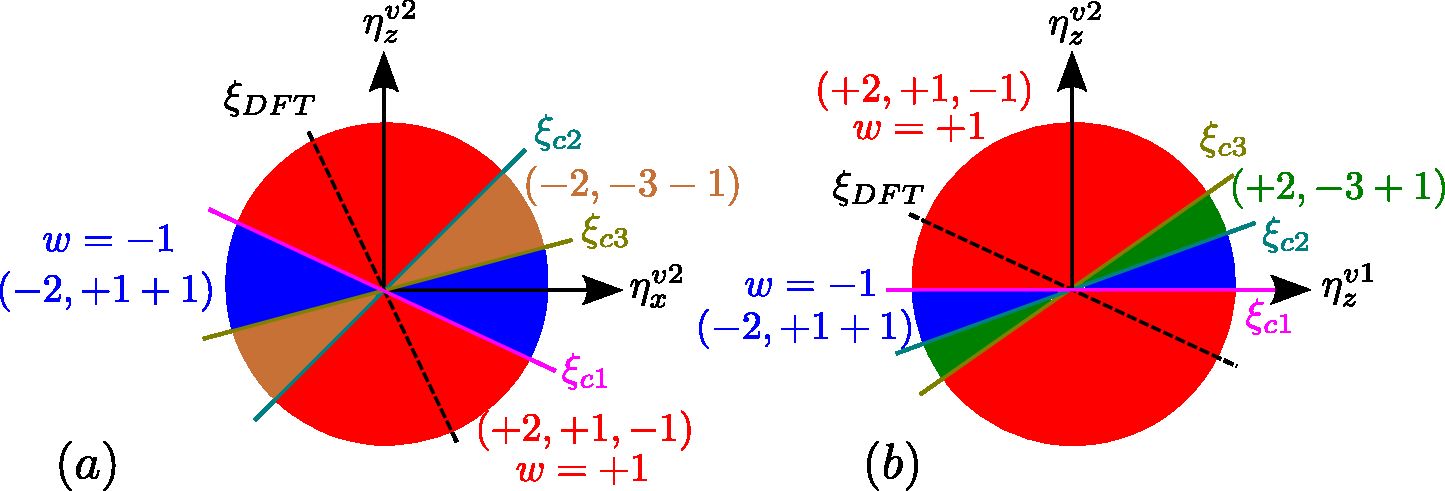}
\caption{Topological crystalline phases as a function of two hybridization parameters,
(a) $\cos\xi=\eta_x^{v2}$, $\sin\xi=\eta_z^{v2}$,
(b) $\cos\xi=\eta_z^{v1}$, $\sin\xi=\eta_z^{v2}$.
The bulk gap closes along $X$-$\Gamma$ at $\xi_{c1}$, along $X$-$M$ at $\xi_{c2}$, along $X$-$R$ at $\xi_{c3}$.
The phases $(+2,+1,-1)$, $w=+1$ and $(-2,+1,+1)$, $w=-1$ always appear;
according to the relative value of $\xi_{c1}$, $\xi_{c2}$, $\xi_{c3}$, a third phase can appear,
which is (a) $(-2,-3,-1)$, corresponding to Fig. \ref{fig_tpt1}, or (b) $(+2,-3,+1)$, corresponding to Fig. \ref{fig_tpt1-2};
these two additional phases are not predicted by the \kdp method, for which $\xi_{c2}=\xi_{c3}$, since they require the knowledge of the full momentum dependence of the hybridization.
In both cases the physical system is in the $(+2,+1,-1)$, $w=+1$, phase, as denoted by $\xi_{DFT}$; we note that the overall sign of the hybridization terms is arbitrary, so $\xi$ is defined modulo $\pi$.
}\label{fig_tpt1b}
\end{figure}

As a second example we consider tuning via $\eta_z^{v1}=\cos\xi$, $\eta_z^{v2}=\sin\xi$.
In this case the \kdp Hamiltonian with 4 orbitals predicts a bulk gap closing at $\xi_{c1}=0$ and $\xi_{c2}=\arctan(1/2)\simeq 0.15\pi$,
as shown in Fig. \ref{fig_tpt1b}(c).
The tight-binding model, Fig. \ref{fig_tpt1b}(a), confirms that $\xi_{c1}=0$, but also gives $\xi_{c2}=0.11\pi$, $\xi_{c3}=0.18\pi$,
with a $(+2,-3,+1)$ phase for $\xi_{c2}<\xi<\xi_{c3}$.
The \kdp Hamiltonian with all 8 orbitals, Fig. \ref{fig_tpt1b}(b), in contrast to the reduced \kdp Hamiltonian with 4 orbitals,
yields $\xi_{c3}\ne\xi_{c2}$, with values only slighly different from the tight-binding solution ($\xi_{c2}=0.10\pi$, $\xi_{c3}=0.21\pi$).
This shows that in this case, to justify $\xi_{c3}\ne\xi_{c2}$ and the presence of the additional $(+2,-3,+1)$ phase,
one has to take into account the coupling to subspace~2. 
We note that it is exactly this coupling, described by parameters $h_{12}^v$, $h_{21}^v$, $h_{17}^v$,
that lowers the cylindrical symmetry of the \kdp Hamiltonian restricted to subspace~1, Eq. \eqref{hkdotp_1}, to the tetragonal symmetry of the full \kdp Hamiltonian, Eq. \eqref{hkdotp},
making the $X$--$M$ and $X$--$R$ directions inequivalent, hence allowing $\xi_{c3}\ne\xi_{c2}$.

The MCNs can be understood by the fact that,
when the gap closes along $X$--$\Gamma$ (so, four times at $k_z=0$ and twice at $k_x=k_y$), MCNs change by $(\pm 4, 0, \pm 2)$,
along $X$--$M$ (four times at $k_z=0$ and $k_z=\pi$) by $(\pm 4, \pm 4, 0)$,
and along $X$--$R$ (four times at $k_z=\pi$ and twice $k_x=k_y$) by $(0, \pm 4, \pm 2)$.\cite{sigrist_tki}
As a consequence, the properties $\C^+_{k_z=0}=2 \mod 4$, $\C^+_{k_z=\pi}=1 \mod 4$, $\C^+_{k_x=k_y}=1 \mod 2$ are always satisfied. We recall that higher MCNs lead, in general, to more surface Dirac cones.

To conclude, in our cylindrical approximation the \kdp method restricted to subspace~1 always gives $\xi_{c2}=\xi_{c3}$, predicting only the $w=\pm 1$ phases.
In contrast, additional intermediate phases $(+2,-3,+1)$ or $(-2,-3,-1)$, with $\xi_{c2}\ne\xi_{c3}$, exist once the symmetry is lowered from cylindrical to tetragonal,
by keeping either all orbitals or more terms in the \kdp expansion;
this is generally achieved in the tight-binding model.
%
Schematic phase diagrams are shown in Fig. \ref{fig_tpt1-2};
similar results are achieved for all other pairs of hybridization terms.

We remark that, using realistic parameters for {\sm} from DFT+Wannier calculations,\cite{pub6,prbr_io_smb6} $\eta_z^{v2}/\eta_x^{v2}\simeq -2.6$, $\eta_z^{v2}/\eta_z^{v1}\simeq -0.3$ (denoted as $\xi_{DFT}$ in Fig. \ref{fig_tpt1-2}),
so in both cases we are deep in the $(+2,+1,-1)$ $(w=+1)$ phase.
This remains true even when considering more hybridization terms, 
so the $\Gamma_8$-only model has $w=+1$ and is most likely not close to a phase transition. 

\begin{figure}[tb]
\includegraphics[width=0.48\textwidth]{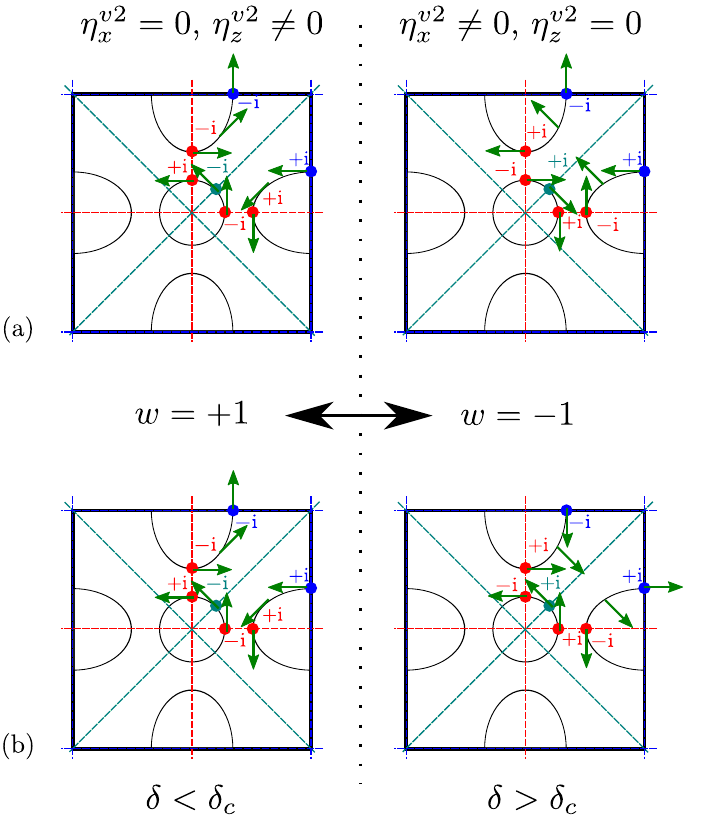}
\caption{
Schematic $w=+1\leftrightarrow w=-1$ transition for surface states on a $(001)$ surface at energies above the Dirac points
(a) varying the hybridization with $\Gamma_8$ states only,
(b) varying the relative on-site energy of $\Gamma_7$ and $\Gamma_8$ states:
when $\delta<\delta_c$, $w=+1$, surface states are mostly $\Gamma_8$, when $\delta>\delta_c$, $w=-1$, they are mostly $\Gamma_7$.
In both cases mirror eigenvalues change in the same way, but, when $w=-1$, the spin in (b) is opposite as in (a),
because $\Gamma_7$ states have opposite spin given the same mirror eigenvalues.
}\label{fig_tpt2}
\end{figure}

From Fig. \ref{fig_tpt2}(a) we see that across the $w=+1\leftrightarrow w=-1$ transition, the SEV on the $\bar\Gamma$ cone is reversed, just like the SEV on the $\bar X$ cone along the $\bar X$--$\bar\Gamma$  direction,
as a consequence of the change of surface mirror eigenvalues. Along the $\bar X$--$\bar M$ direction, instead, the SEV is the same because $\C^+_{k_z=\pi}=+1$
is left invariant. 
As shown in Refs.~\onlinecite{prb_io_tki,smb6_prl_io}, the spin structure of the $\bar X$ cones strongly affects intercone scattering:
as a consequence, in quasiparticle interference (QPI) experiments, the transition from $w=+1$ to $w=-1$ can be observed on a $(001)$ surface by the appearance of peaks due to intercone $\bar X$--$\bar X'$ scattering.

We could repeat the same analysis with the $\Gamma_7$ doublet, with exactly the same results,
the only difference being that the SEV would be everywhere reversed.
From a quantitative point of view, following Refs. \onlinecite{pub6,prbr_io_smb6}, 
the $\Gamma_7$-only model is in the $w=-1$ phase, and presumably far from a transition.

\subsection{Varying the relative multiplet energy}
\label{sec:varysplit}

A topological transition can also be realized by tuning the crystal-field splitting, based on the fact that, for realistic choices of the parameters, the $E_g$-$\Gamma_8$ model is in the $w=+1$ phase, while the $E_g$-$\Gamma_7$ one is in the $w=-1$ phase.
Given that about half a hole is expected in the $j=5/2$ multiplet, large crystal-field splitting will put the hole in the energetically higher of the $\Gamma_7$ and $\Gamma_8$ multiplets which will then determine the value of $w$. In practice, since the crystal-field splitting is comparable to the kinetic energy, both multiplets will be partly unoccupied, and $w$ results from a competition between the two.
The ARPES results of Ref. \onlinecite{smb6_arpes_mesot_spin} indicate that {\sm} is in the $w=+1$ phase, which lead us to conclude that $\Gamma_8$ are more important.\cite{smb6_prl_io}
Consequently, tuning the $\Gamma_7$ orbitals to higher energies can induce a transition to the $w=-1$ phase. Experimentally, this could be achieved e.g. by negatively doping the $B_6$ cages (preserving cubic symmetry), since $\Gamma_7$ orbitals have maxima along their direction, while $\Gamma_8$ have minima.
We note that a change of the crystal-field splitting might even be interaction-induced,\cite{sigrist_tki_prb} such that DFT and more advanced computational methods might predict different phase due to interaction-driven renormalizations; this is beyond the scope of this paper.

We can expect a phase transition when $w$ from Eq. \eqref{w_17} is zero.
We can here put $\beta_1\equiv\cos\xi$, $\beta_7\equiv\sin\xi$,
leading to
\be
w=\sgn[(f_1^v\cos\xi + f_7^v\sin\xi )(h_1^v\cos\xi + h_7^v\sin\xi )],\label{w_17_xi}
\ee
which gives $\tan\xi_{c1}=-f_1^v/f_7^v$ and $\tan\xi_{c2}=-h_1^v/h_7^v$.
To simplify things, we use the results quoted in the Appendix and Refs. \onlinecite{pub6,prbr_io_smb6},
which tell us that $f_1^v,h_1^v,h_7^v>0$, $f_7^v<0$. 
As a consequence, when the coupling $m_{78}$ from Eq.~\eqref{m_78} obeys $m_{78}>0$ (corresponding to the physical system), only $\xi_{c1}$ exists, which corresponds to $f_{17}^v=0$, so to a gap closing along $X$--$\Gamma$.
On the other hand, when $m_{78}<0$, only $\xi_{c2}$ exists, which corresponds to $h_{17}^v=0$, so to a gap closing along $X$--$M$ and $X$--$R$.
Finally, if $m_{78}=0$ we expect a gap closing at $X$.

We can also predict the value of $\Delta_c$ required for the transition:
\bea
\Delta_c&=&m_{78}\frac{(f_1^v)^2-(f_7^v)^2}{|f_1^vf_7^v|}, \hspace{5pt} m_{78}>0,\\
\Delta_c&=&|m_{78}|\frac{(h_1^v)^2-(h_7^v)^2}{h_1^vh_7^v}, \hspace{5pt} m_{78}<0,\\
\Delta_c&=&0, \hspace{5pt} m_{78}=0.
\eea

\begin{figure}[tb]
\includegraphics[width=0.49\textwidth]{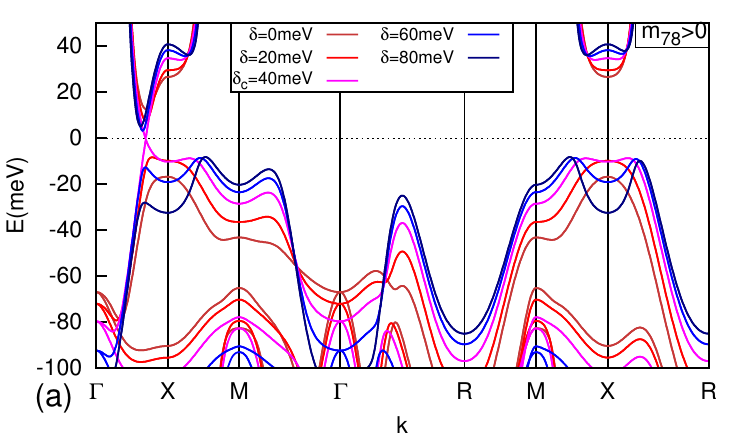}
\includegraphics[width=0.49\textwidth]{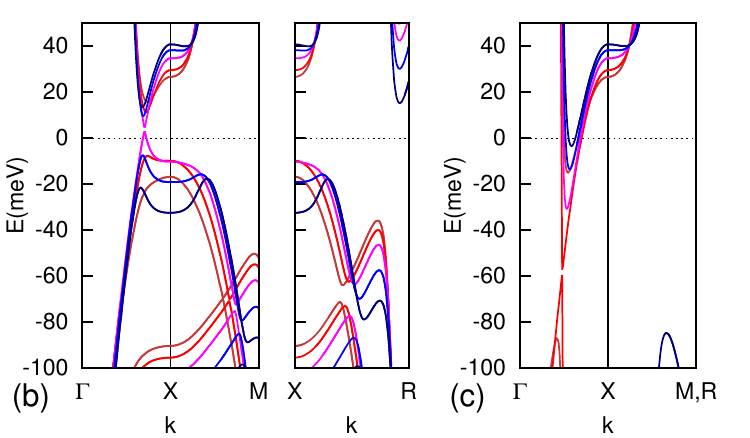}
\caption{Evolution of the bandstructure for the $E_g$ - $\Gamma_7$ - $\Gamma_8$ model as a function of the energy difference $\delta\equiv\epsilon_7-\epsilon_8$ (crystal-field splitting) between the $\Gamma_7$ doublet and the $\Gamma_8$ quadruplet,
(a) for the tight-binding model, (b) for the \kdp Hamiltonian with all 10 orbitals, (c) for the \kdp Hamiltonian with 4 orbitals, that we used in the text to allow for analytical calculations.
In (a) the gap closes at $\delta=\delta_c$ along $X-\Gamma$ denoting the topological phase transition: when $\delta<\delta_c$, $w=+1$, while when $\delta>\delta_c$, $w=-1$;
in (b) the \kdp method captures qualitatively and quantitatively the details of the transition;
in (c) the reduced \kdp method captures qualitatively the details of the transition, but it misses the exact value of $\delta_c$.
Non-zero parameters are $t_c=0.8$eV, $t_f=-0.015$eV, $V=0.05$eV, $\eta_z^{d1}=\eta_z^{f1}=0.8$, $\eta_z^{d2}=-0.3$, $\eta_z^{f2}=-0.5$, $\epsilon_d-\epsilon_8=1.45$eV, $\eta_7^{f2}=0.5$, $\eta_7^{f3}=0.25$, $\eta_{x7}^{f2}=0.16$,
$\eta_z^{v1}=-2.1$, $\eta_z^{v2}=0.6$, $\eta_7^{v2}=0.5$.
}\label{fig_tpt3}
\end{figure}

In Fig. \ref{fig_tpt3}
we show an example with $m_{78}=8\eta_{x7}^{f2}>0$, corresponding to the physical system (additional examples with $m_{78}=0$ and $m_{78}<0$ are shown in the supplement\cite{suppl_tci_long}).
We note that in the tight-binding model we can fix the on-site energy-difference $\delta\equiv\epsilon_7-\epsilon_8$,
with $\Delta$ depending on $\delta$ and on the hopping parameters according to Eqs. \eqref{eps_f1}, \eqref{eps_f7}.
Fig. \ref{fig_tpt3}(a) shows the results for the tight-binding model: the gap closes along $\Gamma$--$X$ at $\delta=\delta_c$, and a direct $w=-1\leftrightarrow +1$ transition can be achieved:
when $\delta<\delta_c$, $w=+1$, while when $\delta>\delta_c$, $w=-1$.
Panel (b) shows the \kdp approximation, which captures all the details of the transition, while panel (c) shows that the \kdp approximation with 4 orbitals
captures qualitatively the transition, but misses the exact value of $\delta_c$.
This shows that the approximation of Subsection \ref{ssec_g78} is qualitatively correct, but not enough to get accurate quantitative results.

\begin{figure}[tb]
\includegraphics[width=0.44\textwidth]{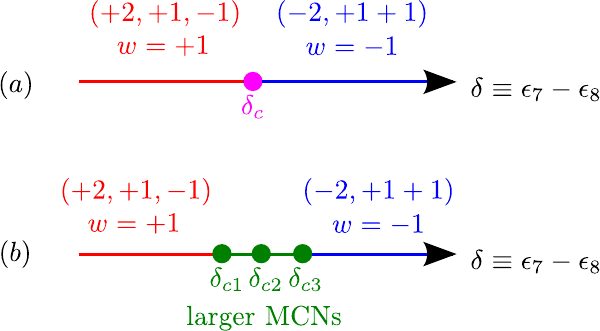}
\caption{The $w=-1\leftrightarrow +1$ transition as a function of the crystal-field splitting $\delta$ between the $\Gamma_7$ and $\Gamma_8$ multiplets
can happen directly, as shown in (a), where at $\delta=\delta_c$ the bulk gap closes along $\Gamma-X$, see Fig. \ref{fig_tpt3}, or via intermediate transitions as shown in (b),
where  at $\delta=\delta_{c1}, \delta_{c2}, \ldots$ the gap closes along $X-\Gamma$, $X-M$, $X-R$, or at low-symmetry points giving rise to intermediate phases with larger MCNs.
}\label{fig_tpt3b}
\end{figure}

We remark that, when taking into account terms in the Hamiltonian coming from further nearest neighbors, like we did in Ref. \onlinecite{prbr_io_smb6},
other phases with larger mirror Chern numbers can appear between the $w=\pm 1$ phases, with the bulk gap closing at points away from high-symmetry directions.
Such transitions change only one MCN at a time: by symmetry considerations,
$\C^+_{k_z=0}$ and $\C^+_{k_z=\pi}$ must change by $\pm 8$ and $\C^+_{k_x=k_y}$ by $\pm 4$.
As a consequence, transitions like $(-2,+1,+1)\leftrightarrow (-2,+1,-3)\leftrightarrow(+6,+1,-3)\leftrightarrow(+2,+1,-1)$ can be observed,
where in the first two cases we have respectively $\Delta \C^+_{k_x=k_y}=-4$ and $\Delta \C^+_{k_z=0}=+8$, while the last one corresponds to closing the gap along $\Gamma-X$ with $\Delta \C^+_{k_z=0}=-4$ and $\Delta \C^+_{k_x=k_y}=+2$;
if the order of the first two transitions is inverted, we have $(-2,+1,+1)\leftrightarrow (+6,+1,+1)\leftrightarrow(+6,+1,-3)\leftrightarrow(+2,+1,-1)$.
The two kinds of transition, with and without intermediate phases, are shown schematically in Fig. \ref{fig_tpt3b}.
However, the appearance of these intermediate phases is not required, depends on the details of the Hamiltonian, and their range of $\delta$ is in any case small.

As shown in Fig. \ref{fig_tpt2}(b), with respect to the case in Section~\ref{sec:varyhyb}, the SEV across the transition changes in a different way because $\Gamma_7$ states have the opposite SEV given the same mirror-symmetry eigenvalue,
so the SEV in the $w=-1$ phase, Fig. \ref{fig_tpt2}(b) right, where surface states have mainly $\Gamma_7$ character, is reversed with respect to Fig. \ref{fig_tpt2}(a) right, where only $\Gamma_8$ states are used,
while in the $w=+1$ phase, Fig. \ref{fig_tpt2}(b) left, surface states have mainly $\Gamma_8$ character, and the SEV is as in Fig. \ref{fig_tpt2}(a) left.
As a consequence, the SEV on the $\bar\Gamma$ cone is the same on both sides of the transition, just like the SEV on the $\bar X$ cone along the $\bar X$--$\bar\Gamma$ direction,
while the SEV on the $\bar X$ cone along the $\bar X$--$\bar M$ direction is now reversed.
We point out that Fig.~\ref{fig_tpt2} is qualitative in the sense that things can become more involved very close to a transition: for example, the SEV, even if small, does not reverse exactly at the transition.

We finally recall that the discussion of this Subsection relies on the assumption that $\Gamma_7$ and $\Gamma_8$ multiplets, when taken alone, give rise to distinct topological phases. This is found to be true in ab-initio calculations for {\pu},\cite{pub6,prbr_io_smb6} whose bandstructure is very similar to that of {\sm}. The only ab-initio data we have for {\sm} is the \kdp expansion from Ref. \onlinecite{yu_smb6_qpi}. Using their data, we were not able to confirm the above assumption, with more details given in the supplement.\cite{suppl_tci_long} We note, however, that Ref.~\onlinecite{yu_smb6_qpi} predicts a spin structure on the $(001)$ surface corresponding to $w=-1$,
in disagreement with experiment,\cite{smb6_arpes_mesot_spin} casting doubts on the accuracy of the description.


\section{Conclusions}\label{sec:concl}

In this paper we have shown how the use of the \kdp theory for {\sm} allows to perturbatively compute surface states and their symmetry properties. A central role is played by the parameters $v$ and $w$ constructed from MCNs, $v\equiv\sgn (\C^+_{k_z=0}\C^+_{k_x=k_y})$ and $w\equiv\sgn(\C^+_{k_z=0}\C^+_{k_z=\pi})$, which determine the topological phase of a particular model for {\sm}.
We have concrete provided predictions for the spin structure on general surfaces. Given the experimental information\cite{smb6_arpes_mesot_spin,smb6_prl_io} $w=+1$, we find all cones to have an in-plane winding number $+1$, and on surfaces of low symmetry a small out-of-plane component of the spin is expected.

We have also proposed a simple physical mechanism for inducing topological phase transition. This exploits the fact that different phases $w=\pm 1$ are realized in models which  retaining only the
$\Gamma_7$ or $\Gamma_8$ multiplets, such that varying the relative energy of these multiplets will lead to a topological phase transition with a sign change of MCNs and hence of $w$. Experimentally, this could be in principle achieved by doping the material in such a way that B$_6$ cages acquire a more negative charge. The topological phase transition is marked by a closing of the bulk gap, and can be observed as a change of the spin structure of surface states.
We also stress that the renormalization of parameters due to the Hubbard repulsion could lead to an interaction-induced topological phase transition.\cite{sigrist_tki_prb}

Our results are based on a number of assumptions, most importantly the validity of a renormalized single-particle picture and the presence of flat non-reconstructed surfaces. A partial discussion of these issues, focussing on $(001)$ surface states, is in Ref.~\onlinecite{smb6_rec_io}, but it is clear that work beyond single-particle approximations is needed to fully validate our analysis.

We close by noting that our results will not only be important for interpreting results from future photoemission and tunneling experiments, which will be able to probe surface states on arbitrary surfaces of {\sm} and related materials, but also for understanding the results of ab-initio DFT calculations: Here, different spin structures have been reported in the DFT literature, but not conclusively assigned to distinct topological phases.


\acknowledgments

We thank O. Rader, S. Wirth, and M. Legner for illuminating discussions.
This research was supported by the DFG through SFB 1143 and GRK 1621
as well as by the Helmholtz association through VI-521.


\appendix

\section{Parameters for the \kdp Hamiltonian}\label{app_par}
Parameters of the \kdp Hamiltonian Eq. \eqref{hkdotp} are defined as follows:
\bea
\epsilon_{d}&=&\langle d^1\uparrow |H|d^1\uparrow\rangle_{000}=\langle d^2\uparrow |H|d^2\uparrow\rangle_{000},\\
\epsilon_8&=&\langle f^1+|H|f^1+\rangle_{000}=\langle f^2+|H|f^2+\rangle_{000},\\
\epsilon_7&=&\langle f^7+|H|f^7+\rangle_{000},
\\
t_d\eta_x^{d1}&=&\langle d^1\uparrow |H|d^1\uparrow\rangle_{001},\\
t_d\eta_z^{d1}&=&\langle d^2\uparrow|H|d^2\uparrow\rangle_{001},\\
t_d\eta_z^{d2}&=&\langle d^2\uparrow|H|d^2\uparrow\rangle_{110},\\
t_f\eta_x^{f1}&=&\langle f^1+|H|f^1+\rangle_{001},\\
t_f\eta_z^{f1}&=&\langle f^2+|H|f^2+\rangle_{001},\\
t_f\eta_z^{f2}&=&\langle f^2+|H|f^2+\rangle_{110},\\
t_f\eta_7^{f1}&=&\langle f^7+|H|f^7+\rangle_{001},\\
t_f\eta_7^{f2}&=&\langle f^7+|H|f^7+\rangle_{110},\\
t_f\eta_7^{f3}&=&\langle f^7+|H|f^7+\rangle_{111},\\
t_f\eta_{78}^{f1}&=&\langle f^1+|H|f^7+\rangle_{001},\label{eta_78_f1}\\
t_f\eta_{x7}^{f2}&=&\langle f^1+|H|f^7+\rangle_{110},\label{eta_x7_f2}
\\
V\eta_x^{v1}&=&\langle d^1\uparrow|H|f^1+\rangle_{001},\\
V\eta_z^{v1}&=&\langle d^2\uparrow|H|f^2+\rangle_{001},\\
V\eta_x^{v2}&=&\langle d^1\uparrow|H|f^1+\rangle_{110},\\
V\eta_z^{v2}&=&\langle d^2\uparrow|H|f^2+\rangle_{110},\\
V\eta_7^{v1}&=&\langle d^1\uparrow|H|f^7+\rangle_{001},\\
V\eta_7^{v2}&=&\langle d^2\uparrow|H|f^7+\rangle_{110},\\
V\eta_{x7}^{v2}&=&\langle d^1\uparrow|H|f^7+\rangle_{110},
\eea
where the subscript $000$, $001$, $110$, or $111$ denotes the direction along which the matrix element is calculated.

Further quantities appearing in Hamiltonian Eq. \eqref{hkdotp} are the energies:
\bea
\epsilon_1^d&=&\epsilon_d-t_d(3\eta_z^{d1}-\eta_x^{d1}-6\eta_z^{d2}),\\
\epsilon_2^d&=&\epsilon_d-t_d(3\eta_x^{d1}-\eta_z^{d1}+2\eta_z^{d2}),\\
\epsilon_1^f&=&\epsilon_8-t_f(3\eta_z^{f1}-\eta_x^{f1}-6\eta_z^{f2}),\label{eps_f1}\\
\epsilon_2^f&=&\epsilon_8-t_f(3\eta_x^{f1}-\eta_z^{f1}+2\eta_z^{f2}),\\
\epsilon_7^f&=&\epsilon_7-t_f(2\eta_7^{f1}-4\eta_7^{f2}-8\eta_7^{f3}),\label{eps_f7}
\eea
the kinetic-energy parameters in $k_z$:
\bea
g_1^d&=&\eta_x^{d1}+3\eta_z^{d2}<0,\\
g_2^d&=&\eta_z^{d1}+\eta_z^{d2}>0,\\
g_1^f&=&\eta_x^{f1}+3\eta_z^{f2}<0,\\
g_2^f&=&\eta_z^{f1}+\eta_z^{f2}>0,\\
g_7^f&=&\eta_7^{f1}+4\eta_7^{f2}+4\eta_7^{f3}<0,
\eea
and those in $k_\parallel$:
\bea
l_1^d&=&\frac{1}{4}(-\eta_x^{d1}-3\eta_z^{d1}+6\eta_z^{d2})<0,\\
l_2^d&=&\frac{1}{4}(-3\eta_x^{d1}-\eta_z^{d1}-6\eta_z^{d2})>0,\\
l_1^f&=&\frac{1}{4}(-\eta_x^{f1}-3\eta_z^{f1}+6\eta_z^{f2})<0,\\
l_2^f&=&\frac{1}{4}(-3\eta_x^{f1}-\eta_z^{f1}-6\eta_z^{f2})>0,\\
l_7^f&=&-\eta_7^{f1}+{4}\eta_7^{f3}<0.
\eea
Finally, hybridization terms other than the ones quoted in the main text are:
\bea
h_{12}^v=h_{21}^v&=&\frac{\sqrt{3}}{2}(-\eta_x^{v1}+\eta_z^{v1}-2\eta_x^{v2}+2\eta_z^{v2}),\\
h_{72}^v&=&\sqrt{3}\eta_7^{v1}-2\eta_7^{v2}+2\sqrt{3}\eta_{x7}^{v2}.
\eea
We note that, when keeping more hybridization terms than the ones used in this work, in general $h_{12}^v\ne h_{21}^v$.

Using values from Ref. \onlinecite{prbr_io_smb6} we find:
\bea
&g_1^d=-0.96,&
l_1^d=-1.02,\\
&g_2^d=0.52,&
l_2^d=1.20,\\
&g_1^f=-8.00,&
l_1^f=-6.19,\\
&g_2^f=1.92,&
l_2^f=3.27,\\
&g_7^f=-14.70,&
l_7^f=-5.14,\\
&f_1^v=3.91,&\label{f1h1_num}
h_1^v=5.42,\\
&f_2^v=-4.52,&
h_2^v=-2.04,\\
&f_7^v=-3.27,&\label{f7h7_num}
h_7^v=2.44,\\
&h_{12}^v=-0.89,&
h_{72}^v=-1.79,
\eea
together with $t_d=1$eV, $t_f=-0.01$eV, $V=0.1$eV.

These values mainly serve as approximate guide, since they refer to {\pu}. A fully accurate microscopic description in any case requires longer-range tight-binding terms to be precise.


\bibliographystyle{apsrev4-1}
\bibliography{tki}

\begin{thebibliography}{10}%
\makeatletter
\providecommand \@ifxundefined [1]{%
 \ifx #1\undefined \expandafter \@firstoftwo
 \else \expandafter \@secondoftwo
\fi
}%
\providecommand \@ifnum [1]{%
 \ifnum #1\expandafter \@firstoftwo
 \else \expandafter \@secondoftwo
\fi
}%
\providecommand \enquote [1]{``#1''}%
\providecommand \bibnamefont  [1]{#1}%
\providecommand \bibfnamefont [1]{#1}%
\providecommand \citenamefont [1]{#1}%
\providecommand\href[0]{\@sanitize\@href}%
\providecommand\@href[1]{\endgroup\@@startlink{#1}\endgroup\@@href}%
\providecommand\@@href[1]{#1\@@endlink}%
\providecommand \@sanitize [0]{\begingroup\catcode`\&12\catcode`\#12\relax}%
\@ifxundefined \pdfoutput {\@firstoftwo}{%
 \@ifnum{\z@=\pdfoutput}{\@firstoftwo}{\@secondoftwo}%
}{%
 \providecommand\@@startlink[1]{\leavevmode\special{html:<a href="#1">}}%
 \providecommand\@@endlink[0]{\special{html:</a>}}%
}{%
 \providecommand\@@startlink[1]{%
  \leavevmode
  \pdfstartlink
   attr{/Border[0 0 1 ]/H/I/C[0 1 1]}%
   user{/Subtype/Link/A<</Type/Action/S/URI/URI(#1)>>}%
  \relax
 }%
 \providecommand\@@endlink[0]{\pdfendlink}%
}%
\providecommand \url  [0]{\begingroup\@sanitize \@url }%
\providecommand \@url [1]{\endgroup\@href {#1}{\urlprefix}}%
\providecommand \urlprefix [0]{URL }%
\providecommand \Eprint[0]{\href }%
\@ifxundefined \urlstyle {%
  \providecommand \doi [1]{doi:\discretionary{}{}{}#1}%
}{%
  \providecommand \doi [0]{doi:\discretionary{}{}{}\begingroup
  \urlstyle{rm}\Url }%
}%
\providecommand \doibase [0]{http://dx.doi.org/}%
\providecommand \Doi[1]{\href{\doibase#1}}%
\providecommand \bibAnnote [3]{%
  \BibitemShut{#1}%
  \begin{quotation}\noindent
    \textsc{Key:}\ #2\\\textsc{Annotation:}\ #3%
  \end{quotation}%
}%
\providecommand \bibAnnoteFile [2]{%
  \IfFileExists{#2}{\bibAnnote {#1} {#2} {\input{#2}}}{}%
}%
\providecommand \typeout [0]{\immediate \write \m@ne }%
\providecommand \selectlanguage [0]{\@gobble}%
\providecommand \bibinfo [0]{\@secondoftwo}%
\providecommand \bibfield [0]{\@secondoftwo}%
\providecommand \translation [1]{[#1]}%
\providecommand \BibitemOpen[0]{}%
\providecommand \bibitemStop [0]{}%
\providecommand \bibitemNoStop [0]{.\EOS\space}%
\providecommand \EOS [0]{\spacefactor3000\relax}%
\providecommand \BibitemShut [1]{\csname bibitem#1\endcsname}%
\bibitem{takimoto}%
  \BibitemOpen
  \bibfield{author}{%
  \bibinfo {author} {\bibfnamefont{T.}~\bibnamefont{Takimoto}},\ }%
  \bibfield{journal}{%
  \bibinfo {journal} {J. Phys. Soc. Jpn.}\ }%
  \textbf{\bibinfo {volume} {80}},\ \bibinfo {pages} {123710} (\bibinfo {year}
  {2011})%
  \bibAnnoteFile{NoStop}{takimoto}%
\bibitem{lu_smb6_gutz}%
  \BibitemOpen
  \bibfield{author}{%
  \bibinfo {author} {\bibfnamefont{F.}~\bibnamefont{Lu}}, \bibinfo {author}
  {\bibfnamefont{J.}~\bibnamefont{Zhao}}, \bibinfo {author}
  {\bibfnamefont{H.}~\bibnamefont{Weng}}, \bibinfo {author}
  {\bibfnamefont{Z.}~\bibnamefont{Fang}},\ and\ \bibinfo {author}
  {\bibfnamefont{X.}~\bibnamefont{Dai}},\ }%
  \bibfield{journal}{%
  \bibinfo {journal} {Phys. Rev. Lett.}\ }%
  \textbf{\bibinfo {volume} {110}},\ \bibinfo {pages} {096401} (\bibinfo {year}
  {2013})%
  \bibAnnoteFile{NoStop}{lu_smb6_gutz}%
\bibitem{tki_cubic}%
  \BibitemOpen
  \bibfield{author}{%
  \bibinfo {author} {\bibfnamefont{V.}~\bibnamefont{Alexandrov}}, \bibinfo
  {author} {\bibfnamefont{M.}~\bibnamefont{Dzero}},\ and\ \bibinfo {author}
  {\bibfnamefont{P.}~\bibnamefont{Coleman}},\ }%
  \bibfield{journal}{%
  \bibinfo {journal} {Phys. Rev. Lett.}\ }%
  \textbf{\bibinfo {volume} {111}},\ \bibinfo {pages} {226403} (\bibinfo {year}
  {2013})%
  \bibAnnoteFile{NoStop}{tki_cubic}%
\bibitem{tki1}%
  \BibitemOpen
  \bibfield{author}{%
  \bibinfo {author} {\bibfnamefont{M.}~\bibnamefont{Dzero}}, \bibinfo {author}
  {\bibfnamefont{K.}~\bibnamefont{Sun}}, \bibinfo {author}
  {\bibfnamefont{V.}~\bibnamefont{Galitski}},\ and\ \bibinfo {author}
  {\bibfnamefont{P.}~\bibnamefont{Coleman}},\ }%
  \bibfield{journal}{%
  \bibinfo {journal} {Phys. Rev. Lett.}\ }%
  \textbf{\bibinfo {volume} {104}},\ \bibinfo {pages} {106408} (\bibinfo {year}
  {2010})%
  \bibAnnoteFile{NoStop}{tki1}%
\bibitem{wolgast_smb6}%
  \BibitemOpen
  \bibfield{author}{%
  \bibinfo {author} {\bibfnamefont{S.}~\bibnamefont{Wolgast}}, \bibinfo
  {author} {\bibfnamefont{C.}~\bibnamefont{Kurdak}}, \bibinfo {author}
  {\bibfnamefont{K.}~\bibnamefont{Sun}}, \bibinfo {author}
  {\bibfnamefont{J.~W.}\ \bibnamefont{Allen}}, \bibinfo {author}
  {\bibfnamefont{D.-J.}\ \bibnamefont{Kim}},\ and\ \bibinfo {author}
  {\bibfnamefont{Z.}~\bibnamefont{Fisk}},\ }%
  \bibfield{journal}{%
  \bibinfo {journal} {Phys. Rev. B}\ }%
  \textbf{\bibinfo {volume} {88}},\ \bibinfo {pages} {180405(R)} (\bibinfo
  {year} {2013})%
  \bibAnnoteFile{NoStop}{wolgast_smb6}%
\bibitem{fisk_smb6_topss}%
  \BibitemOpen
  \bibfield{author}{%
  \bibinfo {author} {\bibfnamefont{D.~J.}\ \bibnamefont{Kim}}, \bibinfo
  {author} {\bibfnamefont{J.}~\bibnamefont{Xia}},\ and\ \bibinfo {author}
  {\bibfnamefont{Z.}~\bibnamefont{Fisk}},\ }%
  \bibfield{journal}{%
  \bibinfo {journal} {Nature Mat.}\ }%
  \textbf{\bibinfo {volume} {13}},\ \bibinfo {pages} {466} (\bibinfo {year}
  {2014})%
  \bibAnnoteFile{NoStop}{fisk_smb6_topss}%
\bibitem{smb6_junction_prx}%
  \BibitemOpen
  \bibfield{author}{%
  \bibinfo {author} {\bibfnamefont{X.}~\bibnamefont{Zhang}}, \bibinfo {author}
  {\bibfnamefont{N.~P.}\ \bibnamefont{Butch}}, \bibinfo {author}
  {\bibfnamefont{P.}~\bibnamefont{Syers}}, \bibinfo {author}
  {\bibfnamefont{S.}~\bibnamefont{Ziemak}}, \bibinfo {author}
  {\bibfnamefont{R.~L.}\ \bibnamefont{Greene}},\ and\ \bibinfo {author}
  {\bibfnamefont{J.}~\bibnamefont{Paglione}},\ }%
  \bibfield{journal}{%
  \bibinfo {journal} {Phys. Rev. X}\ }%
  \textbf{\bibinfo {volume} {3}},\ \bibinfo {pages} {011011} (\bibinfo {year}
  {2013})%
  \bibAnnoteFile{NoStop}{smb6_junction_prx}%
\bibitem{neupane_smb6}%
  \BibitemOpen
  \bibfield{author}{%
  \bibinfo {author} {\bibfnamefont{M.}~\bibnamefont{Neupane}}, \bibinfo
  {author} {\bibfnamefont{N.}~\bibnamefont{Alidoust}}, \bibinfo {author}
  {\bibfnamefont{S.}~\bibnamefont{Xu}}, \bibinfo {author}
  {\bibfnamefont{T.}~\bibnamefont{Kondo}}, \bibinfo {author}
  {\bibfnamefont{Y.}~\bibnamefont{Ishida}}, \bibinfo {author}
  {\bibfnamefont{D.-J.}\ \bibnamefont{Kim}}, \bibinfo {author}
  {\bibfnamefont{C.}~\bibnamefont{Liu}}, \bibinfo {author}
  {\bibfnamefont{I.}~\bibnamefont{Belopolski}}, \bibinfo {author}
  {\bibfnamefont{Y.}~\bibnamefont{Jo}}, \bibinfo {author}
  {\bibfnamefont{T.-R.}\ \bibnamefont{Chang}}, \bibinfo {author}
  {\bibfnamefont{H.-T.}\ \bibnamefont{Jeng}}, \bibinfo {author}
  {\bibfnamefont{T.}~\bibnamefont{Durakiewicz}}, \bibinfo {author}
  {\bibfnamefont{L.}~\bibnamefont{Balicas}}, \bibinfo {author}
  {\bibfnamefont{H.}~\bibnamefont{Lin}}, \bibinfo {author}
  {\bibfnamefont{A.}~\bibnamefont{Bansil}}, \bibinfo {author}
  {\bibfnamefont{S.}~\bibnamefont{Shin}}, \bibinfo {author}
  {\bibfnamefont{Z.}~\bibnamefont{Fisk}},\ and\ \bibinfo {author}
  {\bibfnamefont{M.~Z.}\ \bibnamefont{Hasan}},\ }%
  \bibfield{journal}{%
  \bibinfo {journal} {Nature Comm.}\ }%
  \textbf{\bibinfo {volume} {4}},\ \bibinfo {pages} {2991} (\bibinfo {year}
  {2013})%
  \bibAnnoteFile{NoStop}{neupane_smb6}%
\bibitem{mesot_smb6}%
  \BibitemOpen
  \bibfield{author}{%
  \bibinfo {author} {\bibfnamefont{N.}~\bibnamefont{Xu}}, \bibinfo {author}
  {\bibfnamefont{X.}~\bibnamefont{Shi}}, \bibinfo {author}
  {\bibfnamefont{P.~K.}\ \bibnamefont{Biswas}}, \bibinfo {author}
  {\bibfnamefont{C.~E.}\ \bibnamefont{Matt}}, \bibinfo {author}
  {\bibfnamefont{R.~S.}\ \bibnamefont{Dhaka}}, \bibinfo {author}
  {\bibfnamefont{Y.}~\bibnamefont{Huang}}, \bibinfo {author}
  {\bibfnamefont{N.~C.}\ \bibnamefont{Plumb}}, \bibinfo {author}
  {\bibfnamefont{M.}~\bibnamefont{Radovic}}, \bibinfo {author}
  {\bibfnamefont{J.~H.}\ \bibnamefont{Dil}}, \bibinfo {author}
  {\bibfnamefont{E.}~\bibnamefont{Pomjakushina}}, \bibinfo {author}
  {\bibfnamefont{K.}~\bibnamefont{Conder}}, \bibinfo {author}
  {\bibfnamefont{A.}~\bibnamefont{Amato}}, \bibinfo {author}
  {\bibfnamefont{Z.}~\bibnamefont{Salman}}, \bibinfo {author}
  {\bibfnamefont{D.~M.}\ \bibnamefont{Paul}}, \bibinfo {author}
  {\bibfnamefont{J.}~\bibnamefont{Mesot}}, \bibinfo {author}
  {\bibfnamefont{H.}~\bibnamefont{Ding}},\ and\ \bibinfo {author}
  {\bibfnamefont{M.}~\bibnamefont{Shi}},\ }%
  \bibfield{journal}{%
  \bibinfo {journal} {Phys. Rev. B}\ }%
  \textbf{\bibinfo {volume} {88}},\ \bibinfo {pages} {121102} (\bibinfo {year}
  {2013})%
  \bibAnnoteFile{NoStop}{mesot_smb6}%
\bibitem{smb6_arpes_feng}%
  \BibitemOpen
  \bibfield{author}{%
  \bibinfo {author} {\bibfnamefont{J.}~\bibnamefont{Jiang}}, \bibinfo {author}
  {\bibfnamefont{S.}~\bibnamefont{Li}}, \bibinfo {author}
  {\bibfnamefont{T.}~\bibnamefont{Zhang}}, \bibinfo {author}
  {\bibfnamefont{Z.}~\bibnamefont{Sun}}, \bibinfo {author}
  {\bibfnamefont{F.}~\bibnamefont{Chen}}, \bibinfo {author}
  {\bibfnamefont{Z.}~\bibnamefont{Ye}}, \bibinfo {author}
  {\bibfnamefont{M.}~\bibnamefont{Xu}}, \bibinfo {author}
  {\bibfnamefont{Q.}~\bibnamefont{Ge}}, \bibinfo {author}
  {\bibfnamefont{S.}~\bibnamefont{Tan}}, \bibinfo {author}
  {\bibfnamefont{X.}~\bibnamefont{Niu}}, \bibinfo {author}
  {\bibfnamefont{M.}~\bibnamefont{Xia}}, \bibinfo {author}
  {\bibfnamefont{B.}~\bibnamefont{Xie}}, \bibinfo {author}
  {\bibfnamefont{Y.}~\bibnamefont{Li}}, \bibinfo {author}
  {\bibfnamefont{X.}~\bibnamefont{Chen}}, \bibinfo {author}
  {\bibfnamefont{H.}~\bibnamefont{Wen}},\ and\ \bibinfo {author}
  {\bibfnamefont{D.}~\bibnamefont{Feng}},\ }%
  \bibfield{journal}{%
  \bibinfo {journal} {Nature Comm.}\ }%
  \textbf{\bibinfo {volume} {4}},\ \bibinfo {pages} {3010} (\bibinfo {year}
  {2013})%
  \bibAnnoteFile{NoStop}{smb6_arpes_feng}%
\bibitem{smb6_arpes_reinert}%
  \BibitemOpen
  \bibfield{author}{%
  \bibinfo {author} {\bibfnamefont{C.-H.}\ \bibnamefont{Min}}, \bibinfo
  {author} {\bibfnamefont{P.}~\bibnamefont{Lutz}}, \bibinfo {author}
  {\bibfnamefont{S.}~\bibnamefont{Fiedler}}, \bibinfo {author}
  {\bibfnamefont{B.}~\bibnamefont{Kang}}, \bibinfo {author}
  {\bibfnamefont{B.}~\bibnamefont{Cho}}, \bibinfo {author}
  {\bibfnamefont{H.-D.}\ \bibnamefont{Kim}}, \bibinfo {author}
  {\bibfnamefont{H.}~\bibnamefont{Bentmann}},\ and\ \bibinfo {author}
  {\bibfnamefont{F.}~\bibnamefont{Reinert}},\ }%
  \bibfield{journal}{%
  \bibinfo {journal} {Phys. Rev. Lett.}\ }%
  \textbf{\bibinfo {volume} {112}},\ \bibinfo {pages} {226402} (\bibinfo {year}
  {2014})%
  \bibAnnoteFile{NoStop}{smb6_arpes_reinert}%
\bibitem{smb6_past_allen}%
  \BibitemOpen
  \bibfield{author}{%
  \bibinfo {author} {\bibfnamefont{J.~D.}\ \bibnamefont{Denlinger}}, \bibinfo
  {author} {\bibfnamefont{J.~W.}\ \bibnamefont{Allen}}, \bibinfo {author}
  {\bibfnamefont{J.-S.}\ \bibnamefont{Kang}}, \bibinfo {author}
  {\bibfnamefont{K.}~\bibnamefont{Sun}}, \bibinfo {author}
  {\bibfnamefont{B.-I.}\ \bibnamefont{Min}}, \bibinfo {author}
  {\bibfnamefont{D.-J.}\ \bibnamefont{Kim}},\ and\ \bibinfo {author}
  {\bibfnamefont{Z.}~\bibnamefont{Fisk}},\ }%
  \bibinfo {journal} {preprint arXiv:1312.6636}%
  \bibAnnoteFile{NoStop}{smb6_past_allen}%
\bibitem{smb6_arpes_mesot_spin}%
  \BibitemOpen
\bibfield{journal}{%
    }%
  \bibfield{author}{%
  \bibinfo {author} {\bibfnamefont{N.}~\bibnamefont{Xu}}, \bibinfo {author}
  {\bibfnamefont{P.~K.}\ \bibnamefont{Biswas}}, \bibinfo {author}
  {\bibfnamefont{J.~H.}\ \bibnamefont{Dil}}, \bibinfo {author}
  {\bibfnamefont{R.~S.}\ \bibnamefont{Dhaka}}, \bibinfo {author}
  {\bibfnamefont{G.}~\bibnamefont{Landolt}}, \bibinfo {author}
  {\bibfnamefont{S.}~\bibnamefont{Muff}}, \bibinfo {author}
  {\bibfnamefont{C.~E.}\ \bibnamefont{Matt}}, \bibinfo {author}
  {\bibfnamefont{X.}~\bibnamefont{Shi}}, \bibinfo {author}
  {\bibfnamefont{N.~C.}\ \bibnamefont{Plumb}}, \bibinfo {author}
  {\bibfnamefont{M.}~\bibnamefont{Radovic}}, \bibinfo {author}
  {\bibfnamefont{E.}~\bibnamefont{Pomjakushina}}, \bibinfo {author}
  {\bibfnamefont{K.}~\bibnamefont{Conder}}, \bibinfo {author}
  {\bibfnamefont{A.}~\bibnamefont{Amato}}, \bibinfo {author}
  {\bibfnamefont{S.}~\bibnamefont{Borisenko}}, \bibinfo {author}
  {\bibfnamefont{R.}~\bibnamefont{Yu}}, \bibinfo {author}
  {\bibfnamefont{H.-M.}\ \bibnamefont{Weng}}, \bibinfo {author}
  {\bibfnamefont{Z.}~\bibnamefont{Fang}}, \bibinfo {author}
  {\bibfnamefont{X.}~\bibnamefont{Dai}}, \bibinfo {author}
  {\bibfnamefont{J.}~\bibnamefont{Mesot}}, \bibinfo {author}
  {\bibfnamefont{H.}~\bibnamefont{Ding}},\ and\ \bibinfo {author}
  {\bibfnamefont{M.}~\bibnamefont{Shi}},\ }%
  \bibfield{journal}{%
  \bibinfo {journal} {Nature Comm.}\ }%
  \textbf{\bibinfo {volume} {5}},\ \bibinfo {pages} {4566} (\bibinfo {year}
  {2014})%
  \bibAnnoteFile{NoStop}{smb6_arpes_mesot_spin}%
\bibitem{sawatzky_smb6}%
  \BibitemOpen
  \bibfield{author}{%
  \bibinfo {author} {\bibfnamefont{Z.-H.}\ \bibnamefont{Zhu}}, \bibinfo
  {author} {\bibfnamefont{A.}~\bibnamefont{Nicolaou}}, \bibinfo {author}
  {\bibfnamefont{G.}~\bibnamefont{Levy}}, \bibinfo {author}
  {\bibfnamefont{N.~P.}\ \bibnamefont{Butch}}, \bibinfo {author}
  {\bibfnamefont{P.}~\bibnamefont{Syers}}, \bibinfo {author}
  {\bibfnamefont{X.~F.}\ \bibnamefont{Wang}}, \bibinfo {author}
  {\bibfnamefont{J.}~\bibnamefont{Paglione}}, \bibinfo {author}
  {\bibfnamefont{G.~A.}\ \bibnamefont{Sawatzky}}, \bibinfo {author}
  {\bibfnamefont{I.~S.}\ \bibnamefont{Elfimov}},\ and\ \bibinfo {author}
  {\bibfnamefont{A.}~\bibnamefont{Damascelli}},\ }%
  \bibfield{journal}{%
  \bibinfo {journal} {Phys. Rev. Lett.}\ }%
  \textbf{\bibinfo {volume} {111}},\ \bibinfo {pages} {216402} (\bibinfo {year}
  {2013})%
  \bibAnnoteFile{NoStop}{sawatzky_smb6}%
\bibitem{smb6_prx_arpes}%
  \BibitemOpen
  \bibfield{author}{%
  \bibinfo {author} {\bibfnamefont{E.}~\bibnamefont{Frantzeskakis}}, \bibinfo
  {author} {\bibfnamefont{N.}~\bibnamefont{de~Jong}}, \bibinfo {author}
  {\bibfnamefont{B.}~\bibnamefont{Zwartsenberg}}, \bibinfo {author}
  {\bibfnamefont{Y.~K.}\ \bibnamefont{Huang}}, \bibinfo {author}
  {\bibfnamefont{Y.}~\bibnamefont{Pan}}, \bibinfo {author}
  {\bibfnamefont{X.}~\bibnamefont{Zhang}}, \bibinfo {author}
  {\bibfnamefont{J.~X.}\ \bibnamefont{Zhang}}, \bibinfo {author}
  {\bibfnamefont{F.~X.}\ \bibnamefont{Zhang}}, \bibinfo {author}
  {\bibfnamefont{L.~H.}\ \bibnamefont{Bao}}, \bibinfo {author}
  {\bibfnamefont{O.}~\bibnamefont{Tegus}}, \bibinfo {author}
  {\bibfnamefont{A.}~\bibnamefont{Varykhalov}}, \bibinfo {author}
  {\bibfnamefont{A.}~\bibnamefont{de~Visser}},\ and\ \bibinfo {author}
  {\bibfnamefont{M.~S.}\ \bibnamefont{Golden}},\ }%
  \bibfield{journal}{%
  \bibinfo {journal} {Phys. Rev. X}\ }%
  \textbf{\bibinfo {volume} {3}},\ \bibinfo {pages} {041024} (\bibinfo {year}
  {2013})%
  \bibAnnoteFile{NoStop}{smb6_prx_arpes}%
\bibitem{smb6_trivial}%
  \BibitemOpen
  \bibfield{author}{%
  \bibinfo {author} {\bibfnamefont{P.}~\bibnamefont{Hlawenka}}, \bibinfo
  {author} {\bibfnamefont{K.}~\bibnamefont{Siemensmeyer}}, \bibinfo {author}
  {\bibfnamefont{E.}~\bibnamefont{Weschke}}, \bibinfo {author}
  {\bibfnamefont{A.}~\bibnamefont{Varykhalova}}, \bibinfo {author}
  {\bibfnamefont{J.}~\bibnamefont{S\'{a}nchez-Barriga}}, \bibinfo {author}
  {\bibfnamefont{N.~Y.}\ \bibnamefont{Shitsevalova}}, \bibinfo {author}
  {\bibfnamefont{A.~V.}\ \bibnamefont{Dukhnenko}}, \bibinfo {author}
  {\bibfnamefont{V.~B.}\ \bibnamefont{Filipov}}, \bibinfo {author}
  {\bibfnamefont{S.}~\bibnamefont{Gabáni}}, \bibinfo {author}
  {\bibfnamefont{K.}~\bibnamefont{Flachbart}}, \bibinfo {author}
  {\bibfnamefont{O.}~\bibnamefont{Rader}},\ and\ \bibinfo {author}
  {\bibfnamefont{E.~D.~L.}\ \bibnamefont{Rienks}},\ }%
  \bibinfo {journal} {preprint arXiv:1502.01542}%
  \bibAnnoteFile{NoStop}{smb6_trivial}%
\bibitem{smb6_sebastian}%
  \BibitemOpen
\bibfield{journal}{%
    }%
  \bibfield{author}{%
  \bibinfo {author} {\bibfnamefont{B.~S.}\ \bibnamefont{Tan}}, \bibinfo
  {author} {\bibfnamefont{Y.-T.}\ \bibnamefont{Hsu}}, \bibinfo {author}
  {\bibfnamefont{B.}~\bibnamefont{Zeng}}, \bibinfo {author}
  {\bibfnamefont{M.}~\bibnamefont{Ciomaga~Hatnean}}, \bibinfo {author}
  {\bibfnamefont{N.}~\bibnamefont{Harrison}}, \bibinfo {author}
  {\bibfnamefont{Z.}~\bibnamefont{Zhu}}, \bibinfo {author}
  {\bibfnamefont{M.}~\bibnamefont{Hartstein}}, \bibinfo {author}
  {\bibfnamefont{M.}~\bibnamefont{Kiourlappou}}, \bibinfo {author}
  {\bibfnamefont{A.}~\bibnamefont{Srivastava}}, \bibinfo {author}
  {\bibfnamefont{M.~D.}\ \bibnamefont{Johannes}}, \bibinfo {author}
  {\bibfnamefont{T.~P.}\ \bibnamefont{Murphy}}, \bibinfo {author}
  {\bibfnamefont{J.-H.}\ \bibnamefont{Park}}, \bibinfo {author}
  {\bibfnamefont{L.}~\bibnamefont{Balicas}}, \bibinfo {author}
  {\bibfnamefont{G.~G.}\ \bibnamefont{Lonzarich}}, \bibinfo {author}
  {\bibfnamefont{G.}~\bibnamefont{Balakrishnan}},\ and\ \bibinfo {author}
  {\bibfnamefont{S.~E.}\ \bibnamefont{Sebastian}},\ }%
  \bibfield{journal}{%
  \bibinfo {journal} {Science}\ }%
  \textbf{\bibinfo {volume} {349}},\ \bibinfo {pages} {287} (\bibinfo {year}
  {2015})%
  \bibAnnoteFile{NoStop}{smb6_sebastian}%
\bibitem{smb6_tci}%
  \BibitemOpen
  \bibfield{author}{%
  \bibinfo {author} {\bibfnamefont{M.}~\bibnamefont{Ye}}, \bibinfo {author}
  {\bibfnamefont{J.~W.}\ \bibnamefont{Allen}},\ and\ \bibinfo {author}
  {\bibfnamefont{K.}~\bibnamefont{Sun}},\ }%
  \bibinfo {journal} {preprint arXiv:1307.7191}%
  \bibAnnoteFile{NoStop}{smb6_tci}%
\bibitem{fu_tci}%
  \BibitemOpen
\bibfield{journal}{%
    }%
  \bibfield{author}{%
  \bibinfo {author} {\bibfnamefont{L.}~\bibnamefont{Fu}},\ }%
  \bibfield{journal}{%
  \bibinfo {journal} {Phys. Rev. Lett.}\ }%
  \textbf{\bibinfo {volume} {106}},\ \bibinfo {pages} {106802} (\bibinfo {year}
  {2011})%
  \bibAnnoteFile{NoStop}{fu_tci}%
\bibitem{smb6_prl_io}%
  \BibitemOpen
  \bibfield{author}{%
  \bibinfo {author} {\bibfnamefont{P.~P.}\ \bibnamefont{Baruselli}}\ and\
  \bibinfo {author} {\bibfnamefont{M.}~\bibnamefont{Vojta}},\ }%
  \bibfield{journal}{%
  \bibinfo {journal} {Phys. Rev. Lett.}\ }%
  \textbf{\bibinfo {volume} {115}},\ \bibinfo {pages} {156404} (\bibinfo {year}
  {2015})%
  \bibAnnoteFile{NoStop}{smb6_prl_io}%
\bibitem{sigrist_tki}%
  \BibitemOpen
  \bibfield{author}{%
  \bibinfo {author} {\bibfnamefont{M.}~\bibnamefont{Legner}}, \bibinfo {author}
  {\bibfnamefont{A.}~\bibnamefont{R\"uegg}},\ and\ \bibinfo {author}
  {\bibfnamefont{M.}~\bibnamefont{Sigrist}},\ }%
  \bibfield{journal}{%
  \bibinfo {journal} {Phys. Rev. Lett.}\ }%
  \textbf{\bibinfo {volume} {115}},\ \bibinfo {pages} {156405} (\bibinfo {year}
  {2015})%
  \bibAnnoteFile{NoStop}{sigrist_tki}%
\bibitem{smb6_magnres}%
  \BibitemOpen
  \bibfield{author}{%
  \bibinfo {author} {\bibfnamefont{F.}~\bibnamefont{Chen}}, \bibinfo {author}
  {\bibfnamefont{C.}~\bibnamefont{Shang}}, \bibinfo {author}
  {\bibfnamefont{Z.}~\bibnamefont{Jin}}, \bibinfo {author}
  {\bibfnamefont{D.}~\bibnamefont{Zhao}}, \bibinfo {author}
  {\bibfnamefont{Y.~P.}\ \bibnamefont{Wu}}, \bibinfo {author}
  {\bibfnamefont{Z.~J.}\ \bibnamefont{Xiang}}, \bibinfo {author}
  {\bibfnamefont{Z.~C.}\ \bibnamefont{Xia}}, \bibinfo {author}
  {\bibfnamefont{A.~F.}\ \bibnamefont{Wang}}, \bibinfo {author}
  {\bibfnamefont{X.~G.}\ \bibnamefont{Luo}}, \bibinfo {author}
  {\bibfnamefont{T.}~\bibnamefont{Wu}},\ and\ \bibinfo {author}
  {\bibfnamefont{X.~H.}\ \bibnamefont{Chen}},\ }%
  \bibfield{journal}{%
  \bibinfo {journal} {Phys. Rev. B}\ }%
  \textbf{\bibinfo {volume} {91}},\ \bibinfo {pages} {205133} (\bibinfo {year}
  {2015})%
  \bibAnnoteFile{NoStop}{smb6_magnres}%
\bibitem{smb6_stm_110}%
  \BibitemOpen
  \bibfield{author}{%
  \bibinfo {author} {\bibfnamefont{S.}~\bibnamefont{Roessler}}, \bibinfo
  {author} {\bibfnamefont{L.}~\bibnamefont{Jiao}}, \bibinfo {author}
  {\bibfnamefont{D.-J.}\ \bibnamefont{Kim}}, \bibinfo {author}
  {\bibfnamefont{S.}~\bibnamefont{Seiro}}, \bibinfo {author}
  {\bibfnamefont{K.}~\bibnamefont{Rasim}}, \bibinfo {author}
  {\bibfnamefont{F.}~\bibnamefont{Steglich}}, \bibinfo {author}
  {\bibfnamefont{L.~H.}\ \bibnamefont{Tjeng}}, \bibinfo {author}
  {\bibfnamefont{Z.}~\bibnamefont{Fisk}},\ and\ \bibinfo {author}
  {\bibfnamefont{S.}~\bibnamefont{Wirth}},\ }%
  \bibinfo {journal} {preprint arXiv:1510.06476}%
  \bibAnnoteFile{NoStop}{smb6_stm_110}%
\bibitem{smb6_arpes_110}%
  \BibitemOpen
\bibfield{journal}{%
    }%
  \bibfield{author}{%
  \bibinfo {author} {\bibfnamefont{J.~D.}\ \bibnamefont{Denlinger}}, \bibinfo
  {author} {\bibfnamefont{S.}~\bibnamefont{Jang}}, \bibinfo {author}
  {\bibfnamefont{G.}~\bibnamefont{Li}}, \bibinfo {author}
  {\bibfnamefont{L.}~\bibnamefont{Chen}}, \bibinfo {author}
  {\bibfnamefont{B.~J.}\ \bibnamefont{Lawson}}, \bibinfo {author}
  {\bibfnamefont{T.}~\bibnamefont{Asaba}}, \bibinfo {author}
  {\bibfnamefont{C.}~\bibnamefont{Tinsman}}, \bibinfo {author}
  {\bibfnamefont{F.}~\bibnamefont{Yu}}, \bibinfo {author}
  {\bibfnamefont{K.}~\bibnamefont{Sun}}, \bibinfo {author}
  {\bibfnamefont{J.~W.}\ \bibnamefont{Allen}}, \bibinfo {author}
  {\bibfnamefont{C.}~\bibnamefont{Kurdak}}, \bibinfo {author}
  {\bibfnamefont{D.-J.}\ \bibnamefont{Kim}}, \bibinfo {author}
  {\bibfnamefont{Z.}~\bibnamefont{Fisk}},\ and\ \bibinfo {author}
  {\bibfnamefont{L.}~\bibnamefont{Li}},\ }%
  \bibinfo {journal} {preprint arXiv:1601.07408}%
  \bibAnnoteFile{NoStop}{smb6_arpes_110}%
\bibitem{liu_ti_model}%
  \BibitemOpen
\bibfield{journal}{%
    }%
  \bibfield{author}{%
  \bibinfo {author} {\bibfnamefont{C.-X.}\ \bibnamefont{Liu}}, \bibinfo
  {author} {\bibfnamefont{X.-L.}\ \bibnamefont{Qi}}, \bibinfo {author}
  {\bibfnamefont{H.}~\bibnamefont{Zhang}}, \bibinfo {author}
  {\bibfnamefont{X.}~\bibnamefont{Dai}}, \bibinfo {author}
  {\bibfnamefont{Z.}~\bibnamefont{Fang}},\ and\ \bibinfo {author}
  {\bibfnamefont{S.-C.}\ \bibnamefont{Zhang}},\ }%
  \bibfield{journal}{%
  \bibinfo {journal} {Phys. Rev. B}\ }%
  \textbf{\bibinfo {volume} {82}},\ \bibinfo {pages} {045122} (\bibinfo {year}
  {2010})%
  \bibAnnoteFile{NoStop}{liu_ti_model}%
\bibitem{smb6_korea}%
  \BibitemOpen
  \bibfield{author}{%
  \bibinfo {author} {\bibfnamefont{K.}~\bibnamefont{Chang-Jong}}, \bibinfo
  {author} {\bibfnamefont{K.}~\bibnamefont{Junwon}}, \bibinfo {author}
  {\bibfnamefont{K.}~\bibnamefont{Kyoo}}, \bibinfo {author}
  {\bibfnamefont{J.-S.}\ \bibnamefont{Kang}}, \bibinfo {author}
  {\bibfnamefont{J.~D.}\ \bibnamefont{Denlinger}},\ and\ \bibinfo {author}
  {\bibfnamefont{B.~I.}\ \bibnamefont{Min}},\ }%
  \bibinfo {journal} {preprint arXiv:1312.5898}%
  \bibAnnoteFile{NoStop}{smb6_korea}%
\bibitem{pub6}%
  \BibitemOpen
\bibfield{journal}{%
    }%
  \bibfield{author}{%
  \bibinfo {author} {\bibfnamefont{X.}~\bibnamefont{Deng}}, \bibinfo {author}
  {\bibfnamefont{K.}~\bibnamefont{Haule}},\ and\ \bibinfo {author}
  {\bibfnamefont{G.}~\bibnamefont{Kotliar}},\ }%
  \bibfield{journal}{%
  \bibinfo {journal} {Phys. Rev. Lett.}\ }%
  \textbf{\bibinfo {volume} {111}},\ \bibinfo {pages} {176404} (\bibinfo {year}
  {2013})%
  \bibAnnoteFile{NoStop}{pub6}%
\bibitem{prbr_io_smb6}%
  \BibitemOpen
  \bibfield{author}{%
  \bibinfo {author} {\bibfnamefont{P.~P.}\ \bibnamefont{Baruselli}}\ and\
  \bibinfo {author} {\bibfnamefont{M.}~\bibnamefont{Vojta}},\ }%
  \bibfield{journal}{%
  \bibinfo {journal} {Phys. Rev. B}\ }%
  \textbf{\bibinfo {volume} {90}},\ \bibinfo {pages} {201106} (\bibinfo {year}
  {2014})%
  \bibAnnoteFile{NoStop}{prbr_io_smb6}%
\bibitem{kane_bisb}%
  \BibitemOpen
  \bibfield{author}{%
  \bibinfo {author} {\bibfnamefont{J.~C.~Y.}\ \bibnamefont{Teo}}, \bibinfo
  {author} {\bibfnamefont{L.}~\bibnamefont{Fu}},\ and\ \bibinfo {author}
  {\bibfnamefont{C.~L.}\ \bibnamefont{Kane}},\ }%
  \bibfield{journal}{%
  \bibinfo {journal} {Phys. Rev. B}\ }%
  \textbf{\bibinfo {volume} {78}},\ \bibinfo {pages} {045426} (\bibinfo {year}
  {2008})%
  \bibAnnoteFile{NoStop}{kane_bisb}%
\bibitem{suppl_tci_long}%
  \BibitemOpen
  \bibinfo {note} {See Supplementary Material for details about the calculation
  at $\bar \Gamma$, the spatial component of surface states, the third order
  correction to velocities, the numerical \kdp method for a slab, the spin
  expectation value (SEV), the calculation on a generic surface, topological
  phase transitions, and DFT results.}%
  \bibAnnoteFile{Stop}{suppl_tci_long}%
\bibitem{hewson}%
  \BibitemOpen
  \bibfield{author}{%
  \bibinfo {author} {\bibfnamefont{A.~C.}\ \bibnamefont{Hewson}},\ }%
  \emph{\bibinfo {title} {The Kondo Problem to Heavy Fermions}}\ (\bibinfo
  {publisher} {Cambridge University Press, Cambridge},\ \bibinfo {year}
  {1993})%
  \bibAnnoteFile{NoStop}{hewson}%
\bibitem{assaad_review}%
  \BibitemOpen
  \bibfield{author}{%
  \bibinfo {author} {\bibfnamefont{M.}~\bibnamefont{Hohenadler}}\ and\ \bibinfo
  {author} {\bibfnamefont{F.~F.}\ \bibnamefont{Assaad}},\ }%
  \bibfield{journal}{%
  \bibinfo {journal} {Journal of Physics: Condensed Matter}\ }%
  \textbf{\bibinfo {volume} {25}},\ \bibinfo {pages} {143201} (\bibinfo {year}
  {2013})%
  \bibAnnoteFile{NoStop}{assaad_review}%
\bibitem{smb6_korea2}%
  \BibitemOpen
  \bibfield{author}{%
  \bibinfo {author} {\bibfnamefont{J.}~\bibnamefont{Kim}}, \bibinfo {author}
  {\bibfnamefont{K.}~\bibnamefont{Kim}}, \bibinfo {author}
  {\bibfnamefont{C.-J.}\ \bibnamefont{Kang}}, \bibinfo {author}
  {\bibfnamefont{S.}~\bibnamefont{Kim}}, \bibinfo {author}
  {\bibfnamefont{H.~C.}\ \bibnamefont{Choi}}, \bibinfo {author}
  {\bibfnamefont{J.-S.}\ \bibnamefont{Kang}}, \bibinfo {author}
  {\bibfnamefont{J.~D.}\ \bibnamefont{Denlinger}},\ and\ \bibinfo {author}
  {\bibfnamefont{B.~I.}\ \bibnamefont{Min}},\ }%
  \bibfield{journal}{%
  \bibinfo {journal} {Phys. Rev. B}\ }%
  \textbf{\bibinfo {volume} {90}},\ \bibinfo {pages} {075131} (\bibinfo {year}
  {2014})%
  \bibAnnoteFile{NoStop}{smb6_korea2}%
\bibitem{yu_smb6_qpi}%
  \BibitemOpen
  \bibfield{author}{%
  \bibinfo {author} {\bibfnamefont{R.}~\bibnamefont{Yu}}, \bibinfo {author}
  {\bibfnamefont{H.}~\bibnamefont{Weng}}, \bibinfo {author}
  {\bibfnamefont{X.}~\bibnamefont{Hu}}, \bibinfo {author}
  {\bibfnamefont{Z.}~\bibnamefont{Fang}},\ and\ \bibinfo {author}
  {\bibfnamefont{X.}~\bibnamefont{Dai}},\ }%
  \bibfield{journal}{%
  \bibinfo {journal} {New J. Phys.}\ }%
  \textbf{\bibinfo {volume} {17}},\ \bibinfo {pages} {023012} (\bibinfo {year}
  {2015})%
  \bibAnnoteFile{NoStop}{yu_smb6_qpi}%
\bibitem{dzero_pert}%
  \BibitemOpen
  \bibfield{author}{%
  \bibinfo {author} {\bibfnamefont{B.}~\bibnamefont{Roy}}, \bibinfo {author}
  {\bibfnamefont{J.~D.}\ \bibnamefont{Sau}}, \bibinfo {author}
  {\bibfnamefont{M.}~\bibnamefont{Dzero}},\ and\ \bibinfo {author}
  {\bibfnamefont{V.}~\bibnamefont{Galitski}},\ }%
  \bibfield{journal}{%
  \bibinfo {journal} {Phys. Rev. B}\ }%
  \textbf{\bibinfo {volume} {90}},\ \bibinfo {pages} {155314} (\bibinfo {year}
  {2014})%
  \bibAnnoteFile{NoStop}{dzero_pert}%
\bibitem{smb6_korea_dft}%
  \BibitemOpen
  \bibfield{author}{%
  \bibinfo {author} {\bibfnamefont{C.-J.}\ \bibnamefont{Kang}}, \bibinfo
  {author} {\bibfnamefont{J.}~\bibnamefont{Kim}}, \bibinfo {author}
  {\bibfnamefont{K.}~\bibnamefont{Kim}}, \bibinfo {author}
  {\bibfnamefont{J.}~\bibnamefont{Kang}}, \bibinfo {author}
  {\bibfnamefont{J.~D.}\ \bibnamefont{Denlinger}},\ and\ \bibinfo {author}
  {\bibfnamefont{B.~I.}\ \bibnamefont{Min}},\ }%
  \bibfield{journal}{%
  \bibinfo {journal} {J. Phys. Soc. Jpn.}\ }%
  \textbf{\bibinfo {volume} {84}},\ \bibinfo {pages} {024722} (\bibinfo {year}
  {2015})%
  \bibAnnoteFile{NoStop}{smb6_korea_dft}%
\bibitem{smb6_dmft_slab}%
  \BibitemOpen
  \bibfield{author}{%
  \bibinfo {author} {\bibfnamefont{R.}~\bibnamefont{Peters}}, \bibinfo {author}
  {\bibfnamefont{Y.}~\bibnamefont{Tsuneya}}, \bibinfo {author}
  {\bibfnamefont{S.}~\bibnamefont{Hirofumi}},\ and\ \bibinfo {author}
  {\bibfnamefont{N.}~\bibnamefont{Kawakami}},\ }%
  \bibinfo {journal} {preprint arXiv:1510.06476}%
  \bibAnnoteFile{NoStop}{smb6_dmft_slab}%
\bibitem{pnas_smb6_stm}%
  \BibitemOpen
\bibfield{journal}{%
    }%
  \bibfield{author}{%
  \bibinfo {author} {\bibfnamefont{S.}~\bibnamefont{R\"o{\ss}ler}}, \bibinfo
  {author} {\bibfnamefont{T.-H.}\ \bibnamefont{Jang}}, \bibinfo {author}
  {\bibfnamefont{D.-J.}\ \bibnamefont{Kim}}, \bibinfo {author}
  {\bibfnamefont{L.~H.}\ \bibnamefont{Tjeng}}, \bibinfo {author}
  {\bibfnamefont{Z.}~\bibnamefont{Fisk}}, \bibinfo {author}
  {\bibfnamefont{F.}~\bibnamefont{Steglich}},\ and\ \bibinfo {author}
  {\bibfnamefont{S.}~\bibnamefont{Wirth}},\ }%
  \bibfield{journal}{%
  \bibinfo {journal} {Proc. Nat. Acad. Sci.}\ }%
  \textbf{\bibinfo {volume} {111}},\ \bibinfo {pages} {4798} (\bibinfo {year}
  {2014})%
  \bibAnnoteFile{NoStop}{pnas_smb6_stm}%
\bibitem{hoffman_smb6}%
  \BibitemOpen
  \bibfield{author}{%
  \bibinfo {author} {\bibfnamefont{M.~M.}\ \bibnamefont{Yee}}, \bibinfo
  {author} {\bibfnamefont{Y.}~\bibnamefont{He}}, \bibinfo {author}
  {\bibfnamefont{A.}~\bibnamefont{Soumyanarayanan}}, \bibinfo {author}
  {\bibfnamefont{D.-J.}\ \bibnamefont{Kim}}, \bibinfo {author}
  {\bibfnamefont{Z.}~\bibnamefont{Fisk}},\ and\ \bibinfo {author}
  {\bibfnamefont{J.~E.}\ \bibnamefont{Hoffman}},\ }%
  \bibinfo {journal} {preprint arXiv:1308.1085}%
  \bibAnnoteFile{NoStop}{hoffman_smb6}%
\bibitem{smb6_rec_io}%
  \BibitemOpen
\bibfield{journal}{%
    }%
  \bibfield{author}{%
  \bibinfo {author} {\bibfnamefont{P.~P.}\ \bibnamefont{Baruselli}}\ and\
  \bibinfo {author} {\bibfnamefont{M.}~\bibnamefont{Vojta}},\ }%
  \bibfield{journal}{%
  \bibinfo {journal} {2D Mater.}\ }%
  \textbf{\bibinfo {volume} {2}},\ \bibinfo {pages} {044011} (\bibinfo {year}
  {2015})%
  \bibAnnoteFile{NoStop}{smb6_rec_io}%
\bibitem{prb_io_tki}%
  \BibitemOpen
  \bibfield{author}{%
  \bibinfo {author} {\bibfnamefont{P.~P.}\ \bibnamefont{Baruselli}}\ and\
  \bibinfo {author} {\bibfnamefont{M.}~\bibnamefont{Vojta}},\ }%
  \bibfield{journal}{%
  \bibinfo {journal} {Phys. Rev. B}\ }%
  \textbf{\bibinfo {volume} {89}},\ \bibinfo {pages} {205105} (\bibinfo {year}
  {2014})%
  \bibAnnoteFile{NoStop}{prb_io_tki}%
\bibitem{sigrist_tki_prb}%
  \BibitemOpen
  \bibfield{author}{%
  \bibinfo {author} {\bibfnamefont{M.}~\bibnamefont{Legner}}, \bibinfo {author}
  {\bibfnamefont{A.}~\bibnamefont{R\"uegg}},\ and\ \bibinfo {author}
  {\bibfnamefont{M.}~\bibnamefont{Sigrist}},\ }%
  \bibfield{journal}{%
  \bibinfo {journal} {Phys. Rev. B}\ }%
  \textbf{\bibinfo {volume} {89}},\ \bibinfo {pages} {085110} (\bibinfo {year}
  {2014})%
  \bibAnnoteFile{NoStop}{sigrist_tki_prb}%
\end{thebibliography}%


\begin{thebibliography}{10}%
\makeatletter
\providecommand \@ifxundefined [1]{%
 \ifx #1\undefined \expandafter \@firstoftwo
 \else \expandafter \@secondoftwo
\fi
}%
\providecommand \@ifnum [1]{%
 \ifnum #1\expandafter \@firstoftwo
 \else \expandafter \@secondoftwo
\fi
}%
\providecommand \enquote [1]{``#1''}%
\providecommand \bibnamefont  [1]{#1}%
\providecommand \bibfnamefont [1]{#1}%
\providecommand \citenamefont [1]{#1}%
\providecommand\href[0]{\@sanitize\@href}%
\providecommand\@href[1]{\endgroup\@@startlink{#1}\endgroup\@@href}%
\providecommand\@@href[1]{#1\@@endlink}%
\providecommand \@sanitize [0]{\begingroup\catcode`\&12\catcode`\#12\relax}%
\@ifxundefined \pdfoutput {\@firstoftwo}{%
 \@ifnum{\z@=\pdfoutput}{\@firstoftwo}{\@secondoftwo}%
}{%
 \providecommand\@@startlink[1]{\leavevmode\special{html:<a href="#1">}}%
 \providecommand\@@endlink[0]{\special{html:</a>}}%
}{%
 \providecommand\@@startlink[1]{%
  \leavevmode
  \pdfstartlink
   attr{/Border[0 0 1 ]/H/I/C[0 1 1]}%
   user{/Subtype/Link/A<</Type/Action/S/URI/URI(#1)>>}%
  \relax
 }%
 \providecommand\@@endlink[0]{\pdfendlink}%
}%
\providecommand \url  [0]{\begingroup\@sanitize \@url }%
\providecommand \@url [1]{\endgroup\@href {#1}{\urlprefix}}%
\providecommand \urlprefix [0]{URL }%
\providecommand \Eprint[0]{\href }%
\@ifxundefined \urlstyle {%
  \providecommand \doi [1]{doi:\discretionary{}{}{}#1}%
}{%
  \providecommand \doi [0]{doi:\discretionary{}{}{}\begingroup
  \urlstyle{rm}\Url }%
}%
\providecommand \doibase [0]{http://dx.doi.org/}%
\providecommand \Doi[1]{\href{\doibase#1}}%
\providecommand \bibAnnote [3]{%
  \BibitemShut{#1}%
  \begin{quotation}\noindent
    \textsc{Key:}\ #2\\\textsc{Annotation:}\ #3%
  \end{quotation}%
}%
\providecommand \bibAnnoteFile [2]{%
  \IfFileExists{#2}{\bibAnnote {#1} {#2} {\input{#2}}}{}%
}%
\providecommand \typeout [0]{\immediate \write \m@ne }%
\providecommand \selectlanguage [0]{\@gobble}%
\providecommand \bibinfo [0]{\@secondoftwo}%
\providecommand \bibfield [0]{\@secondoftwo}%
\providecommand \translation [1]{[#1]}%
\providecommand \BibitemOpen[0]{}%
\providecommand \bibitemStop [0]{}%
\providecommand \bibitemNoStop [0]{.\EOS\space}%
\providecommand \EOS [0]{\spacefactor3000\relax}%
\providecommand \BibitemShut [1]{\csname bibitem#1\endcsname}%
\bibitem{liu_ti_model}%
  \BibitemOpen
  \bibfield{author}{%
  \bibinfo {author} {\bibfnamefont{C.-X.}\ \bibnamefont{Liu}}, \bibinfo
  {author} {\bibfnamefont{X.-L.}\ \bibnamefont{Qi}}, \bibinfo {author}
  {\bibfnamefont{H.}~\bibnamefont{Zhang}}, \bibinfo {author}
  {\bibfnamefont{X.}~\bibnamefont{Dai}}, \bibinfo {author}
  {\bibfnamefont{Z.}~\bibnamefont{Fang}},\ and\ \bibinfo {author}
  {\bibfnamefont{S.-C.}\ \bibnamefont{Zhang}},\ }%
  \bibfield{journal}{%
  \bibinfo {journal} {Phys. Rev. B}\ }%
  \textbf{\bibinfo {volume} {82}},\ \bibinfo {pages} {045122} (\bibinfo {year}
  {2010})%
  \bibAnnoteFile{NoStop}{liu_ti_model}%
\bibitem{dzero_pert}%
  \BibitemOpen
  \bibfield{author}{%
  \bibinfo {author} {\bibfnamefont{B.}~\bibnamefont{Roy}}, \bibinfo {author}
  {\bibfnamefont{J.~D.}\ \bibnamefont{Sau}}, \bibinfo {author}
  {\bibfnamefont{M.}~\bibnamefont{Dzero}},\ and\ \bibinfo {author}
  {\bibfnamefont{V.}~\bibnamefont{Galitski}},\ }%
  \bibfield{journal}{%
  \bibinfo {journal} {Phys. Rev. B}\ }%
  \textbf{\bibinfo {volume} {90}},\ \bibinfo {pages} {155314} (\bibinfo {year}
  {2014})%
  \bibAnnoteFile{NoStop}{dzero_pert}%
\bibitem{lu_smb6_gutz}%
  \BibitemOpen
  \bibfield{author}{%
  \bibinfo {author} {\bibfnamefont{F.}~\bibnamefont{Lu}}, \bibinfo {author}
  {\bibfnamefont{J.}~\bibnamefont{Zhao}}, \bibinfo {author}
  {\bibfnamefont{H.}~\bibnamefont{Weng}}, \bibinfo {author}
  {\bibfnamefont{Z.}~\bibnamefont{Fang}},\ and\ \bibinfo {author}
  {\bibfnamefont{X.}~\bibnamefont{Dai}},\ }%
  \bibfield{journal}{%
  \bibinfo {journal} {Phys. Rev. Lett.}\ }%
  \textbf{\bibinfo {volume} {110}},\ \bibinfo {pages} {096401} (\bibinfo {year}
  {2013})%
  \bibAnnoteFile{NoStop}{lu_smb6_gutz}%
\bibitem{smb6_korea_dft}%
  \BibitemOpen
  \bibfield{author}{%
  \bibinfo {author} {\bibfnamefont{C.-J.}\ \bibnamefont{Kang}}, \bibinfo
  {author} {\bibfnamefont{J.}~\bibnamefont{Kim}}, \bibinfo {author}
  {\bibfnamefont{K.}~\bibnamefont{Kim}}, \bibinfo {author}
  {\bibfnamefont{J.}~\bibnamefont{Kang}}, \bibinfo {author}
  {\bibfnamefont{J.~D.}\ \bibnamefont{Denlinger}},\ and\ \bibinfo {author}
  {\bibfnamefont{B.~I.}\ \bibnamefont{Min}},\ }%
  \bibfield{journal}{%
  \bibinfo {journal} {J. Phys. Soc. Jpn.}\ }%
  \textbf{\bibinfo {volume} {84}},\ \bibinfo {pages} {024722} (\bibinfo {year}
  {2015})%
  \bibAnnoteFile{NoStop}{smb6_korea_dft}%
\bibitem{kondo_breakdown}%
  \BibitemOpen
  \bibfield{author}{%
  \bibinfo {author} {\bibfnamefont{V.}~\bibnamefont{Alexandrov}}, \bibinfo
  {author} {\bibfnamefont{P.}~\bibnamefont{Coleman}},\ and\ \bibinfo {author}
  {\bibfnamefont{O.}~\bibnamefont{Erten}},\ }%
  \bibfield{journal}{%
  \bibinfo {journal} {Phys. Rev. Lett.}\ }%
  \textbf{\bibinfo {volume} {114}},\ \bibinfo {pages} {177202} (\bibinfo {year}
  {2015})%
  \bibAnnoteFile{NoStop}{kondo_breakdown}%
\bibitem{smb6_rec_io}%
  \BibitemOpen
  \bibfield{author}{%
  \bibinfo {author} {\bibfnamefont{P.~P.}\ \bibnamefont{Baruselli}}\ and\
  \bibinfo {author} {\bibfnamefont{M.}~\bibnamefont{Vojta}},\ }%
  \bibfield{journal}{%
  \bibinfo {journal} {2D Mater.}\ }%
  \textbf{\bibinfo {volume} {2}},\ \bibinfo {pages} {044011} (\bibinfo {year}
  {2015})%
  \bibAnnoteFile{NoStop}{smb6_rec_io}%
\bibitem{prbr_io_smb6}%
  \BibitemOpen
  \bibfield{author}{%
  \bibinfo {author} {\bibfnamefont{P.~P.}\ \bibnamefont{Baruselli}}\ and\
  \bibinfo {author} {\bibfnamefont{M.}~\bibnamefont{Vojta}},\ }%
  \bibfield{journal}{%
  \bibinfo {journal} {Phys. Rev. B}\ }%
  \textbf{\bibinfo {volume} {90}},\ \bibinfo {pages} {201106} (\bibinfo {year}
  {2014})%
  \bibAnnoteFile{NoStop}{prbr_io_smb6}%
\bibitem{yu_smb6_qpi}%
  \BibitemOpen
  \bibfield{author}{%
  \bibinfo {author} {\bibfnamefont{R.}~\bibnamefont{Yu}}, \bibinfo {author}
  {\bibfnamefont{H.}~\bibnamefont{Weng}}, \bibinfo {author}
  {\bibfnamefont{X.}~\bibnamefont{Hu}}, \bibinfo {author}
  {\bibfnamefont{Z.}~\bibnamefont{Fang}},\ and\ \bibinfo {author}
  {\bibfnamefont{X.}~\bibnamefont{Dai}},\ }%
  \bibfield{journal}{%
  \bibinfo {journal} {New J. Phys.}\ }%
  \textbf{\bibinfo {volume} {17}},\ \bibinfo {pages} {023012} (\bibinfo {year}
  {2015})%
  \bibAnnoteFile{NoStop}{yu_smb6_qpi}%
\bibitem{pub6}%
  \BibitemOpen
  \bibfield{author}{%
  \bibinfo {author} {\bibfnamefont{X.}~\bibnamefont{Deng}}, \bibinfo {author}
  {\bibfnamefont{K.}~\bibnamefont{Haule}},\ and\ \bibinfo {author}
  {\bibfnamefont{G.}~\bibnamefont{Kotliar}},\ }%
  \bibfield{journal}{%
  \bibinfo {journal} {Phys. Rev. Lett.}\ }%
  \textbf{\bibinfo {volume} {111}},\ \bibinfo {pages} {176404} (\bibinfo {year}
  {2013})%
  \bibAnnoteFile{NoStop}{pub6}%
\bibitem{smb6_arpes_mesot_spin}%
  \BibitemOpen
  \bibfield{author}{%
  \bibinfo {author} {\bibfnamefont{N.}~\bibnamefont{Xu}}, \bibinfo {author}
  {\bibfnamefont{P.~K.}\ \bibnamefont{Biswas}}, \bibinfo {author}
  {\bibfnamefont{J.~H.}\ \bibnamefont{Dil}}, \bibinfo {author}
  {\bibfnamefont{R.~S.}\ \bibnamefont{Dhaka}}, \bibinfo {author}
  {\bibfnamefont{G.}~\bibnamefont{Landolt}}, \bibinfo {author}
  {\bibfnamefont{S.}~\bibnamefont{Muff}}, \bibinfo {author}
  {\bibfnamefont{C.~E.}\ \bibnamefont{Matt}}, \bibinfo {author}
  {\bibfnamefont{X.}~\bibnamefont{Shi}}, \bibinfo {author}
  {\bibfnamefont{N.~C.}\ \bibnamefont{Plumb}}, \bibinfo {author}
  {\bibfnamefont{M.}~\bibnamefont{Radovic}}, \bibinfo {author}
  {\bibfnamefont{E.}~\bibnamefont{Pomjakushina}}, \bibinfo {author}
  {\bibfnamefont{K.}~\bibnamefont{Conder}}, \bibinfo {author}
  {\bibfnamefont{A.}~\bibnamefont{Amato}}, \bibinfo {author}
  {\bibfnamefont{S.}~\bibnamefont{Borisenko}}, \bibinfo {author}
  {\bibfnamefont{R.}~\bibnamefont{Yu}}, \bibinfo {author}
  {\bibfnamefont{H.-M.}\ \bibnamefont{Weng}}, \bibinfo {author}
  {\bibfnamefont{Z.}~\bibnamefont{Fang}}, \bibinfo {author}
  {\bibfnamefont{X.}~\bibnamefont{Dai}}, \bibinfo {author}
  {\bibfnamefont{J.}~\bibnamefont{Mesot}}, \bibinfo {author}
  {\bibfnamefont{H.}~\bibnamefont{Ding}},\ and\ \bibinfo {author}
  {\bibfnamefont{M.}~\bibnamefont{Shi}},\ }%
  \bibfield{journal}{%
  \bibinfo {journal} {Nature Comm.}\ }%
  \textbf{\bibinfo {volume} {5}},\ \bibinfo {pages} {4566} (\bibinfo {year}
  {2014})%
  \bibAnnoteFile{NoStop}{smb6_arpes_mesot_spin}%
\end{thebibliography}%


\end{document}


\title{Supplemental material:\\
Spin textures on general surfaces of the correlated topological insulator {\sm}
}

\author{Pier Paolo Baruselli}
\author{Matthias Vojta}
\affiliation{Institut f\"ur Theoretische Physik, Technische Universit\"at Dresden, 01062 Dresden, Germany}
\date{\today}
\maketitle

In this Supplemental Material we provide further information about the calculation at $\bar \Gamma$,
the spatial component of surface states, the third order correction to velocities, the numerical \kdp method for a slab, the spin expectation value (SEV), the calculation on a generic surface, topological phase transitions, and DFT results.

\section{Details of the calculation at $\bar\Gamma$}\label{app_gamma}

Here we illustrate how to get the doublet of surface states at $\bar\Gamma$, following Refs. \onlinecite{liu_ti_model} and \onlinecite{dzero_pert}.

We start from Eq. {\eho} of the main text.
In a semi-infinite slab geometry, we want to find surface states at $\bar\Gamma$, so we transform $k_z\rightarrow -\ii d/dz$, and we pose $\psi_{+}^n(z)=e^{\lambda z}\psi_+^n(\lambda)=e^{\lambda z}(\alpha_+^n(\lambda),\beta_+^n(\lambda))$, so that
\be
H_{0}^{+(n)}(k_z\rightarrow -\ii \lambda)\psi^n_+(\lambda)=E^n_+(\lambda)\psi^n_+(\lambda), \hspace{5pt} n=1,2,
\ee
together with $\psi(z=0)=\psi(z\rightarrow-\infty)=0$.
We can now write
\bea
(B-\lambda^2)\alpha_+^1-D_1\lambda \beta_+^1&=&0\\
D_2\lambda \alpha_+^1+(C-\lambda^2)\beta_+^1&=&0
\eea
with (we set $E=E^1_+(\lambda)$)
\bea
B=\frac{\epsilon_1^d-E}{-t_d g_1^d},\hspace{10pt}
C=\frac{\epsilon_1^f-E}{-t_f g_1^f},\\
D_1=\frac{V f_1^v}{-t_d g_1^d},\hspace{10pt}
D_2=\frac{V f_1^v}{-t_f g_1^f},\label{d1d2}
\eea
from which
\be\label{lambda4}
\lambda^4-\lambda^2(B+C-D_1D_2)+BC=0,
\ee
which determines four solutions $\lambda_i(E)$, $i=1,2,3,4$ in the form $\pm \sqrt{\Lambda_i}$, $i=1,2$, where $\lambda^2=\Lambda$.
In order to have surface states we require two solutions $\lambda_1$ and $\lambda_2$ to have
a positive real part. This is equivalent to saying that $\Lambda_i$ are both real positive,
which leads to $\lambda_i=+\sqrt{\Lambda_i}$, or complex conjugate, which leads to $\lambda_1$ and $\lambda_2$ being complex conjugate, too.
So
\be
\psi_+^1(z)=e^{\lambda_1 z} (\alpha_+^{1'}, \beta_+^{1'}) + e^{\lambda_2 z} (\alpha_+^{1''}, \beta_+^{1''}),
\ee
where, imposing the boundary condition at $z=0$ we get $\alpha_+^{1'}=-\alpha_+^{1''}\equiv \alpha$, $\beta_+^{1'}=-\beta_+^{1''}\equiv \beta$, and
\bea
\frac{D_2\lambda_1}{C-\lambda_1^2}&=&\frac{D_2\lambda_2}{C-\lambda_2^2},\\
\frac{B-\lambda_1^2}{D_1\lambda_1}&=&\frac{B-\lambda_2^2}{D_1\lambda_2},
\eea
from which $B=C=-\lambda_1\lambda_2<0$,
\bea
E=\frac{\epsilon_1^f t_d g_1^d-\epsilon_1^d t_f g_1^f}{t_dg_1^d-t_fg_1^f},\label{e_gamma}\\
B=C=\frac{\epsilon_1^f-\epsilon_1^d}{t_dg_1^d-t_fg_1^f},\label{bc}
\eea
and finally, after integrating in $dz$  from $-\infty$ to 0 
\bea
\psi_+^1&=&N\left(\frac{1}{\lambda_1}-\frac{1}{\lambda_2}\right)(1, D_2/(\lambda_1+\lambda_2))\nonumber\\
&=&N\left(\frac{1}{\lambda_1}-\frac{1}{\lambda_2}\right)(1, -(\lambda_1+\lambda_2)/D_1),
\eea
where $N$ is an unimportant normalization factor. We also find that $D_1D_2=-(\lambda_1+\lambda_2)^2<0$,
which leads to $B<0$, $C<0$, and $D_1D_2<0$ as conditions for having surface states. 

For $\Gamma_8^{(2)}$ states we note that, while $\epsilon_1^d<\epsilon_1^f$, now $\epsilon_2^d>\epsilon_2^f$, which means that in this subspace no inversion is achieved,
and topological surface states are not allowed.
Indeed, we observe that the kinetic energy is such that $d_{z^2}$ and $\Gamma_8^{(2)}$ bands never cross along $\Gamma$-$X$,
while $d_{x^2-y^2}$ and $\Gamma_8^{(1)}$ do.\cite{lu_smb6_gutz,smb6_korea_dft}

So surface states at $\bar\Gamma$ only exist in the $d_{x^2-y^2}$ - $\Gamma_8^{(1)}$ subspace, and we can write after the $dz$ integration:
\bea
|\psi_+\rangle&=&\alpha|d^1\uparrow\rangle+\beta|f^1+\rangle,\nonumber\\
|\psi_-\rangle&=&\alpha|d^1\downarrow\rangle-\beta|f^1-\rangle,
\eea
which corresponds to Eq. {\eqpsi} of the main text;
$|\psi_-\rangle$ can be obtained with the same procedure for $H_0^{-(1)}$, or simply as $|\psi_-\rangle=T|\psi_+\rangle$, and:
\bea
\alpha&=&\sqrt{\frac{t_fg_1^f}{t_fg_1^f-t_dg_1^d}},\\
\beta&=&\alpha\frac{V f_1^v}{-t_f g_1^f(\lambda_1+\lambda_2)}=\alpha\frac{t_d g_1^d(\lambda_1+\lambda_2)}{V f_1^v}=\nonumber\\
&=&-\sgn(Vf_1^v)\sqrt{\frac{t_dg_1^d}{t_dg_1^d-t_fg_1^f}},
\eea
where we have used $t_d>0$, $g_1^d<0$, or, equivalently, $t_f<0$, $g_1^f<0$, and $\lambda_1+\lambda_2>0$ because they are both positive, or complex conjugate with a positive real part.
These correspond to Eqs. {\eqalpha} and {\eqbeta} of the main text.

After finding surface states at $k_{x,y}=0$, we can look at the perturbing Hamiltonian $H_P$, which is,
in the basis $|d_1\uparrow\rangle$, $|d_1\downarrow\rangle$, $|f_1+\rangle$, $|f_1-\rangle$:
\begin{align}\label{h_p}
H_P=
\left( \begin{array}{llll}
-t_dk_\parallel^2l_1^d &0 & 0& -iV h_1^v k_-\\
0 &-t_dk_\parallel^2l_1^d &  -iV h_1^v k_+&0\\
 0& iV h_1^v k_-&-t_fk_\parallel^2l_1^f&0\\
iVh_1^vk_+ &0 &0& -t_fk_\parallel^2l_1^f
  \end{array}\right)&,
\end{align}
from which we get Eqs. {\eqheff} of the main text at the linear order.

When we retain $\Gamma_7$ and $\Gamma_8$ states, we have:
\begin{align}\label{h1pp_17}
H_0^{+(1)}=
\left( \begin{array}{lll}
\epsilon_1^d-t_dk_z^2 g_1^d & -\ii Vk_z f_1^v&-\ii Vk_z f_7^v \\
\ii Vk_z f_1^v &\epsilon_1^f-t_fk_z^2 g_1^f&m_{78}\\
\ii Vk_z f_7^v &m_{78} & \epsilon_7^f-t_fk_z^2 g_7^f
  \end{array}\right)&,
\end{align}
which we reduce to Eq. {\eqred} of the main text by keeping $|d^1\uparrow\rangle$ and $|f^{17}_p+\rangle$.

\section{Spatial dependence of surface states}\label{app_length}
We now consider the spatial dependence of surface states.

Knowing that $B=C$, Eq. \eqref{bc}, we can write Eq. \eqref{lambda4} as
\bea
0&=&\lambda^4-\lambda^2(2B-D_1D_2)+B^2\\
&=&(\lambda^2-\sqrt{-D_1D_2}\lambda-B)(\lambda^2+\sqrt{-D_1D_2}\lambda-B)\nonumber
\eea
which reduces to
\be
\lambda^2-\sqrt{-D_1D_2}\lambda-B=0
\ee
if we require the real part of $\lambda_{1,2}$ to be positive, with solution
\bea
2\lambda_{1,2}&=&\sqrt{-D_1D_2}\pm\sqrt{-D_1D_2+4B}\nonumber\\
&=&\sqrt{|D_1D_2|}\pm\sqrt{|D_1D_2|-4|B|},
\eea
where $\lambda_{1,2}$ can be real or complex conjugate.
The spatial part of the wavefunction is:
\be
\psi(z)=N(e^{\lambda_1 z}-e^{\lambda_2 z}),
\ee
where we take
\be
\int_{-\infty}^0 \psi^2(z)dz=1.
\ee
We observe that $\psi(z)$ can always be taken as real;
useful integrals are:
\bea
\int_{-\infty}^0 \psi'(z)\psi(z)dz&=&0,\label{psip_psi}\\
\int_{-\infty}^0 \psi''(z)\psi(z)dz&=&-\lambda_1\lambda_2=B,\label{psipp_psi}
\eea
which hold for both real and complex conjugate $\lambda_{1,2}$.

We note that, for a slave-boson like renormalization $t_f\rightarrow b^2 t_f$, $V \rightarrow bV$, $b<1$,
$D_1D_2$ is not affected, see Eq. \eqref{d1d2}, while $B$ only slightly, and in the limit $b^2t_f\rightarrow 0$ it becomes
$B=(\epsilon_1^f-\epsilon_1^d)/t_dg_1^d$, see Eq. \eqref{bc}.
If we insert numerical values from Appendix A, and we take $\epsilon_1^f-\epsilon_1^d=1.6$eV,
corresponding to the minimum of the $d$ band with respect to the Fermi energy, we get $D_1D_2=-1.99$, $B=-1.67$.
This leads to $\lambda_{1,2}=(0.71\pm\ii 1.08)$,
which shows that we are in the regime in which the solutions are complex conjugate, hence the spatial part of the wavefunction oscillates, and is:
\be
\psi_{\bar\Gamma}(z)=2.01 e^{0.71 z} \sin(1.08 z),
\ee
where $z$ is expressed in units of $a_0$, the lattice spacing, that we set to 1 everywhere.
This holds at $\bar\Gamma$ for a $(001)$ surface.

At $\bar X$, as shown in the main text, we have to substitute $f_1^v\rightarrow h_1^v$, $g_1^a\rightarrow l_1^a$, $a=d/f$.
This leads to the new values of $D_1D_2=-4.65$, $B=-1.57$,
$\lambda_{1,2}=(1.08\pm\ii 0.64)$, and:
\be
\psi_{\bar X}(z)=4.08 e^{1.08 z} \sin(0.64 z)
\ee

If, instead, we use $\Gamma_7$ states, at $\bar\Gamma$, $f_1^v\rightarrow f_7^v$, $g_1^f\rightarrow g_7^f$,
we find $D_1D_2=-0.76$, $B=-1.67$, $\lambda_{1,2}=(0.43\pm i 1.21)$,
\be
\psi_{\bar\Gamma}^{\Gamma_7}(z)=1.40 e^{0.43 z} \sin(1.21 z),
\ee
and at $\bar X$, $f_1^v\rightarrow h_7^v$, $g_1^f\rightarrow l_7^f$, $D_1D_2=-1.14$, $B=-1.57$, $\lambda_{1,2}=(0.53\pm i 1.13)$,
\be
\psi_{\bar\Gamma}^{\Gamma_7}(z)=1.61 e^{0.53 z} \sin(1.13 z).
\ee

\begin{figure}[tb]
\includegraphics[width=0.48\textwidth]{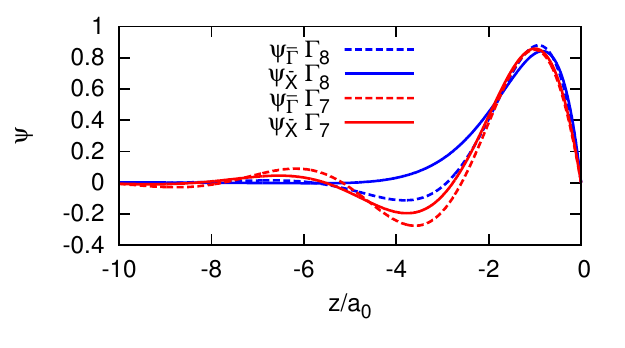}
\caption{Spatial part of the wavefunction $\psi(z)$ at $\bar\Gamma$ and $\bar X$ on a $(001)$ surface,  in our approximation with realistic parameters from Ref. \onlinecite{prbr_io_smb6};
we consider either the $\Gamma_8$ or the $\Gamma_7$ multiplets.
}\label{fig_psi}
\end{figure}

These are shown in Fig. \ref{fig_psi}.
We notice that in all cases the wavefunction has a maximum at about $z=-1$, and decays quickly within a few layers.
This justifies the argument that a perturbation in the first layer, for example Kondo breakdown\cite{kondo_breakdown} or surface reconstruction,\cite{smb6_rec_io}
should have a significant impact on the wavefunction and of the dispersion of surface states.
However, we stress that the exact $z$-dependence of the wavefunction does not affect the results of the main text.

\section{Third-order corrections to the velocities}\label{app_third}
We note that our theory is in general not enough to give the exact value of the velocities $v_0$, $v_1$, $v_2$.
We can realize this by noting that cubic terms proportional to $k_z^2 k_{x,y}$ in the hybridization, that we have ignored in the main text, will give rise to linear terms in $k_{x,y}$ once $k_z\rightarrow -id/dz$,
and the same holds for $k_x^2 k_{y,z}$ once $k_x\rightarrow -id/dx$. The same will happen for higher powers in the momentum $\bk$.
These higher-order terms come from next-nearest neighbors (NNs): $1st$ NN terms only give rise to linear terms (for which our theory is exact) and to terms of the form $k_{x,y,z}^{2n+1}$ (which do not create linear correction terms),
while $2nd$ NN to the mentioned mixed cubic terms, and so on.
This makes a general treatment quite involved. We can, however, restrict our analysis up to $2nd$ NN hybridization, and in particular to terms $\eta_x^{v1}$, $\eta_z^{v1}$, $\eta_z^{v2}$,
to find:
\bea
f_1^v&\rightarrow& 2\eta_x^{v1}+6\eta_z^{v2}\left(1-\frac{k^2_\parallel}{4}\right)\equiv f_1^v+f_1' k^2_\parallel,\label{f1etav_new}\\
h_1^v&\rightarrow& -\frac{1}{2}\eta_x^{v1}-\frac{3}{2}\eta_z^{v1}+3\eta_z^{v2}\left(1-\frac{k^2_z}{2}\right)\equiv h_1^v+h_1' k^2_z,\label{h1etav_new}
\eea
with $f_1'=h_1'=-3\eta_z^{v2}/2$.
These expressions, which come respectively from the expansion of $(\cos k_x+\cos k_y)$ and $\cos k_z$, see Ref. \onlinecite{prbr_io_smb6}, must be inserted into Eq. {\eqhlarge} of the main text.
Here we ignore term $\eta_x^{v2}$ since it introduces an even more complicated  momentum dependence in $(k_+^2-k_-^2)k_+$ for $h_1^v$, and it is anyway smaller than the other three terms \cite{prbr_io_smb6}.
Now, for the $\bar\Gamma$ cone on the $(001)$ surface, $k_z\rightarrow -id/dz$, so we can ignore the quadratic term in $k^2_\parallel$ in $f_1^v$, while, in $h_1^v$, $k_z^2$ becomes:
\be
k_z^2\rightarrow -\frac{d^2}{dz^2}=-B>0,
\ee
from Eq. \eqref{psipp_psi}, with $B\sim -1$ from Eq. \eqref{bc}, so we get:
\be
h_1^v\rightarrow -\frac{1}{2}\eta_x^{v1}-\frac{3}{2}\eta_z^{v1}+3\eta_z^{v2}\left(1+\frac{B}{2}\right)=h_1^v-h_1'B,\label{h1etav_new_gamma}
\ee
which must be inserted in Eq. {\eqvzero} of the main text to get the right value of the velocity $v_0$.
For the $\bar X$ cone on the same surface, $k_x\rightarrow -id/dx$, so we can ignore the quadratic terms in $k^2_z$ in $h_1^v$ and in $k^2_y$ in $f_1^v$, while, in $f_1^v$, $k_x^2$ becomes:
\be
k_x^2\rightarrow -\frac{d^2}{dz^2}=-B'>0,
\ee
where $B'$ is obtained by $B$ from Eq. \eqref{bc}  with the usual substitution $g_1^{d,f}\rightarrow l_1^{d,f}$.
Hence the value of $v_2$, Eq. {\eqvtwo} of the main text, is unchanged, while for $v_1$, Eq. {\eqvone}, we have to substitute
\bea
f_1^v&\rightarrow& 2\eta_x^{v1}+6\eta_z^{v2}\left(1+\frac{B'}{4}\right)=f_1^v-f_1'B'.\label{f1etav_new_x}
\eea
With these corrections we get the exact values of the velocities, as drawn in Fig. 3 of the main text;
we note that, since $B,B'<0$, these corrections reduce the value of the velocities, by a factor which can be as large as~$\sim 2$.

For a general surface, things are more complicated.
On general grounds, linear terms in $H_P$ proportional to $\bar k_{x,y}$ will be of the form $\bar k_{x,y}\sum_n c_n \bar k_z^n$,
with even powers of $\bar k_z$ coming from the hybridization, and odd powers from the kinetic energy.
Once $\bar k_z \rightarrow -i d/dz$ the $n=1$ term vanishes from Eq. \eqref{psip_psi}, but all the other terms survive, and all contribute to the Dirac velocity.
In addition, we will have terms $\sum_n d_n \bar k_z^n$ which belong to $H_0$.

These terms come from the same terms we just considered, but, with respect to the $(001)$ surface,
we will also have terms coming from higher terms in the expansion of $\sin k_{i}=\pm(k_i-k_i^3/6+\ldots)$, $i=x,y,z$,
which give rise to linear terms in $\bar k_{x,y}$ when rotated.
As a consequence, a general theory becomes rather involved, and we stick to the linear order in Figs. 10, 11 and 12 of the main text. 



\section{Numerical \kdp method for slab geometry}\label{app_kdp_slab}

In Figure 3 of the main text, together with the numerical solution of the tight-binding model and the analytical results described in detail in the paper,
we also present the numerical solution of the \kdp Hamiltonian in a slab configuration with $N$ layers for the $E_g$ - $\Gamma_8$ basis. Here we quickly describe this numerical bulk \kdp calculation, which loosely follows Ref. \onlinecite{yu_smb6_qpi}.

Around the $\bar\Gamma$ cone, with the usual substitution $k_z\rightarrow -\ii d/dz$ in Eq. {\eqh} of the main text, the operator $d/dz$ now acts on the basis $\psi_n(z)=\sin [\pi n z /(N+1)]$ (up to a normalization factor),
where $z=1,...,N$ is the layer index and $n=1,..,N$ is a quantum number.
The kinetic energy gives a diagonal term $\langle\psi_n|-d^2/dz^2|\psi_n\rangle=2-2\cos[\pi(n+1)/(N+1)]$,
while the hybridization gives non-diagonal terms $\propto \langle\psi_m|d/dz|\psi_n\rangle$ which are evaluated numerically.
This grants that the energies at $\bar \Gamma$ are exactly the same as those of the tight-binding model.
We then take terms in $k_{x,y}$ as a perturbation.
We finally numerically diagonalize the resulting Hamiltonian, of size $8N$ ($10N$ if one had to consider $\Gamma_7$ states as well), at each momentum close to $\bar \Gamma$ to obtain the dispersion.
We note that to obtain the exact value of the Dirac velocities we need to retain the third-order terms as described in the previous Section.

Around the $\bar X$ cone, we follow the same method but with $k_x\rightarrow -\ii d/dx$, and terms in $k_{y,z}$ as a perturbation.

\section{Calculation of the SEV}\label{app_sev}
To determine the SEV $\langle \vec {\hat S} \rangle$ on our basis, we must trace out the orbital degree of freedom; this is trivial in the $d$ shell, since states are direct product
of a spin and an orbital part, which is no more true in the $f$ shell, due to the spin-obit coupling.
In this latter case we obtain the results of Table \ref{table_spin}.\cite{prbr_io_smb6}
In Table \ref{table_pspin} we report our definition of the pseudospin $\vec{\hat\sigma}$.

The same can be done for the angular momentum $\langle \vec  {\hat L} \rangle$: for $d$ states it is completely quenched, since we only take the $E_g$ multiplet;
while for $f$ states we obtain that its expectation value is $-8$ times the SEV.
This is true in any basis, for example for $j=5/2$, $j_z=-5/2,\dots,5/2$ states:
\bea
|j_z\rangle=\sqrt{\frac{1}{2}-\frac{j_z}{7}}|j_z-\frac{1}{2},\uparrow\rangle - \sqrt{\frac{1}{2}+\frac{j_z}{7}}|j_z+\frac{1}{2},\downarrow\rangle
\eea
we find:
\bea
\langle j_z|  {\hat L}_z|j_z \rangle=\frac{8}{7}j_z,\hspace{10pt}
\langle j_z|  {\hat S}_z|j_z \rangle=-\frac{1}{7}j_z,
\eea
and similar (matrix) relations hold for $ {\hat L}_x$, $ {\hat S}_x$ and $ {\hat L}_y$, $ {\hat S}_y$ (one can build a matrix like the one in Table \ref{table_spin} substituting $\vec  {\hat S}$ with $\vec  {\hat L}$ and multiplying all entries  by $-8$).

As a consequence, for $d$ states we find the following rules:
\bea
\langle \vec  {\hat L}^d \rangle&=&0,\\
\langle \vec  {\hat J}^d \rangle=\langle \vec  {\hat L}^d \rangle+\langle \vec  {\hat S}^d \rangle&=&\langle \vec  {\hat S}^d \rangle,\\
\langle \vec  {\hat L}^d+2\vec  {\hat S}^d \rangle&=&2 \langle \vec  {\hat S}^d \rangle,
\eea
while for $f$ states:
\bea
\langle \vec  {\hat L}^f \rangle&=&-8 \langle \vec  {\hat S}^f \rangle,\\
\langle \vec  {\hat J}^f \rangle&=&-7\langle \vec  {\hat S}^f \rangle,\\
\langle \vec  {\hat L}^f+2\vec  {\hat S}^f \rangle&=&-6 \langle \vec  {\hat S}^f \rangle.
\eea
For a generic state with $\alpha^2$ weight on $d$ states, and $\beta^2$ weight on $f$ states ($\alpha^2+\beta^2=1$):
\bea
\langle \vec  {\hat S} \rangle&=& \langle \vec  {\hat S}^d \rangle \alpha^2 + \langle \vec  {\hat S}^f \rangle \beta^2,\\
\langle \vec  {\hat L} \rangle&=& -8 \langle \vec  {\hat S}^f \rangle \beta^2,\\
\langle \vec  {\hat J} \rangle&=&2 \langle \vec  {\hat S}^d \rangle \alpha^2 -7 \langle \vec  {\hat S}^f \rangle \beta^2,\\
\langle \vec  {\hat L}+2\vec  {\hat S} \rangle&=&2 \langle \vec  {\hat S}^d \rangle \alpha^2 -6 \langle \vec  {\hat S}^f \rangle \beta^2,
\eea
which shows that the expectation value of operator $\vec  {\hat L}+2\vec  {\hat S}$ must be comprised between $-6\langle\vec  {\hat S} \rangle$ (pure $f$ states)
and $2\langle\vec  {\hat S} \rangle$ (pure $d$ states).

In Ref. \onlinecite{yu_smb6_qpi} these quantities were computed for surface states using an ab-initio approach;
for in plane components authors find on the $\bar\Gamma$ cone:
\bea
\langle  {\hat L}_{x,y}^{\bar\Gamma}+2 {\hat S}_{x,y}^{\bar\Gamma} \rangle&=&-5.98 \langle  {\hat S}_{x,y}^{\bar\Gamma} \rangle,
\eea
which is in excellent agreement with our results for $f$ states, and the small difference is likely due to $d$ weight on surface states;
on the $\bar X$ cone:
\bea
\langle  {\hat L}_{x}^{\bar X}+2 {\hat S}_{x}^{\bar X} \rangle&=&-5.11 \langle  {\hat S}_{x}^{\bar X} \rangle,\\
\langle  {\hat L}_{y}^{\bar X}+2 {\hat S}_{y}^{\bar X} \rangle&=&-5.43 \langle  {\hat S}_{y}^{\bar X} \rangle,
\eea
where the larger disagreement with $f$ state results can be interpreted as a larger $d$ weight on surface states,
or as the influence of the $j=7/2$ multiplet, for which spin and angular momentum are parallel instead of antiparallel.
For the out-of-plane component things are similar on the $\bar X$ cone:
\bea
\langle  {\hat L}_z^{\bar X}+2 {\hat S}_z^{\bar X} \rangle&=&-5.27 \langle  {\hat S}_z^{\bar X} \rangle,
\eea
but the disagreement is larger on the $\bar \Gamma$ cone:
\bea
\langle  {\hat L}_z^{\bar \Gamma}+2 {\hat S}_z^{\bar \Gamma} \rangle&=&-7.47 \langle  {\hat S}_z^{\bar \Gamma} \rangle,
\eea
and cannot be explained by our theory, probably due to the small basis used.

We remark that all these calculations involve single-particle states, hence do not include many-particles effects.

\begin{table}
$$
\begin{array}{|c|cc|}\hline
(\hat\sigma_x,\hat\sigma_y, \hat\sigma_z)&\Gamma_8^{(1)}+ & \Gamma_8^{(1)}-\\\hline
\Gamma_8^{(1)}+ &(0,0,-1) & (-1,i,0) \\
\Gamma_8^{(1)}- &(-1,-i,0) & (0,0,1) \\\hline
  \end{array}
$$
$$
\begin{array}{|c|cc|}\hline
(\hat\sigma_x,\hat\sigma_y,\hat\sigma_z)& \Gamma_7+ & \Gamma_7-\\\hline
\Gamma_7+ &(0,0,-1) & (-1,i,0) \\
\Gamma_7- &(-1,-i,0) & (0,0,1) \\\hline
  \end{array}
$$
\caption{Definition of the pseudospin in the $\Gamma_8^{(1)}$ basis (above) and in the $\Gamma_7$ basis (below).}\label{table_pspin}
\end{table}

\begin{table*}
$$
\begin{array}{|c|cccc|cc|}\hline
2( {\hat S}^x, {\hat S}^y,  {\hat S}^z)&\Gamma_8^{(1)}+ & \Gamma_8^{(1)}- & \Gamma_8^{(2)}+ & \Gamma_8^{(2)}- & \Gamma_7 +& \Gamma_7 -\\\hline
\Gamma_8^{(1)}+&\frac{11}{21}(0,0,-1) & \frac{5}{21}(-1,i,0) & (0,0,0) & \frac{2\sqrt{3}}{21}(-1,-i,0) & \frac{4\sqrt{5}}{21}(0,0-1)&\frac{{2\sqrt{5}}}{21} (1,-i,0)\\
\Gamma_8^{(1)}-&\frac{5}{21}(-1,-i,0) & \frac{11}{21} (0,0,1) & \frac{2\sqrt{3}}{21} (-1,i,0) & (0,0,0)&\frac{2\sqrt{5}}{21}(1,i,0)&\frac{4\sqrt{5}}{21}(0,0,1)\\
\Gamma_8^{(2)}+&(0,0,0) & \frac{2\sqrt{3}}{21}(-1,-i,0) & \frac{1}{7} (0,0,-1) & \frac{3}{7}(-1,i,0)& (0,0,0)& \frac{{2}\sqrt{15}}{21}(1,i,0)\\
\Gamma_8^{(2)}-&\frac{{2}\sqrt{3}}{21}(-1,i,0)&(0,0,0)&\frac{3}{7} (-1,-i,0)&\frac{1}{7} (0,0,1) &\frac{{2}\sqrt{15}}{21}(1,-i,0)&(0,0,0)\\\hline
\Gamma_7+&\frac{4\sqrt{5}}{21}(0,0,-1)&\frac{2\sqrt{5}}{21} (1,-i,0) &(0,0,0)&\frac{{2}\sqrt{15}}{21}(1,i,0)&\frac{5}{21}(0,0,1)&\frac{5}{21}(1,-i,0)\\
\Gamma_7-&\frac{2\sqrt{5}}{21}(1,i,0)&\frac{4\sqrt{5}}{21}(0,0,1)&\frac{{2}\sqrt{15}}{21}(1,-i,0)&(0,0,0)&\frac{5}{21}(1,i,0)& \frac{5}{21}(0,0,-1)\\\hline
  \end{array}
$$
\caption{Expectation value of the spin in the $\Gamma_7-\Gamma_8$ basis.
}\label{table_spin}
\end{table*}


\section{Surface states for a generic surface}
In this Section we give details about the effective Hamiltonian for a generic surface.
\subsection{Rotations}
Given the $(lm/n)$ triplet and angles $\theta$ and $\phi$ from Eqs. {\eqtheta} and {\eqphi} of the main text,
we can perform a rotation in momentum space with Euler angles $\omega$, $\omega'=\theta$, $\omega''=\phi$,
to obtain ($c_\theta\equiv \cos\theta$, $s_\theta\equiv \sin\theta$ and similar for $\omega$ and $\phi$):
\bea
\bar k_x &=& (c_\omega c_\theta c_\phi-s_\omega s_\phi)k_x+(c_\omega c_\theta s_\phi+s_\omega c_\phi)k_y-c_\omega s_\theta k_z, \nonumber\\
\bar k_y &=& -(c_\omega s_\phi+s_\omega c_\theta c_\phi)k_x+(c_\omega c_\phi-s_\omega c_\theta s_\phi)k_y+s_\omega s_\theta k_z, \nonumber\\
\bar k_z &=& s_\theta c_\phi k_x + s_\theta s_\phi k_y + c_\theta k_z,\label{kbarkomega}
\eea
which can be inverted to give:
\bea
k_x &=& (c_\omega c_\theta c_\phi-s_\omega s_\phi)\bar k_x-(c_\omega s_\phi+s_\omega c_\theta c_\phi)\bar k_y+s_\theta c_\phi \bar k_z, \nonumber\\
k_y &=& (c_\omega c_\theta s_\phi+s_\omega c_\phi)\bar k_x+(c_\omega c_\phi-s_\omega c_\theta s_\phi)\bar k_y+s_\theta s_\phi \bar k_z, \nonumber\\
k_z &=& -c_\omega s_\theta \bar k_x + s_\omega s_\theta \bar k_y + c_\theta \bar k_z.
\eea
In the main text we take take $\omega=0$ to get:
\bea
\bar k_x &=& k_x\cos \theta \cos \phi +k_y\cos\theta \sin\phi -k_z\sin\theta, \nonumber\\
\bar k_y &=& -k_x\sin\phi + k_y\cos\phi , \nonumber\\
\bar k_z &=& k_x\sin\theta \cos\phi  + k_y\sin\theta \sin\phi  + k_z\cos\theta,\label{kbark}
\eea
and:
\bea
k_x &=& \bar k_x\cos \theta \cos \phi - \bar k_y\sin\phi +\bar k_z\sin\theta\cos\phi , \nonumber\\
k_y &=& \bar k_x\cos\theta\sin\phi + \bar k_y\cos\phi +\bar k_z\sin\theta\sin\phi , \nonumber\\
k_z &=& -\bar k_x\sin\theta   + \bar k_z\cos\theta .\label{kkbar}
\eea

This shows that the $X=(0,0,\pi)$ point is projected at
\be
\bar\bk_X=(-\pi\sin\theta,0),\label{barbk}
\ee
so $\bar k_x$ is the direction which joins $\bar \Gamma$ to the position of the cone, unless $\theta=0$, which corresponds to the $\bar\Gamma$ cone on the (001) surface,
for which $\bar k_x$ and $\bar k_y$ directions are equivalent.

We can also notice from Eq. \eqref{kbark} that $X'=(\pi,0,0)$ and $X''=(0,\pi,0)$
are projected respectively to:
\bea
\bar\bk_{X'}&=&\pi(\cos\theta\cos\phi,-\sin\phi),\label{barbkp}\\
\bar\bk_{X''}&=&\pi(\cos\theta\sin\phi,\cos\phi),\label{barbkpp}
\eea
which are the positions of the two other Dirac cones in the BZ when $\omega=0$.

However, 
we do not want to describe these two other cones by projecting $X'$ and $X''$,
but, instead, always projecting $X$, since this allows to safely neglect subspace~2.

As a consequence, we now require  $X$ to be projected at $\bar\bk_{X'}$ with angles $\omega',\theta',\phi'$,
and at $\bar\bk_{X''}$ with angles $\omega'',\theta'',\phi''$;
from Eq. \eqref{kbarkomega} we find that, when $\omega\ne 0$, $X$ is projected at $\pi(-\cos\omega\sin\theta,\sin\omega\sin\theta)$.
So for $\bar\bk_{X'}$ we get
\bea
-\cos\omega'\sin\theta'&=&\cos\theta \cos\phi,\nonumber\\
\sin\omega'\sin\theta'&=&-\sin\phi,
\eea
which gives:
\bea
\tan^2 \omega'=\tan^2\phi(\tan^2\theta\!+\!1)=\frac{m^2(l^2+m^2+n^2)}{l^2n^2},\\
\tan^2 \theta'=\tan^2\phi\cot^2\theta+\tan^2\phi+\cot^2\theta=\frac{m^2+n^2}{l^2}
\eea
with  $-\pi\le\omega'\le-\pi/2$, $0\le\theta'\le\pi/2$.
Using our notation, this corresponds to a $(mn/l)$ triplet.

For $\bar\bk_{X''}$ we get
\bea
-\cos\omega''\sin\theta''&=&\cos\theta \sin\phi,\\
\sin\omega''\sin\theta''&=&\cos\phi,
\eea
which gives:
\bea
\tan^2 \omega''=\cot^2\phi(\tan^2\theta+1)=\frac{l^2(l^2+m^2+n^2)}{m^2n^2},\nonumber\\
\tan^2 \theta''=\cot^2\phi\cot^2\theta+\cot^2\phi+\cot^2\theta=\frac{n^2+l^2}{m^2}
\eea
with $\pi/2\le\omega''\le\pi$ , $0\le\theta''\le\pi/2$.
This corresponds to a $(nl/m)$ triplet.



This means that we can then apply the theory of the following Subsection using $\theta$ to get the effective Hamiltonian at $\bar\bk_X$,
$\theta'$ for $\bar\bk_{X'}$ and $\theta''$ for $\bar\bk_{X''}$.


Angles $\omega=0$, $\omega'$ and $\omega''$ tell how we have to rotate our system in the surface plane to get the same
coordinate system for all cones. If we ignore this, every cone will have its own coordinates,
with $\bar k_x$ joining the position of the cone to the center of the BZ.


\begin{figure}[!t]
\includegraphics[width=0.49\textwidth]{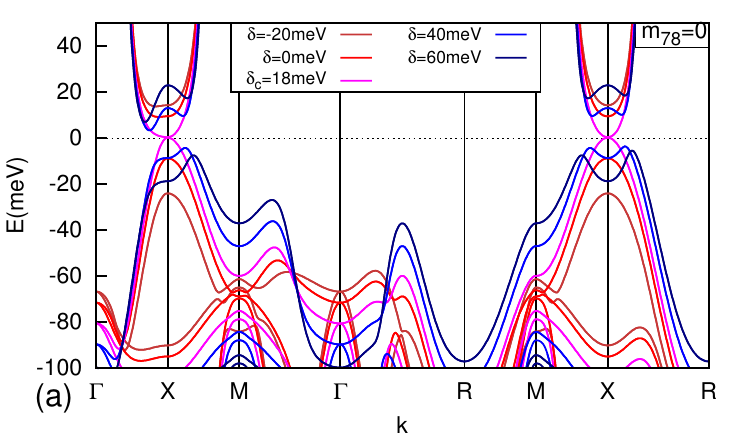}
\includegraphics[width=0.49\textwidth]{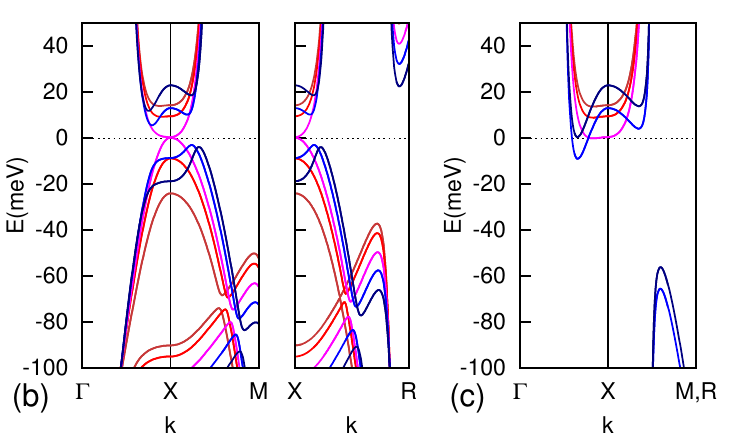}
\caption{Same as Fig. 17 of the main text but with $\eta_{x7}^{f2}=0$, i.e. without any direct coupling between the $\Gamma_7$ and $\Gamma_8$ subspaces, which leads to $m_{78}=0$ in the \kdp Hamiltonian.
In this case the gap closes at $X$.
}\label{fig_tpt4}
\end{figure}

\begin{figure}[!t]
\includegraphics[width=0.49\textwidth]{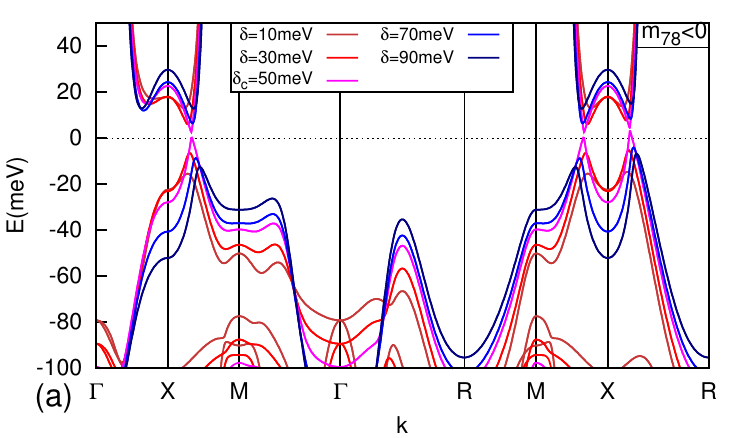}
\includegraphics[width=0.49\textwidth]{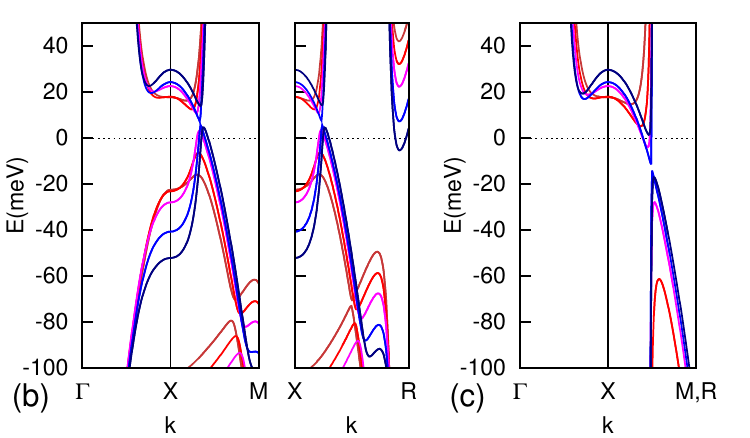}
\caption{Same as Fig. 17 of the main text but with $\eta_{x7}^{f2}=-0.16$, which leads to $m_{78}<0$ in the \kdp Hamiltonian.
In this case the gap closes along $X$-$M$ and $X$-$R$ approximatively at the same energy, which is slightly incorrect in the \kdp method.
}\label{fig_tpt5}
\end{figure}

\subsection{Details of the calculation}
In this Subsection we find the effective surface Hamiltonian for a given $(lm/n)$ triplet;
this Hamiltonian is valid for small momenta around the surface point $\bar\bk_X$ on which the bulk $X=(0,0,\pi)$ point is projected.

We concentrate on subspace~1, so ignoring any coupling to subspace~2, and use only $\Gamma_8$ states,
so the bulk \kdp Hamiltonian is $4\times 4$,  and in the $|d^1\uparrow\rangle$, $|d^1\downarrow\rangle$, $|f^1+\rangle$, $|f^1-\rangle$ basis reads:
\bea
H=
\left(
\begin{array}{cc}
\epsilon^d_1(\bk)\sigma_0 & -\ii V H_h\\
\ii V H_h^\dagger &\epsilon^f_1(\bk)\sigma_0
\end{array}
\right),\label{hkdotp_details}\\
H_h=h_1^v (k_x\sigma_x + k_y\sigma_y) + f_1^v k_z\sigma_z.
\eea

When we perform the rotation from $\bk$ to $\bar\bk$ the kinetic energy becomes:
\be
\epsilon^a_1[\bk(\bar\bk)]=\epsilon_1^a-t_a(g_1^a k^2_z(\bar \bk)+ l_1^a k^2_\parallel(\bar \bk)),\hspace{5pt} a=d/f,\label{kinkbar}
\ee
and the hybridization:
\be
H_h=h_1^v [k_x(\bar\bk)\sigma_x + k_y(\bar\bk)\sigma_y] + f_1^v k_z(\bar\bk)\sigma_z,\label{hybrkbar}
\ee
where one has to express $\bk$ as a function of $\bar \bk$.

When we use Eq. \eqref{kkbar}, for the kinetic energy we find:
\bea
\epsilon^a_1[\bk(\bar\bk)]&=&\epsilon_1^a-t_a[g_1^a (\bar k_x\sin\theta-\bar k_z\cos\theta)^2\nonumber\\
 &+&l_1^a(\bar k_y^2+ (\bar k_x\cos\theta+\bar k_z\sin\theta)^2)].\label{hkinbar}
\eea

Upon substitution of Eqs. \eqref{kkbar} into Eq. \eqref{hybrkbar}, we see that the hybridization term in $\bar k_z$ contains terms in $\sigma_x$, $\sigma_y$, and $\sigma_z$.
However, we would like it to be proportional just to $\sigma_z$, so that we can follow what we did in Section \ref{app_gamma}.
We see that this is achieved if we perform a Wigner transformation in the pseudospin space with angles $\omega$, $\omega_1'=\arctan[\tan\theta (h_1^v/f_1^v)]$, $\omega''=\phi$;
we will denote this transformation matrix as $U$.
This corresponds to the same rotation that we did in the momentum space, except for angle $\omega_1'$, which now depends on $f_1^v$ and $h_1^v$, and that we take between 0 and $\pi$.
The total rotation matrix is
\be
U_t=
\left(
\begin{array}{cc}
U & 0\\
0 &U
\end{array}
\right),\label{u_tot}
\ee
with
\be
U=
\left(
\begin{array}{cc}
 e^{\ii \phi/2}\cos(\omega_1'/2)&e^{-\ii \phi/2}\sin(\omega_1'/2) \\
 -e^{\ii \phi/2}\sin(\omega_1'/2)&e^{-\ii \phi/2}\cos(\omega_1'/2)
\end{array}
\right).\label{u_rot}
\ee


In the new basis the hybridization Eq. \eqref{hybrkbar} takes the simpler form:
\bea
H_h=(\bar h_{1xx}^v \sigma_x+\bar h_{1xz}^v \sigma_z)\bar k_x + h_1^v\sigma_y \bar k_y+\bar f_1^v \sigma_z\bar k_z
\eea
with
\bea
\bar h_{1xx}^v&=&\frac{f_1^v |h_1^v|}{\sqrt{(f_1^v)^2\cos^2\theta+(h_1^v)^2\sin^2\theta}},\\
\bar h_{1xz}^v&=&\frac{[(h_1^v)^2-(f_1^v)^2]\sin\theta\cos\theta}{\sqrt{(f_1^v)^2\cos^2\theta+(h_1^v)^2\sin^2\theta}}\sgn(h_1^v),\\
\bar f_{1}^v&=&\sqrt{(f_1^v)^2\cos^2\theta+(h_1^v)^2\sin^2\theta}\sgn(h_1^v),\label{barf1v}
\eea
while the kinetic energy, being the identity in the (pseudo)spin index, is left unchanged.
In this way, when we set $\bar k_x=\bar k_y=0$,
the only hybridization term left is in $\sigma_z$,
the Hamiltonian $H_0$ splits into two $2\times 2$ blocks,
and we can repeat the treatment of Section \ref{app_gamma}.

In the kinetic energy part Eq. \eqref{hkinbar}, we obtain that the terms multiplying $\bar k_z^2$ are:
\be
\bar g_1^a=g_1^a\cos^2\theta+l_1^a \sin^2\theta<0, \hspace{5pt} a=d/f,\label{bar_g1a}
\ee
and the only information we needed is that they are both negative, $\bar g_1^a<0$, since $g_1^a<0$, $l_1^a<0$ (see Section \ref{app_gamma}).

With the substitutions $g_1^a\rightarrow \bar g_1^a$, $f_1^v\rightarrow \bar f_1^v$,
the basis at $\bar k_\parallel=0$ is
\bea
|\bar\psi_+\rangle&=&\alpha|\bar d^1\uparrow\rangle+\beta|\bar f^1+\rangle,\nonumber\\
|\bar\psi_-\rangle&=&\alpha|\bar d^1\downarrow\rangle-\beta|\bar f^1-\rangle,\label{psipmbar}
\eea
with $|\bar d\rangle= U |d\rangle$, $|\bar f\rangle= U |f\rangle$.
The effective Hamiltonian up to the linear term in $\bar \bk_\parallel$, Eq. \eqeffh of the main text, is:
\bea
H^{eff}_{\theta}&=&-2\alpha\beta V h_1^v\left(\frac{f_1^v}{\bar f_1^v}\bar k_x \hat s_y - \bar k_y \hat s_x \right)\nonumber\\
&\equiv&v_1 \bar k_x \hat s_y - v_2 \bar k_y \hat s_x\nonumber\\
&=&|v_1|w  \bar k_x \hat s_y - |v_2| \bar k_y \hat s_x.\label{hefflmn}
\eea

There is now an additional term coming from $\bar k_z \bar k_x$ of the kinetic energy Eq. \eqref{hkinbar}:
however, with the substitution $\bar k_z \rightarrow -i d/dz$, it gives rise to an integral of the form $\int_{-\infty}^0 \psi(z) \psi'(z)dz\propto \psi^2(0)- \psi^2(-\infty)=0$,
see Eq. \eqref{psip_psi}.


\section{Further topological phase transitions}
In Figs. \ref{fig_tpt4} and \ref{fig_tpt5} we present the analogous of Fig. 17 of the main text but with $m_{78}=0$, and $m_{78}<0$.
It can be observed that this parameter controls where the gap closes, i.e. along $X-\Gamma$ for $m_{78}>0$,
along $X-R$ and $X-M$ for $m_{78}<0$, and at $X$ for $m_{78}=0$.



\section{Remarks on DFT results}

As noted in the main text, the available ab-initio results do not appear fully consistent regarding the question of whether the $\Gamma_7$ and $\Gamma_8$ multiplets, when taken alone, yield distinct topological phases. Hence, varying the relative multiplet energy drives a topological transition as discussed in Section~VII~B of the main text.

While this applies to the ab-initio results obtained for {\pu},\cite{pub6,prbr_io_smb6} the ab-initio-based \kdp expansion for {\sm} from Ref. \onlinecite{yu_smb6_qpi} appears problematic. We have analyzed their results, and upon varying the relative multiplet energy we do {\em not} find topological phase transition. This is what we expect when both multiplets, taken alone, realize the same topological phase, i.e. $\sgn(f_1^vh_1^v)=\sgn(f_7^vh_7^v)$,
and $m_{78}$ is such that $w$ from Eq. {\eqwoneseven} of the main text never vanishes (if $m_{78}$ has the opposite sign, a double transition $w=\pm 1 \leftrightarrow w=\mp 1 \leftrightarrow w=\pm 1$ could be in principle achieved).
Which phase is exactly realized depends on the exact knowledge of the basis used in Ref. \onlinecite{yu_smb6_qpi} (including phase factors); based on the reported spin structure (which disagrees with experiment\cite{smb6_arpes_mesot_spin})  we guess it is the $w=-1$ phase.

Possible explanations are that hybridization terms in {\sm} and {\pu} are so different as to place the two materials into two different topological phases when considering just one or both multiplets;
or that in particular cases the \kdp expansion is not good enough to describe the topological phase transition, for example because more terms must be kept in the small-momentum expansion;
it also possible that a particular choice of complex phases of the basis states has led us to wrong conclusions about that work.

For completeness we quote numerical values which we extracted from Ref.~\onlinecite{yu_smb6_qpi}:
$|f_1^v|=|c_2\cos\phi  + c_3\sin\phi|/2\pi V=0.79$,
$|h_1^v|=|c_1\cos\phi  + c_4\sin\phi |/2\pi V=0.73$,
$|f_7^v|=|-c_2\sin\phi  + c_3\cos\phi|/2\pi V=0.52$,
$|h_7^v|=|-c_1\sin\phi + c_4\cos\phi |/2\pi V=0.56$,
($\cos\phi=\sqrt{5/6}$, $\sin\phi=\sqrt{1/6}$, $c_i$'s are given numerically there),
when measured in units $V=0.1$eV. These are about one order of magnitude smaller than our values for {\pu} given in the Appendix of the main text, Eqs. {\eqfone}, {\eqfsix}.
This difference in the absolute value is most likely due to the renormalization by the Gutzwiller method and the fact that {\pu} has a larger $f$ kinetic energy than {\sm}, hence is likely to have also a larger hybridization. However, we stress that absolute values do {\em not} affect the spin structure. The latter instead hinges on signs, and this is where the above numbers would require a cross-check.

\bibliographystyle{apsrev4-1}
\bibliography{tki}